\documentclass[fleqn,usenatbib]{mnras}

\usepackage{newtxtext,newtxmath}

\usepackage[T1]{fontenc}

\DeclareRobustCommand{\VAN}[3]{#2}
\let\VANthebibliography\thebibliography
\def\thebibliography{\DeclareRobustCommand{\VAN}[3]{##3}\VANthebibliography}

\usepackage{graphicx}	
\usepackage{amsmath}	

\newcommand{\NII}{{[N\,{\sc ii}]}}

\newcommand{\OIII}{{[O\,{\sc iii}]}}

\newcommand{\OIIIs}{{[O\,{\sc iii}]\,}}

\newcommand{\FeII}{{[Fe\,{\sc ii}]\,}}
\newcommand{\Ha}{H$\alpha$}
\newcommand{\Has}{H$\alpha$\,}
\newcommand{\Hb}{H$\beta$}

\newcommand{\Hbs}{H$\beta$\,}

\title[The absorbers census]{An (in)complete NIRSpec census of Balmer absorption in Type 1 AGN -- radiation-driven outflows in little red dots, quasars and variable stars}

\author[Juodžbalis et al.]{
Ignas Juodžbalis,$^{1,2}$\thanks{E-mail: ij284@cam.ac.uk}
Xihan Ji,$^{1,2}$
Francesco D'Eugenio,$^{1,2}$
Jan Scholtz,$^{1,2}$
Roberto Maiolino,$^{1,2,3}$
\newauthor
Amanda Stoffers,$^{1,2}$
Alessandro Marconi,$^{4,5}$
Elena Bertola$^{5}$,
Andrew J.\ Bunker,$^{6}$
Stefano Carniani,$^{7}$
\newauthor
Giovanni Cresci,$^{5}$
Emma Curtis-Lake,$^{8}$
Zheng Ma,$^{9}$
Cosimo Marconcini,$^{4,5}$
Eleonora Parlanti,$^{7}$
\newauthor
Pierluigi Rinaldi,$^{10}$
Brant Robertson,$^{11}$
Hannah \"Ubler,$^{12}$
Giacomo Venturi,$^{5}$
Junyu Zhang$^{9}$
\\
$^{1}$Kavli Institute for Cosmology, University of Cambridge, Madingley Road, Cambridge, CB3 0HA, UK\\
$^{2}$Cavendish Laboratory, University of Cambridge, 19 JJ Thomson Avenue, Cambridge, CB3 0HE, UK\\
$^{3}$ Department of Physics and Astronomy, University College London, Gower Street, London WC1E 6BT, UK \\
$^{4}$ Dipartimento di Fisica e Astronomia, Università degli Studi di Firenze, Via G. Sansone 1,I-50019, Sesto Fiorentino, Firenze, Italy \\
$^{5}$ INAF - Osservatorio Astrofisico di Arcetri, Largo E. Fermi 5, I-50125, Firenze, Italy \\
$^{6}$ Department of Physics, University of Oxford, Denys Wilkinson Building, Keble Road, Oxford OX1 3RH, UK \\
$^{7}$ Scuola Normale Superiore, Piazza dei Cavalieri 7, I-56126 Pisa, Italy \\
$^{8}$ Centre for Astrophysics Research, Department of Physics, Astronomy and Mathematics, University of Hertfordshire, Hatfield AL10 9AB, UK \\
$^{9}$ Steward Observatory, University of Arizona, 933 N. Cherry Avenue, Tucson, AZ 85721, USA \\
$^{10}$ Space Telescope Science Institute, 3700 San Martin Drive, Baltimore, Maryland 21218, USA \\
$^{11}$ Department of Astronomy and Astrophysics University of California, Santa Cruz, 1156 High Street, Santa Cruz CA 96054, USA \\
$^{12}$ Max-Planck-Institut f\"ur extraterrestrische Physik (MPE), Gie{\ss}enbachstra{\ss}e 1, 85748 Garching, Germany
}

\date{Accepted XXX. Received YYY; in original form ZZZ}

\pubyear{\the\year{}}

\begin{document}
\label{firstpage}
\pagerange{\pageref{firstpage}--\pageref{lastpage}}
\maketitle

\begin{abstract}
A notable achievement of the first generation of JWST surveys was the discovery of an abundance of faint high-redshift ($z > 4$) broad-line active galactic nuclei (AGN) in the $L_{\rm bol} < 10^{45}$~erg~s$^{-1}$ regime, previously only accessible at $z < 1$. The high prevalence of absorption features in their broad hydrogen lines is one of the peculiarities of a significant fraction of this new population. In this paper, we conduct a broad census of Balmer absorption in a sample of 47 Type-1 AGN spanning $2 < z < 7$. Accounting for incompleteness of JWST spectroscopic data, we estimate that $44_{-6}^{+21}$~\% of Little Red Dots (LRDs) have absorption in \Has while the incidence rate is $< 25$~\% (at 2$\sigma$) in Little Blue Dots (LBDs). Additionally, R1000 JWST/NIRSpec data is strongly resolution limited, implying an inability to disentangle Doppler broadening in the absorption from optical depth effects. This is alleviated with R2700 observations, although some degeneracies remain. We find a significant correlation between the velocity of the Balmer absorption and the [OIII]5007 narrow line luminosity, suggesting a common driving mechanism. We do not find any other significant correlations (and do not confirm previously claimed correlations) between Balmer absorption and other spectral properties of LRDs (except for the expected correlation between \Has and \Hbs absorption velocities).
Comparing LRDs, quasars and stellar \Has absorbers, we establish that absorption in both LRDs and broad absorption line quasars is consistent with radiatively driven outflows, echoing the physics of variable stars yet occurring at vastly different scales.
\end{abstract}





\section{Introduction}
The discovery of a large population of faint active galactic nuclei (AGN) at high redshifts \citep[e.g.][]{Kocevski_AGN, Matthee2024, Maiolino24_GN-z11, Harikane2023} was one of the first accomplishments of early James Webb Space Telescope (JWST) surveys. However, this faint population quickly revealed itself to be peculiar relative to their more luminous counterparts. Despite showing clear broad hydrogen lines contributing well over 60\% of the total \Has flux \citep{Matthee2024, Juodzbalis2026, DEugenio2025b}, these sources appear to be nearly universally X-ray weak with non detections even in the deepest Chandra fields \citep{Ananna2024, Maiolino_xray_weak, Sacchi2025, Yue2024}. In addition, a substantial subset of these AGN exhibit `v' shaped continua -- with blue slopes in the UV contrasting with the red optical emission with the turnover always occurring around the Balmer break \citep{Setton2025}. Dubbed `Little Red Dots' (LRDs), these objects became a source of significant debate with early models proposing a mix of old stellar populations and different dust attenuation to explain the unusual continuum shapes \citep{Kocevski_AGN, Casey2024, Labbe2024}. However, these initial models struggled with the inferred compact sizes and large masses of LRDs, which implied extreme stellar densities if the `v' shaped continua were interpreted as stellar Balmer breaks. This was partially resolved by \cite{Inayoshi2024} demonstrating that any obscuring hydrogen gas at sufficiently high ($n_H \sim 10^9$~cm$^{-3}$) densities could produce a Balmer break in the transmitted continuum, while attenuating the X-rays produced by the central engine. This framework was quickly demonstrated to be successful at modeling LRD spectra without invoking massive stellar populations \citep{Ji2025}. This family of models attempt to reproduce LRD emission through invoking dense gas shells blanketing the central engine and resembling stellar atmospheres, dubbed the black hole star \citep[BH*;][]{Naidu2025, deGraaff2025}. While successful in reproducing LRD continuum shapes, these models raised their own tensions -- the implied extreme column densities ($N_H \approx 10^{25}$~cm$^{-2}$) suggest that the observed broad-lines widths may be the result of electron scattering effects rather than virial broadening, although this is in tension with the claim of a black-body-like emitting pseudo-atmosphere, unless this is clumpy or has openings
\citep{Ji2026QSO,Tang2026,Geris2026,Sok2026}. The finding that a significant fraction of LRDs (about 50\%) have exponential \Has profiles  \citep{Rusakov2025,Scholtz2026} has been used to suggest that black hole (BH) masses may be overestimated by an order of magnitude if standard virial calibrations are used \citep{Rusakov2025}. However, recent findings that exponential line profiles can be produced by stratification in the broad line region (BLR) by  \cite{Scholtz2026, Madau2026_wings} and a dynamical BH mass measurement of \cite{Juodzbalis2026qso1} contradict the idea that scattering in an optically thick medium is the main line broadening mechanism in LRDs. Nevertheless, the prevalence of Balmer absorption features in LRD spectra \citep{Matthee2026, Juodzbalis2024b, Lin2026} clearly indicates the presence of an optically thick medium in these sources. While this dense gas has often been associated with a pseudo-stellar atmosphere embedding the growing black hole, simpler models invoking the classical equatorial dusty obscuring medium around a highly accreting black hole have also been proposed \citep{Madau2026_LRD_LBD, Madau2026_LRD_LBD2}.

While X-ray attenuation in a Compton-thick medium remains an attractive explanation for the X-ray weakness of LRDs \citep{Comastri2026}, their blue counterparts -- Little Blue Dots (LBDs), defined in \cite{Brazzini2026, Geris2026} -- resemble regular Type 1 AGN in their UV--optical spectra yet share the extreme X-ray weakness of LRDs while appearing to lack the deep Balmer absorption lines of the latter. In addition, while the incidence rate of \Has absorption in LRDs is generally high, with recent estimates placing it as high as $60$\% \citep{Matthee2026, Lin2026}, deriving a precise occurrence rate is non-trivial due to inherent incompleteness in JWST spectroscopic data, most of which is taken at R1000 or even lower resolution which can miss rest-frame absorbers \citep{DEugenio2026}. In this paper we construct a sample of \Has absorbers from archival JWST data and obtain a completeness corrected estimate of the absorber incidence rate across both LRD and LBD populations. We then move on to investigating the ability of JWST data to recover the properties of the detectable \Has absorption features and review the demographics of our absorber sample, discussing their implications for the BH* and alternative models of LRDs. Lastly, we investigate the apparent similarities between LRDs and other broad-line emitters with Balmer absorption: 
FeLoBALs - a subset of broad absorption line (BAL) quasars (QSOs) \citep{Aoki2006,Choi2022, Leighly2025} and emission-line stars \citep{Leitherer1987, Genderen2001}.

The paper is organized as follows -- in Section~\ref{sec:data} we provide an overview of the data used as well as our absorber selection criteria, BLR fitting and LRD - LBD classification. Section~\ref{sec:incidence} presents our completeness simulations and parameter recovery tests as well as the completeness corrected \Has absorption incidence rates for LRDs, LBDs and the combined AGN population. Section~\ref{sec:demographics} presents an overview of the absorber demographics and their comparison to the BH* model predictions along with the observed non trivial correlations between the absorber properties. Section~\ref{sec:discussion} contains the extended discussion on the similarities and differences of LRDs and FeLoBALs as well as parallels between \Has absorption in AGN and emission line stars. Lastly, Section~\ref{sec:conclusions} presents our conclusions and future outlook. 

Throughout this work we assume a flat $\Lambda$CDM cosmology with $\Omega_m = 0.315$, $H_0 = 67.4$~km~s$^{-1}$~Mpc$^{-1}$ \citep{Planck2020}. All reported magnitudes are in the AB system \citep{oke+gunn1983}.

\section{Data description and sample selection}
\label{sec:data}
The data used in this study originates from three JWST NIRSpec MSA surveys - JADES (PIDs 1210, 1287, 1286, 1181 and 1180, \citealp{JADES_desc, Bunker2023_nirspec, DEugenio2024, Curtis-Lake2026}), CEERS (PID 1345, \citealp{Finkelstein2023}) and RUBIES (PID 4233, \citealp{deGraaf2025}) as well as the NIRSpec IFU surveys BlackTHUNDER (BT; PID 5015) 
and PID 5664 along with NIRSpec/IFU DDT PID 9433. 
In order to ensure homogeneity in the data, all spectra were processed with the JADES GTO pipeline (see \citealp{Scholtz2026_JADES} for details). For the IFU data, we performed optimal aperture extraction to obtain 1D integrated spectra directly comparable to those from MSA. In addition, as the IFU data reduction pipeline generally underestimates uncertainties, we rescaled the errors on the 1D spectra by a ratio between the mean error and the rms of the continuum residuals in line-free parts of the spectra, finding that the IFU errors are generally underestimated by a factor of 2-3.

As the main goal of this work is studying the incidence rate of absorption features in the broad H$\alpha$ line profiles of AGN, we only utilize R1000 and R2700 data. As RUBIES only utilized the F290LP/G395M grating, which covers $\lambda > 2.9$~$\mu$m, no AGN at $z < 3.4$ were selected from this survey. The procedure of selecting broad line emitters closely follows that of \cite{Juodzbalis2026}. We refit all spectra available, modelling the \Has line as a superposition of multiple Gaussians representing the narrow and broad \Has emission and the \NII\ doublet. To select a source as a broad line AGN, we require the Bayesian Information Criterion (${\rm BIC} \equiv \chi^2 + k\ln{n}$, where $k$ is the number of fitting parameters and $n$ is the number of data points) to differ by $>10$ between narrow-only and narrow+broad H$\alpha$ models. In addition, we discard sources which have a broad component in \OIII$\lambda$5007 that is kinematically matched to what is seen in \Ha, since for such objects, the broad line signal is likely produced by outflows. Unlike \cite{Juodzbalis2026}, we implement a stricter signal to noise\footnote{We define S/N as the ratio of the total line flux integrated within the line FWHM to the quadrature-combined error in the same range.} cut of 20 as below this threshold it becomes difficult to detect absorption features in the broad line profiles (see Section \ref{sec:completenes} for details). This selection procedure resulted in a parent sample of 38 AGN from the MSA surveys (20 from JADES, 17 from RUBIES and 1 from CEERS) as well as 9 additional sources from the IFU surveys that were not already covered by equivalent resolution MSA data, for a total of 47 targets, for which we performed a search of absorption features. In addition, we split the parent sample into LRD and non-LRD sources by employing the `v' shape selection criterion consistent with that of \cite{Scholtz2026, Brazzini2026}:
\begin{enumerate}
    \item $\beta_{UV} < -0.2$
    \item $\beta_{opt.} > 0$
\end{enumerate}
The slopes above were computed at rest-frame wavelengths of $1200$--$3600$~\AA\ for $\beta_{UV}$ and $3700$--$7000$~\AA\ for $\beta_{opt.}$ respectively, masking the flux at the locations of strong emission lines such as \Has and \OIII$\lambda\lambda$4959,5007.

In total, 35 sources in the parent sample are identified as LRDs, with the remaining 12 mostly consisting of LBDs, with the exception of JADES-GN-209777 -- a reddened quasar \citep[][Ma, in prep.]{Juodzbalis2026}. This skew towards LRDs likely results from a combination of selection bias, as LRDs are easy to select photometrically. Historically LRDs were selected via stringent photometric cuts requiring very red optical continua \citep{Hviding2025, Rinaldi2026}. As a result, JWST spectroscopic surveys, such as RUBIES, were optimized to target bright red sources rather than the general AGN population.  In addition, the LRD fraction appears to increase at higher AGN luminosities \citep{Hainline2025}.

While in the initial selection above, we approximated the BLR shape as a single Gaussian, high S/N data generally reveals the inherent non-Gaussianity of BLR profiles \citep[e.g.][]{Laor2006, Rusakov2025}. Indeed, for all of the parent sample sources, the single Gaussian fits leave considerable residuals in the wings of the line. Hence, when searching for absorbers, we employ three different more complex BLR models following the methods of \cite{Scholtz2026}:
\begin{itemize}
    \item Double Gaussian (2G) -- the line profile is modeled as a sum of two Gaussians with independent FWHM, but fixed redshift.
    \item Lorentzian (Lor) -- the line profile is modeled as a Lorentzian distribution with FWHM, redshift and normalization left as free parameters.
    \item Electron scattering (Sct) -- such a profile is produced by Thomson scattering off of free electrons and consists of a Gaussian core and exponential wings \citep{Laor2006, Rusakov2025}.  We model this profile as a Gaussian convolved with an exponential of the form:
    \begin{equation}
        E(\lambda) = e^{-|\lambda-\lambda_0|/W},
    \end{equation}
    where $\lambda_0$ is the central wavelength and $W$ the width of the exponential.
\end{itemize}
We note that the detailed exploration of BLR shapes of our objects is beyond the scope of this work and was a topic of \cite{Brazzini2025M, Brazzini2026, Scholtz2026}. Hence, in our final measurements, we characterized the best fit profiles by three parameters - redshift, total luminosity and total FWHM for self-consistent comparison between the best-fitting models. 

The absorption features themselves were modeled assuming a Gaussian optical depth profile \citep{Juodzbalis2024}:
\begin{equation}
\label{eq:depth_prof}
    \tau_{\lambda} = \tau_0\exp{\left[-0.5\left(\frac{\lambda - \lambda_0e^{\Delta v/c}}{\sigma}\right)^{2}\right]},
\end{equation}
where $\tau_0$ is the optical depth at the core of the line, $\Delta v$ is the velocity shift of the absorber and $\sigma$ is its velocity dispersion, coupled with a standard attenuation model:
\begin{equation}
\label{eq:absorption}
    f_{\lambda} = 1 - C_f + C_fe^{-\tau_{\lambda}},
\end{equation}
where $C_f$ is the covering fraction of the absorbing medium.

The fitting was carried out utilizing Bayesian methods, with the posteriors sampled using the \texttt{emcee} sampler \citep{emcee}. We utilized uninformative uniform priors on most of the fitted quantities with the exception of narrow and broad line velocity shifts, which were constrained to be within 100~km~s$^{-1}$ and 500~km~s$^{-1}$ of the values determined by prior visual inspection respectively. 

The absorbers were selected from the parent sample following the same $\Delta \rm BIC$ method as the initial AGN selection, requiring absorber models to improve the $\rm BIC$ by more than 5 to be selected. This was followed by visual inspection of the selected targets to exclude any data artifacts masquerading as absorption features like cosmic ray hits causing single pixel downturns that automated fitting could flag as absorption. In addition, we visually inspected the remainder of the sample to ensure that no absorption features were missed by the fitting procedure. The preferred BLR model was likewise determined by the BIC criterion, however, in cases where multiple BLR models had $\Delta\rm BIC < 5$ relative to each other, the posterior samples of all acceptable BLR models were combined in estimating the final properties of the absorbers and broad line profiles. Hence, the uncertainties on the measurements of some of our sources are considerably higher - reflecting the degeneracies induced by competing BLR models.

\begin{figure}
    \centering
    \includegraphics[width=\linewidth]{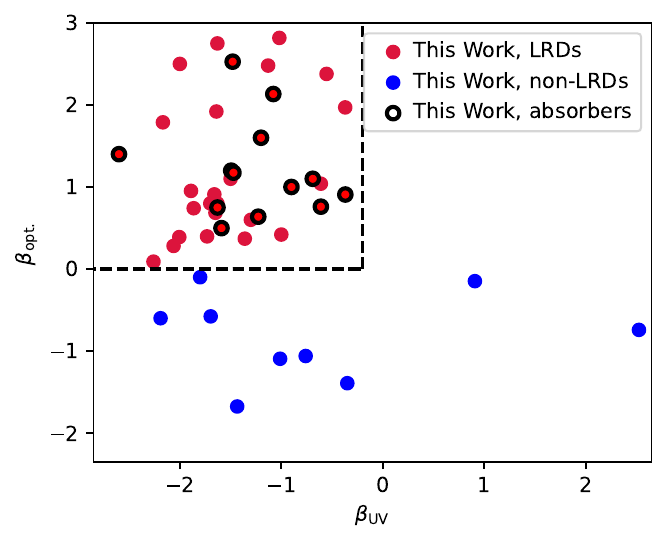}
    \caption{Showcase of the UV and optical slopes of the parent sample. The dashed lines denote the LRD region. Sources identified as LRDs are shown in red while LBDs - in blue. The objects with absorption features are highlighted with black edges. It can be seen that all absorbers lie in the LRD region.}
    \label{fig:parent_absorbers}
\end{figure}

The absorber selection procedure described above yielded 13 sources with robust detections of H$\alpha$ absorption features from the parent sample (all of which are LRDs; \autoref{fig:parent_absorbers}) which were subsequently used for estimating the absorber incidence rate. In addition to these targets, we also include additional independently discovered AGN with absorption features -- The Cliff \citep{deGraaff2025}, for which we use the R2700 IFU observations from DDT PID 9433
(\citealp{Ivey2026}) and QSO1 \citep{Furtak2023_AGN}, which is a strongly lensed source that would be undetectable in field surveys, to boost the overall sample statistics by including the most BH*-like LRDs. Hence, the sample used to analyze absorber properties consists of a total of 15 targets. Aside from the main set of absorbers, we identify a tentative absorption feature in the spectrum of BT-G23-4286, while appearing robust in the raw aperture spectrum (\autoref{fig:tentative_absorber}), the feature fails to pass the robustness criteria once the underestimated IFU errors are taken into account. Hence, we report the best fit values for BT-G23-4286, but do not include it in further analysis.

\begin{figure*}
    \centering
    \includegraphics[width=0.32\linewidth]{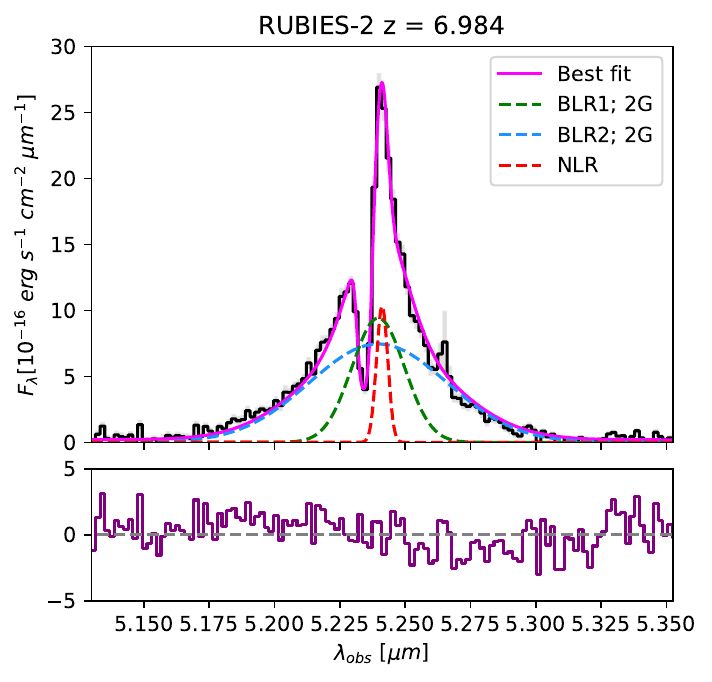}
    \includegraphics[width=0.32\linewidth]{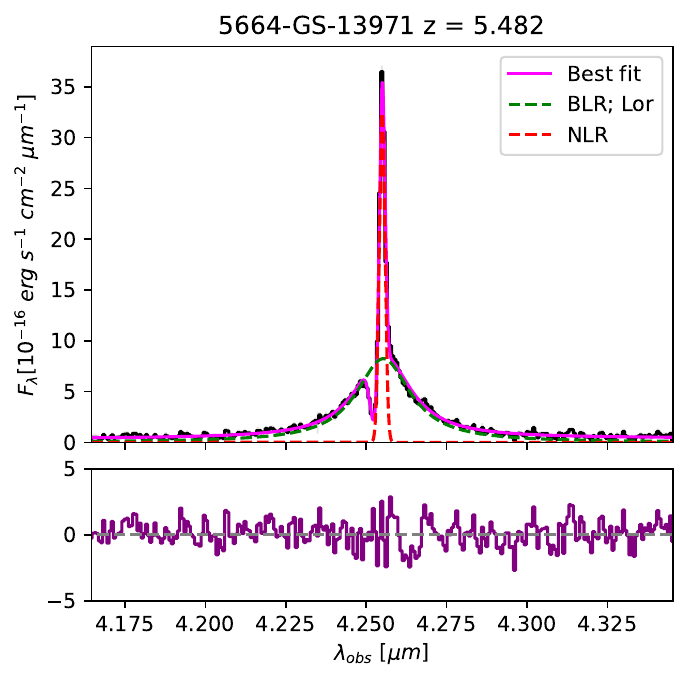}
    \includegraphics[width=0.32\linewidth]{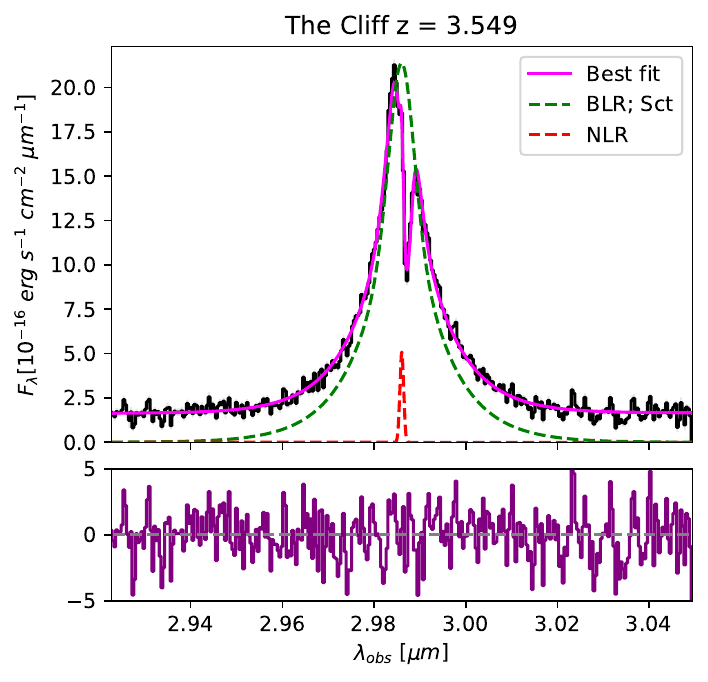}
    \caption{Example fits to the broad \Has absorbers in the sample. Left panel shows a source fit best by a double Gaussian, middle panel - Lorentzian profile and the right panel - electron scattering. The data is shown in black, with gray shading indicating errors, the best fit model is displayed in magenta. The dashed green line shows the BLR profile, the dashed blue line shows the second BLR component in case of a double Gaussian fit. The red line shows the narrow \Ha. The panel below each \Has profile fit shows the best fit residuals. In case of two BLR models fitting equally well, we plot the absolute best one in terms of $\Delta BIC$.}
    \label{fig:sample_example}
\end{figure*}

 We also find that 9 of our sample sources show clear detections of corresponding \Hbs absorption, which we fit with similar methods to that of \Ha. The main measured \Has absorber properties for our targets are summarized in \autoref{tab:Absorber_sample} while \autoref{tab:Absorber_sample_Hbeta} shows the properties of \Hbs absorbers. \autoref{fig:parent_absorbers} summarizes the classification of the parent sample into LRDs and LBDs and shows that only LRDs have absorption features. Some example \Has profile fits are shown in \autoref{fig:sample_example}, while the remainder are given in \autoref{fig:sample_fits_Ha}. The \Hbs fits for sources with \Hbs absorbers are shown in \autoref{fig:sample_fits_Hb}.

\begin{table*}
\centering
\setlength{\tabcolsep}{3pt}
    \begin{tabular}{cccccccccccc}
    \hline
        ID & R.A. & Dec & z &BLR model& $\rm FWHM_{H\alpha;br}$ & $L_{\rm H\alpha;br}$& $\sigma_{\rm H\alpha}$& $\tau_{\rm 0;H\alpha}$& $C_{\rm f;H\alpha}$ &$\mathbf{\Delta v_{H\alpha}}$ & $\mathbf{EW_{0;H\alpha}}$\\
         & [$^\circ$] & [$^\circ$] &  & & [km~s$^{-1}$] & [$\times 10^{43}$~erg~s$^{-1}$]& [km~s$^{-1}$] & & & [km~s$^{-1}$] &[\AA]\\
    \hline
    \multicolumn{12}{c}{Main sample} \\
    \hline
    JADES-GN-28074 & 189.06458 & 62.27382 & 2.260 & Sct & $1750_{-6}^{+5}$ & $2.514_{-0.006}^{+0.006}$ & $205_{-1}^{+1}$ & $1.28_{-0.01}^{+0.02}$& $0.994_{-0.008}^{+0.004}$ & $-425_{-1}^{+1}$ &  $8.73_{-0.07}^{+0.07}$  \\
    DH-GS-5070 & 53.09200  & -27.90314 & 3.642 & 2G & $1785_{-14}^{+14}$ & $0.556_{-0.003}^{+0.003}$ & $107_{-3}^{+3}$ & $6.63_{-0.74}^{+1.01}$& $0.93_{-0.06}^{+0.04}$ & $-17_{-3}^{+3}$ &  $10.59_{-0.62}^{+0.44}$  \\
    RUBIES-182791 & 34.21381 & -5.08705 & 4.715 & 2G & $1752_{-96}^{+87}$ & $0.740_{-0.014}^{+0.015}$ & $130_{-24}^{+19}$ & $5.41_{-2.67}^{+2.97}$& $0.42_{-0.05}^{+0.06}$ & $-298_{-54}^{+48}$ &  $5.44_{-0.92}^{+1.07}$  \\
    BT-J1148$^\dagger$ & 177.05796 & 52.86280 & 5.011 & 2G & $956_{-6}^{+6}$ & $0.542_{-0.003}^{+0.003}$ & $58_{-2}^{+2}$ & $4.25_{-0.51}^{+0.58}$& $0.52_{-0.02}^{+0.02}$ & $57_{-2}^{+2}$ &  $2.78_{-0.07}^{+0.08}$  \\
    JADES-GN-68797$^\dagger$ & 189.22914 & 62.14619 & 5.040 & Sct & $2009_{-11}^{+9}$ & $5.443_{-0.018}^{+0.017}$ & $186_{-3}^{+3}$ & $1.91_{-0.11}^{+0.02}$& $0.984_{-0.022}^{+0.012}$ & $-215_{-6}^{+6}$ &  $8.50_{-0.31}^{+0.27}$  \\
    JADES-GS-159717$^\dagger$ & 53.09753 & -27.90126 & 5.078 & 2G,Sct & $1617_{-67}^{+67}$ & $0.859_{-0.023}^{+0.023}$ & $109_{-7}^{+7}$ & $2.73_{-0.92}^{+0.92}$& $0.828_{-0.125}^{+0.125}$ & $-3.5_{-4}^{+4}$ &  $6.54_{-0.45}^{+0.39}$  \\
    5664-GN-15498$^\dagger$ & 189.28554 & 62.28078 & 5.085 & Lor & $1162_{-10}^{+10}$ & $0.529_{-0.004}^{+0.004}$ & $67_{-3}^{+3}$ & $5.96_{-1.12}^{+1.47}$& $0.931_{-0.053}^{+0.044}$ & $-47_{-5}^{+5}$ &  $6.10_{-0.25}^{+0.26}$  \\
    RUBIES-42046 & 214.79537 & 52.788847 & 5.276 & Sct & $2003_{-53}^{+51}$ & $4.866_{-0.069}^{+0.071}$ & $263_{-15}^{+15}$ & $2.03_{-0.33}^{+0.53}$& $0.928_{-0.064}^{+0.050}$ & $-284_{-39}^{+32}$ &  $18.63_{-0.52}^{+0.49}$  \\
    5664-GS-13971$^\dagger$ & 53.13858 & -27.79025  & 5.481 & Lor,Sct & $1384_{-185}^{+185}$ & $0.861_{-0.078}^{+0.078}$ & $100_{-13}^{+13}$ & $2.19_{-0.55}^{+0.55}$& $0.91_{-0.13}^{+0.13}$ & $-146_{-36}^{+36}$ &  $6.31_{-0.23}^{+0.22}$  \\
    5664-GN-9771$^\dagger$ & 189.28100 & 62.24731  & 5.538 & 2G,Sct & $2312_{-299}^{+299}$ & $5.492_{-0.276}^{+0.276}$ & $69_{-16}^{+16}$ & $1.66_{-1.58}^{+1.58}$& $0.466_{-0.51}^{+0.51}$ & $-199_{-57}^{+57}$ &  $3.47_{-0.08}^{+0.08}$  \\
    JADES-GN-38147 & 189.27068 & 62.14842 & 5.869 & 2G & $1675_{-49}^{+55}$ & $0.991_{-0.019}^{+0.020}$ & $72_{-24}^{+27}$ & $2.25_{-1.10}^{+1.82}$& $0.596_{-0.129}^{+0.290}$ & $-358_{-23}^{+31}$ &  $3.72_{-0.63}^{+1.02}$  \\
    RUBIES-49140 & 214.89225 & 52.87741 & 6.686 & 2G & $2475_{-45}^{+46}$ & $6.039_{-0.079}^{+0.086}$ & $124_{-10}^{+13}$ & $6.14_{-1.86}^{+2.33}$& $0.622_{-0.129}^{+0.192}$ & $3.7_{-12}^{+12}$ &  $8.02_{-1.64}^{+2.19}$  \\
    RUBIES-2 & 214.98303 & 52.95600 & 6.984 & 2G & $1922_{-38}^{+41}$ & $4.33_{-0.045}^{+0.045}$ & $74_{-5}^{+5}$ & $2.71_{-0.33}^{+0.33}$& $0.794_{-0.028}^{+0.034}$ & $-305_{-5}^{+4}$ &  $5.45_{-0.18}^{+0.18}$  \\
    \hline
    \multicolumn{12}{c}{Additional sources} \\
    \hline
    The Cliff$^\dagger$ & 34.41075 &  -5.12966 & 3.549 & Sct & $865_{-7}^{+2}$ & $0.400_{-0.002}^{+0.001}$ & $92_{-1}^{+1}$ & $1.12_{-0.04}^{+0.05}$& $0.976_{-0.025}^{+0.017}$ & $70_{-1}^{+1}$ &  $3.62_{-0.06}^{+0.07}$  \\
    A2744-QSO1$^\dagger$ & 3.57984 &-30.40157 & 7.036 & Lor & $816_{-10}^{+11}$ & $0.149_{-0.002}^{+0.02}$ & $71_{-7}^{+8}$ & $4.66_{-1.74}^{+2.24}$& $0.268_{-0.028}^{+0.032}$ & $-28_{-5}^{+5}$ &  $1.68_{-0.17}^{+0.17}$  \\
    \hline
    \multicolumn{12}{c}{Tentative absorber} \\
    \hline
     BT-G23-4286$^\dagger$ & 3.6192  &  -30.4233 & 5.840 & 2G & $928_{-21}^{+76}$ & $0.352_{-0.005}^{+0.083}$  & $35_{-4}^{+10}$ & $2.75_{-1.79}^{+1.35}$ & $0.51_{-0.05}^{+0.31}$  & $-140_{-24}^{+7}$ & $1.61_{-0.13}^{+0.17}$ \\
    \hline
    \end{tabular}
    \caption{Table summarizing the broad \Has line and absorber properties. The first column indicates the object ID formed by adding the name or PID of the survey to the original catalog ID. Columns 2 and 3 give the R.A. and Dec respectively. Column 4 shows the redshifts inferred from the narrow lines. The following column lists the best fit BLR models with multiple models listed indicating comparable performance. Columns 6 and 7 list FWHM of broad \Has and its luminosity respectively. The remaining five columns list the absorber properties in order of Doppler broadening ($\sigma$), peak optical depth $\tau_0$, velocity offset ($\Delta v$) and rest-frame equivalent width ($\rm EW_0$). \Has absorber parameters in bold ($\Delta v$ and $\rm EW_0$) are more reliably constrained by our data than the others. Object IDs with $\dagger$ next to them indicate sources for which R2700 data was available.}
    \label{tab:Absorber_sample}
\end{table*}

\begin{table*}
\centering
\setlength{\tabcolsep}{4pt}
    \begin{tabular}{cccccccc}
    \hline
        ID &$\rm FWHM_{H\beta;br}$ & ${L_{H\beta;br}}$& $\sigma_{H\beta}$& $\tau_{0;H\beta}$& $C_{f;H\beta}$ &$\mathbf{\Delta v_{H\beta}}$ & $\mathbf{EW_{0;H\beta}}$\\
         & [km~s$^{-1}$] & [$\times 10^{43}$~erg~s$^{-1}$]& [km~s$^{-1}$] & & & [km~s$^{-1}$] &[\AA]\\
    \hline
    \multicolumn{8}{c}{Main sample} \\
    \hline
    JADES-GN-28074 & $3227_{-46}^{+47}$ & $0.173_{-0.003}^{+0.003}$ & $167_{-13}^{+12}$ & $2.12_{-0.54}^{+0.75}$& $0.888_{-0.102}^{+0.077}$ & $-234_{-34}^{+31}$ &  $9.40_{-1.33}^{+1.36}$  \\
    DH-GS-5070 &  $3299_{-329}^{+395}$ & $0.035_{-0.003}^{+0.003}$ & $125_{-36}^{+33}$ & $5.14_{-2.98}^{+3.23}$& $0.658_{-0.129}^{+0.144}$ & $-154_{-133}^{+132}$ &  $7.23_{-2.02}^{+2.72}$  \\
    JADES-GN-68797$^\dagger$ & $2708_{-200}^{+209}$ & $0.244_{-0.017}^{+0.017}$ & $106_{-18}^{+15}$ & $5.77_{-2.79}^{+2.89}$& $0.786_{-0.147}^{+0.131}$ & $-146_{-87}^{+87}$ &  $7.47_{-2.36}^{+1.83}$  \\
    RUBIES-42046 & $2692_{-236}^{+250}$ & $0.374_{-0.030}^{+0.031}$ & $153_{-34}^{+40}$ & $4.57_{-2.00}^{+3.20}$& $0.833_{-0.123}^{+0.113}$ & $-420_{-94}^{+97}$ &  $11.54_{-2.27}^{+2.39}$  \\
    5664-GN-9771$^\dagger$  & $4385_{-63}^{+66}$ & $0.457_{-0.009}^{+0.009}$ & $93_{-7}^{+7}$ & $2.98_{-0.44}^{+0.54}$& $0.977_{-0.029}^{+0.017}$ & $-222_{-19}^{+19}$ &  $6.84_{-0.53}^{+0.56}$  \\
    RUBIES-49140 & $3255_{-199}^{+209}$ & $0.596_{-0.032}^{+0.034}$ & $64_{-33}^{+44}$ & $4.27_{-2.35}^{+3.56}$& $0.664_{-0.145}^{+0.187}$ & $173_{-96}^{+88}$ &  $4.26_{-1.11}^{+3.31}$  \\
    RUBIES-2  & $3716_{-206}^{+186}$ & $0.470_{-0.022}^{+0.025}$ & $63_{-39}^{+42}$ & $5.80_{-2.69}^{+2.84}$& $0.856_{-0.150}^{+0.102}$ & $-252_{-100}^{+100}$ &  $5.39_{-1.18}^{+4.98}$  \\
    \hline
    \multicolumn{8}{c}{Additional sources} \\
    \hline
    The Cliff$^\dagger$ & $2804_{-101}^{+123}$ & $0.035_{-0.001}^{+0.001}$ & $167_{-13}^{+12}$ & $2.12_{-0.54}^{+0.75}$& $0.629_{-0.083}^{+0.124}$ & $264_{-51}^{+51}$ &  $4.54_{-0.48}^{+0.40}$  \\
    A2744-QSO1$^\dagger$ &  $2061_{-238}^{+264}$ & $0.013_{-0.006}^{+0.006}$ & $104_{-37}^{+77}$ & $3.61_{-2.68}^{+3.93}$& $0.312_{-0.123}^{+0.188}$ & $178_{-101}^{+91}$ &  $2.22_{-0.82}^{+1.36}$  \\
    \hline
    \end{tabular}
    \caption{Same as \autoref{tab:Absorber_sample}, but for sources with \Hbs absorption.}
    \label{tab:Absorber_sample_Hbeta}
\end{table*}

\section{Incidence rate of Balmer absorbers}
\label{sec:incidence}
While previous studies already hinted at a high ($>20\%$) incidence rate of Balmer absorption in JWST-discovered AGN and LRDs in particular \citep{Matthee2024, Juodzbalis2026, Lin2026}, more comprehensive assessments of this quantity are required. With a parent sample of 47 unique sources of which 13 show absorption signatures, we  estimate an incidence rate of $\sim 28\%$ for our entire sample of AGN, a $37$\% absorber incidence for LRDs and $<8$\% for LBDs. However, this number only represents a lower limit as a simple ratio does not account for inherent incompleteness of R1000 and R2700 data since narrow line infill can hide absorbers with velocities lower than the LSF as discussed in the following section.

\subsection{Completeness simulations}
\label{sec:completenes}
In order to account for this incompleteness, we perform completeness simulations by fitting simulated broad \Has profiles and testing the recovery rates across the simulated parameter space. The broad \Has profiles were simulated to be single Gaussians with widths drawn from a uniform distribution spanning the range from 1000 to 7500~km~s$^{-1}$ -- representative of the FWHM of the general JWST-discovered AGN population \citep{Juodzbalis2026}. We note that adopting a more complex BLR shape affects the inferred absorber properties, but not the selection. The properties of the simulated absorption ($\tau_0$, $\Delta v$, $\sigma$ and $C_f$) were likewise drawn from uniform distributions with boundaries set according to the parameter space spanned by our sample of absorbers, albeit slightly extended to cover weaker features that may be missed even in deep R2700 observations. The assumed distributions were -- $\Delta v \in [-800, 800]$~km~s$^{-1}$, $\sigma \in [50, 300]$~km~s$^{-1}$, $\tau_0 \in [0.8, 10]$ and $C_f \in [0.1, 1.0]$. The normalization of the broad emission line was sampled in flux space with $\log{F_{\rm BLR}} \in [-18, -15.5]$~erg~s$^{-1}$~cm$^{-2}$, corresponding to $\log{L_{\rm BLR}}\in [41, 43.7]$~erg~s$^{-1}$ at the median redshift of our sample ($z = 4$). This, coupled with parametrising the final completeness functions in terms of BLR S/N, ensures that the results are agnostic to the redshift or noise level of the simulated sources. The narrow \Has flux was fixed to 50\% of the BLR flux, consistent with the mean flux ratio for the parent sample sources. The \NII\ line emission was drawn from a Gaussian distribution, ensuring that it is at least 1~dex fainter than the narrow \Ha, consistent with what is seen in high redshift AGN \citep{Maiolino_AGN, Juodzbalis2026}.

We carry out the completeness simulations by convolving the simulated spectra with either R2700 or R1000 line spread functions (LSFs) and injecting Gaussian noise consistent with that of JADES Medium tier observations, since their 0.8~h exposure times per grating \citep{Scholtz2026_JADES} gives noise level  comparable to that of the majority of our sample spectra. In addition, expressing the completeness function in terms of BLR S/N means that the exact value of the noise injected has little effect on our results. In total, we generated 150,000 simulated R1000 and R2700 spectra each. These spectra were then fit using bounded least squares minimization to make the fitting computationally tractable. We used entirely the same criteria for selecting absorbers from the simulated spectra as for the real data set, requiring $\Delta BIC > 5$ between the model without and the one with the absorber. However, as our simulations do not accurately model correlated NIRSpec noise and do not include data artefacts, we add an additional selection criterion for the simulated data. Namely, we require the absorption feature to have S/N $>5$, defined similarly to the BLR S/N except that it is calculated for the feature remaining after subtracting the fitted emission. Accompanying completeness simulations, we fit similar sets of simulated spectra that do not include absorption features to quantify the potential contamination rates, that is, the fraction of sources with no absorption identified as absorbers.

\begin{figure}
    \centering
    \includegraphics[width=1.0\linewidth]{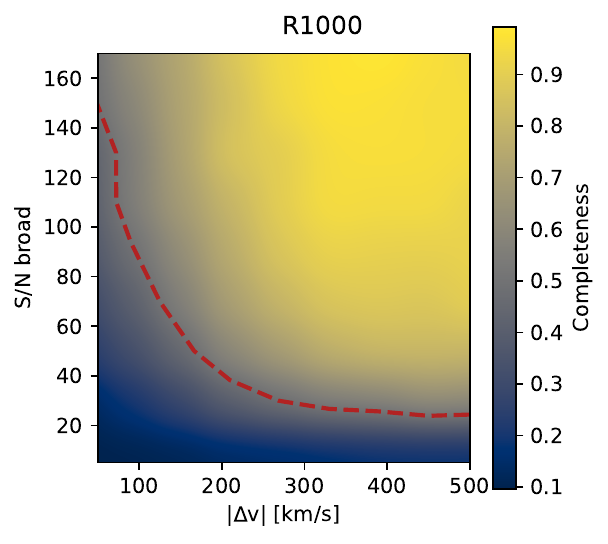}
    \includegraphics[width=1.0\linewidth]{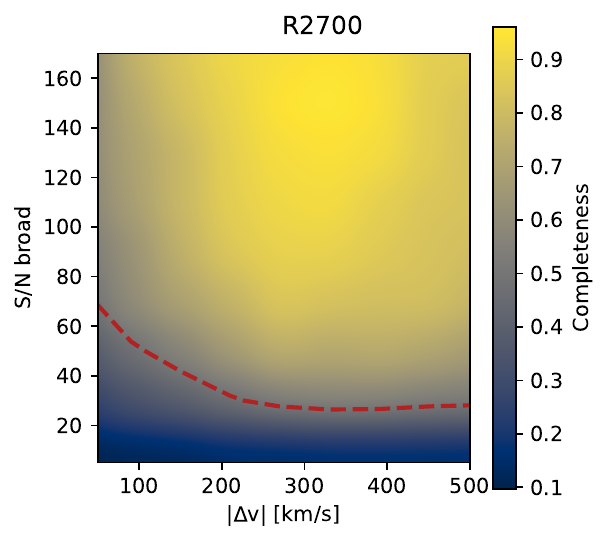}
    \caption{2D completeness maps on the $\Delta v$ --  BLR S/N plane with the dashed lines indicating the 50\% completeness threshold. Top panel shows the simulations for R1000, bottom -- R2700. As can be seen here, the R1000 grating performs significantly worse than R2700 for $|\Delta v| < 200$~km~s$^{-1}$.}
    \label{fig:completeness_v}
\end{figure}

We find that the contamination across both gratings is negligible and completeness drops off to below 50\% for S/N < 20 for both R1000 and R2700 observations (\autoref{fig:completeness_v}). The completeness function strongly depends on the absorber properties such as the velocity shift ($\Delta v$) as illustrated in \autoref{fig:completeness_v} and discussed in Section \ref{sec:comp_correction} (also demonstrated for a single object in \citealp{DEugenio2026}). In addition, R1000 performs substantially worse than R2700 in parameter recovery as shown by \cite{DEugenio2026} and discussed in more detail in the following section. 

\subsection{Parameter recovery tests}
Before continuing towards further analysis of the demographics of the detected absorbers, we utilize the simulated data set to conduct parameter recovery tests in order to determine which parameters of \autoref{eq:depth_prof} and \autoref{eq:absorption} are recovered well in the fitting.

We define the deviation between the simulated `real' and fitted values as the Euclidean distance between the simulated and estimated feature vectors. In order to ensure that each parameter is evenly weighted, we remap the four absorber parameters ($\Delta v$, $\sigma$, $\tau_0$ and $C_f$) to dimensionless features spanning the same dynamic range using the following equation:
\begin{equation}
    \theta_i = \frac{\vartheta_i - \bar{\vartheta_i}}{\max{\vartheta_i}-\min{\vartheta_i}},
\end{equation}
where $\vartheta_i$ is the original parameter while $\theta_i$ is the corresponding feature. The above remapping ensures that $\theta_i \in [-0.5, 0.5]$ for all $\theta_i$. The deviation between the simulated and estimated feature vectors is then defined as follows:
\begin{equation}
\label{eq:deviation}
    D = \sqrt{\frac{\sum_{i=0}^N{\left (\theta_i-\hat{\theta}_i\right)^2}}{N}},
\end{equation}
where $\theta_i$ and $\hat{\theta}_i$ are the ground truth and fitted features respectively and $N$ is the total number of features in the vector. Deviation defined this way maps to $D \in [0, 1]$ with zero representing perfect parameter recovery and one -- a catastrophic failure.

\begin{figure}
    \centering
    \includegraphics[width=\linewidth]{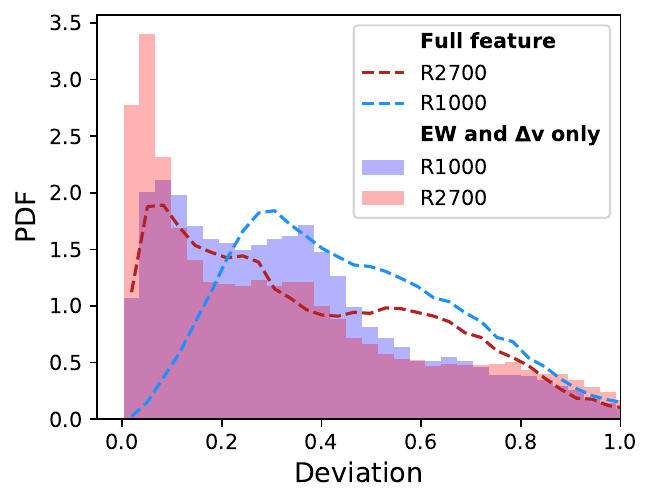}
    \caption{The distribution of deviation, as defined in \autoref{eq:deviation}, for the simulated R2700 and R1000 data when considering the full feature vector (dashed lines) and the simplified version including just $\rm EW_0$ and $\Delta v$ (histograms).}
    \label{fig:deviation}
\end{figure}

We find that the R1000 grating performs significantly worse than R2700 when recovering the full 4-parameter absorber model (dashed lines in \autoref{fig:deviation}). We quantify the difference between the $D$ distributions using the Kullback–Leibler divergence ($D_{KL}$), finding that $D_{KL} = 0.17$ between the simulated R2700 and R1000 data sets with the mode of the distribution equal to $0.1$ and $0.3$ for R2700 and R1000 respectively. It should also be noted that both distributions have significant tails towards high $D$ values with excess kurtosis values of 6 -- 7, indicating a prevalence of catastrophic outliers. While this could partially be attributed to least-squares fitting being prone to getting trapped in local minima, the parameters defining the prominence of an absorption feature -- $\sigma$, $\tau_0$ and $C_f$, are intrinsically degenerate. We attempt to remove these degeneracies by coalescing the three aforementioned parameters into a single equivalent width measure:
\begin{equation}
    {\rm EW_0}(1+z) = \int\left(1-F_{obs}/F_{em}\right)d\lambda \equiv \int C_f\left(1 - e^{-\tau_\lambda}\right)d\lambda,
\end{equation}
where $F_{obs}$ is the observed flux including the absorption feature, $F_{em}$ - the total emission and $z$ - the redshift of the source. The remaining parameters are the same as in \autoref{eq:depth_prof} and \autoref{eq:absorption}. This reduces the total number of features from 4 to 2, reapplying the earlier analysis shows that the simplified feature vector is better recovered by both R2700 and R1000 data (\autoref{fig:deviation}) with reduced catastrophic outlier rates (excess kurtosis of 1 -- 3) and the peaks of the distributions shifted to $D = 0.05, 0.08$. The KL divergence between the performance of R1000 and R2700 likewise reduces to 0.07, indicating a much closer match in the performance of the two gratings. Hence, in subsequent analysis we focus on $\Delta v$ and $\rm EW_0$ as the main parameters of each absorber as these are considerably better recovered than the full model and similar performance of R1000 and R2700 gratings reduces the inhomogeneity in the measurement accuracy as only 17/47 of our targets have R2700 data.

Lastly, we note that the infill by the narrow \Has line causes unrecoverable degeneracies even in the high resolution data. In particular, such infill can make a broad rest-frame absorption feature appear narrower and redshifted. Hence, even the reduced feature vector is not recovered with perfect accuracy. Therefore, the absorber demographics of any JWST samples should be interpreted with caution. 

\subsection{Completeness corrected absorption incidence rate}
\label{sec:comp_correction}
Correcting the directly estimated incidence rate for incompleteness is a non-trivial task as a simple ratio will miss the intricacies introduced by the underlying distribution of absorber properties. For example, even if broad \Has is detected to high S/N and the absorption has a high intrinsic $\rm EW_0$, the feature may nevertheless be missed if it is rest-frame ($\Delta v = 0$) and covered by the narrow \Has emission. Hence, we perform a parameter-aware completeness correction by first defining a completeness function from the simulated data as a function of the broad line S/N and the two absorber parameters -- $C \equiv C(S/N, EW_0, \Delta v)$. This function was constrained by binning the simulated data in 3D bins of the three variables. As all of our sources with detected absorbers have $S/N > 60$, we are in the regime where $C$ varies more with $EW_0$ and $\Delta v$ rather than the brightness of the BLR (see \autoref{fig:completeness_v}). Unfortunately, the small size of our sample makes 2D histograms prone to overbinning. Hence, we construct two 1D completeness functions varying with $\Delta v$ and $EW_0$ respectively by averaging out the remaining two parameters, only considering regions where the parameters being averaged are within the range populated by our sample sources to avoid biasing the completeness. We thus perform two separate completeness corrections, checking for consistency between them.

With the 1D completeness functions in hand, we bin the absorbers sample in the same space as $C$ and perform Bayesian inference to constrain the best-estimate on the incidence rate of Balmer absorption features together with a 95\% credible interval (CI). The inference was performed on a per-bin basis utilizing the following formalism to infer the posteriors.

We denote the measured number of absorbers in the $i$-th bin as $n_i$, the completeness - $C_i$, the real number of absorbers is denoted by $N_i$, the parent sample size - $M$ and the number of bins - $B$. The posterior is then defined by the standard Bayesian formula:
\begin{equation}
    P(N_i|n_i) = \frac{P(n_i|N_i)P(N_i)}{P(n_i)}.
\end{equation}
The likelihood function -- $P(n_i|N_i)$, is then simply a binomial distribution with $N_i$ as the number of trials and $C_i$ as the success probability:
\begin{equation}
    P(n_i|N_i) = \binom{N_i}{n_i}C_i^{n_i}\left(1 - C_i\right)^{N_i-n_i}
\end{equation}
The prior, $P(N_i)$ is constructed to be uniform between $N_i = n_i$ and $N_i = M$. Marginalizing over the likelihood to obtain $P(n_i)$ then yields the following expression for the posterior:
\begin{equation}
    P(N_i|n_i) = \frac{P(n_i|N_i)}{\sum_{N_i=n_i}^{M} P(n_i|N_i)}
\end{equation}
We construct our best estimate for each bin by taking the most probable value given by the posterior, while the CI is defined to contain 95\% of the posterior probability. We note that, while the above procedure does not automatically ensure that $\sum_{i}N_i \leq M$, we find that the completeness corrected source counts and their CIs automatically satisfy this constraint without it being imposed explicitly.

The final incidence rate is thus estimated as:
\begin{equation}
    I = \frac{\sum N_{i;2700}+\sum N_{i;1000}}{M},
\end{equation}
we find that, for the combined sample of LBDs and LRDs, $I = 0.32$ (95\% CI: $0.28$ -- $0.47$) when using completeness defined with respect to $\Delta v$ and $I = 0.30$ (95\% CI: $0.28$ -- $0.45$) when correcting with respect to $EW_0$. Both estimates and their CIs overlap within 2-3\% and are hence functionally equivalent. We thus estimate that the incidence rate of absorbers, if LBDs and LRDs are treated as belonging to the same population, is $I \approx 31\%$ and no higher than 50\%. However, this rate may still be underestimated if a significant fraction of absorbers occurs at rest-frame wavelengths where they can be hidden by the narrow \Has emission.

\begin{figure}
    \centering
    \includegraphics[width=\linewidth]{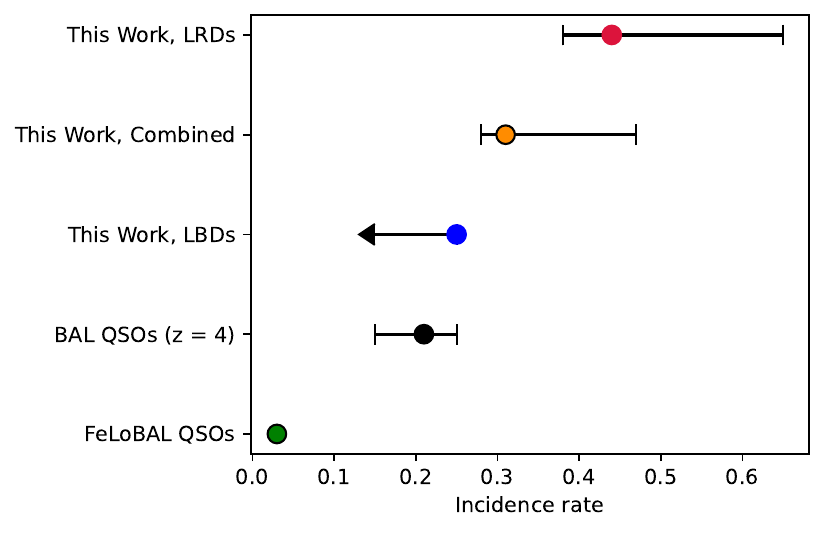}
    \caption{Comparison between Balmer absorption incidence rates for our sample populations and those of BAL and FeLoBAL QSOs \protect\citep{Leighly2025, Bischetti2023}. The colored points show the best estimates while the black error bars represent the 95\% Credible Intervals. As no LBDs in our sample have absorption features, their estimate is an upper limit.}
    \label{fig:Incidence}
\end{figure}

Splitting the sample into LRD and LBD sources results in an inferred $I = 0.44$ (95\% CI: $0.38$ -- $0.65$) for LRDs. While the latter value is somewhat lower than recent estimates by \cite{Matthee2026} and \cite{Lin2026}, who report $\sim60\%$ and $\sim 65\%$ incidence rates for high redshift and local LRDs respectively, these values are still consistent within our CI. The inferred incidence rate is high enough that the scenario in which effectively all LRDs have \Has absorption cannot be ruled out in light of narrow \Has infill effectively hiding rest-frame $|\Delta v| < 100$~km~s$^{-1}$ systems in R1000 data. On the other hand, no non-LRD sources in our parent sample display absorption features. Hence, we infer a 2$\sigma$ upper limit on the incidence rate by assuming the same absorber distribution as seen for LRDs. Depending on whether the completeness is defined with respect to $\Delta v$ or $\rm EW_0$, the incidence of Balmer absorption in LBDs is $I < 0.08 - 0.25$ and conservatively we adopt the higher value. The comparison between our estimated Balmer absorption incidence rates and that for FeLoBAL QSOs \citep{Leighly2025} is shown in \autoref{fig:Incidence}. As can be seen there, the JWST-discovered AGN population taken as a whole has a significantly higher incidence rate of Balmer line absorption than FeLoBAL QSOs (30\% vs 0.3\%). On the other hand, the Balmer absorbers appear to be a defining feature of LRDs with no LBDs in our sample displaying hydrogen absorption. Nevertheless, the conservative $I < 25\%$ upper limit for LBDs leaves significant room for a high absorber occurrence with larger sample sizes required to see if this absorption is truly absent in LBDs. We also note that, while FeLoBAL QSOs, which display prominent Balmer absorption are rare, the overall BAL fraction strongly evolves with redshift and reaches 50\% at $z = 6$ \citep{Bischetti2023}. While the overall BAL fraction at the median $z$ of our sample is lower ($\sim 20$\%), it is still considerably closer to the inferred incidence rates for our sample AGN (\autoref{fig:Incidence}). Therefore, it is reasonable to suggest that LRDs and BAL QSOs may be shaped by similar physical processes, yet occurring at different luminosities. 

\section{Balmer absorption demographics}
\label{sec:demographics}
Having explored the overall incidence rates of \Has absorption, we move to examining the demographics of our absorber sample. As only $\Delta v$ and $\rm EW_0$ are recovered well in R1000 and R2700 data, these are the main quantities on which we focus in the following analysis.
\subsection{Absorption velocity distribution}
\label{sec:vel_dist}
\begin{figure}
    \centering
    \includegraphics[width=\linewidth]{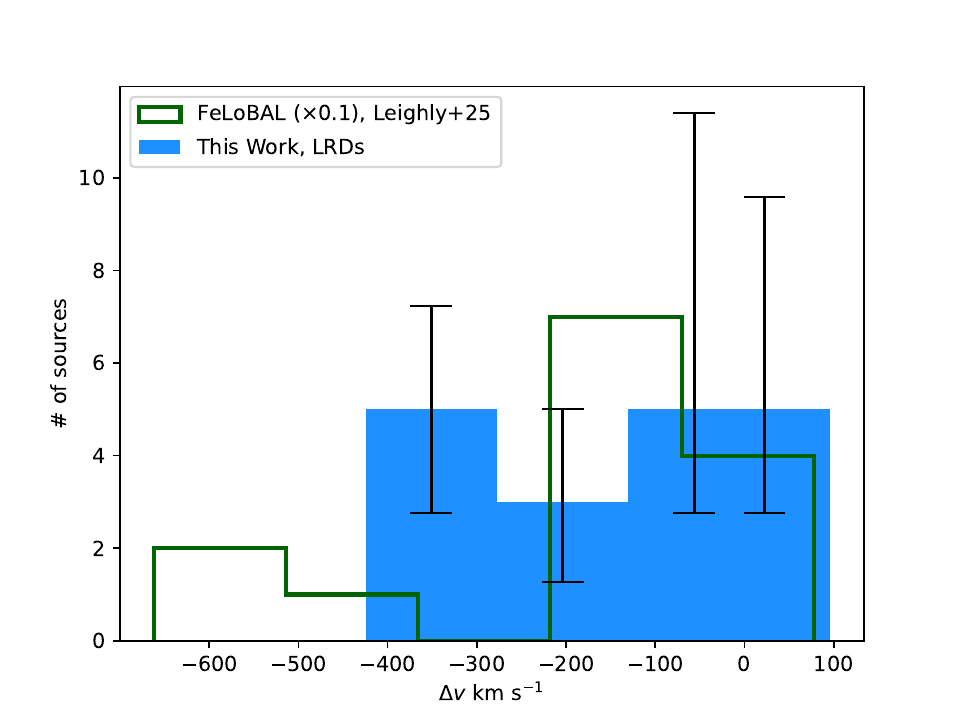}
    \caption{Velocity distribution of the sample absorbers when compared to FeLoBAL QSOs from \protect\cite{Leighly2025}, whose velocities were scaled down by a factor of 10 (green). The blue histogram shows the fiducial estimate while the black errorbars represent the 95\% confidence intervals combined in quadrature with Poissonian counting error.}
    \label{fig:dv_dist}
\end{figure}

The completeness corrected $\Delta v$ distribution of our absorber sample is presented in \autoref{fig:dv_dist}. As shown there, the absorption features are either at systemic redshift ($|\Delta v| < 100$~km~s$^{-1}$) or significantly blue shifted ($\Delta v < -200$~km~s$^{-1}$). Given that $C(\Delta v)$ is symmetric around $\Delta v = 0$, this asymmetry in the velocity distribution can not be accounted for by completeness effects alone. Hence, the dense media blanketing LRDs are generally weakly outflowing with median $\Delta v \approx -100$~km~s$^{-1}$ and largest blueshifts approaching $-400$~km~s$^{-1}$. This argues against scenarios invoking quasi-static gas envelopes such as the BH* model \citep{Naidu2025} as atmospheric pulsations possible in such environments would produce roughly equal amounts of outflowing or inflowing absorbers. 

Nevertheless, the preponderance of absorbers within 100~km~s$^{-1}$ of the \Has rest-frame together with all absorber $\rm FWHM < 600$~km~s$^{-1}$ may suggest that the absorbing gas is not tracing escaping high velocity outflows -- in comparison, \OIIIs outflows in AGN generally exceed $\rm FWHM = 1000$~km~s$^{-1}$ \citep{Escott2025}. While the \Has absorption is only tracing the densest parts of the outflowing gas, it is interesting to note that only two of our objects have any indication of large scale ionized outflows traced by the \OIII$\lambda\lambda$4959,5007 doublets (\autoref{fig:sample_fits_Hb}). This would tentatively support a scenario in which LRDs represent a phase of AGN accretion in which the BH is surrounded by a dense medium, which prevents outflows from breaking out into lower density ISM, causing an abundance of dense gas absorption and a dearth of ionized outflows. In addition, the \Has outflow velocities in our sample resemble those of `warm absorbers' found in $\sim 50\%$ of Seyfert 1 galaxies \citep{Elvis2000, Blustin2005}. These warm absorbers generally share the $\Delta v \sim 100-300$~km~s$^{-1}$ velocity scale of LRDs, yet their lower densities mean that they are highly ionized and do not appear in \Ha, suggesting that the outflowing gas seen in LRDs may be the high density counterpart of warm absorbers. However, our current sample is small, hence dedicated studies are required to establish a connection between the presence of \Has absorption and lack of outflows in the forbidden lines along with the relation between dense \Has absorption in LRDs and warm absorbers seen in Seyfert 1s.

Interestingly, the BAL QSOs that display Balmer absorption (the FeLoBALs) have a qualitatively similar velocity distribution (\autoref{fig:dv_dist}) with some sources clustering around $\Delta v = 0$ plus a significant population of outflowing \Has absorbers. The velocity scales for Balmer absorption in FeLoBALs are greater by about a factor of 10, with $ -6000 < \Delta v < 1000$~km~s$^{-1}$ compared to $-400 < \Delta v < 100$~km~s$^{-1}$ for our LRDs. However, the median bolometric luminosity of FeLoBALs from \cite{Leighly2025} is $\sim 100$ times greater than for our LRD sample ($10^{45.1}$ vs $10^{47}$~erg~s$^{-1}$). Taking a simple assumption of radiation-driven outflows for which the kinetic energy injected into gas surrounding the central engine is proportional to the luminosity of the source, we would naively expect $L_{\rm bol} \propto \Delta v^2$. Hence, a 2~dex difference in luminosity would naturally produce a 1~dex difference in velocity. Therefore, it is possible that the Balmer absorption in LRDs is a physically similar phenomenon to that of  FeLoBALs except occurring on lower energy scales.

\subsection{Equivalent widths and column densities}
Having discussed the $\Delta v$ properties, we move on to the second parameter well constrained by our data - the  $\rm EW_0$. We find that our absorber sample spans a wide range of equivalent widths - from 1 to 15~\AA. Considering that \Has absorption $EW_0$ traces the same n = 2 hydrogen required for a Balmer break, we investigate the relationship between these two parameters in our sample to investigate whether the Balmer breaks in our sample are produced by the same gas as the \Has absorbers as postulated by the BH* models \citep{Ji2025, Naidu2025}. 

\begin{figure*}
    \centering
    \includegraphics[width=0.45\linewidth]{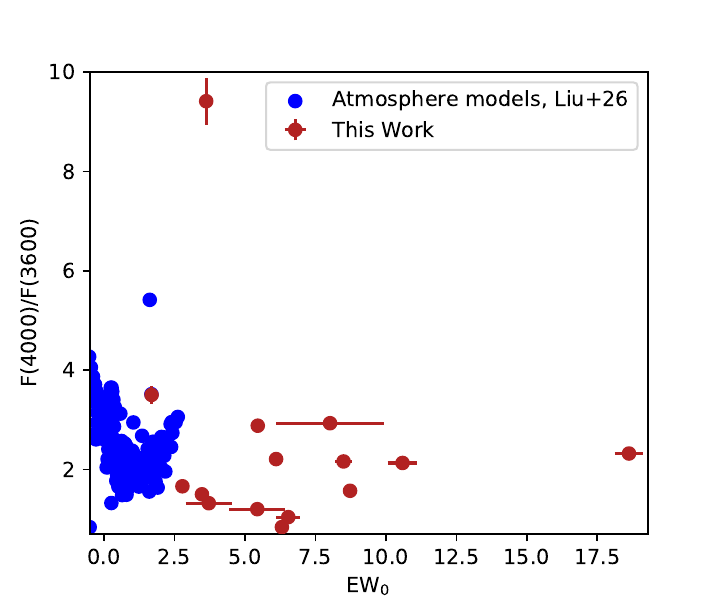}
    \includegraphics[width=0.53\linewidth]{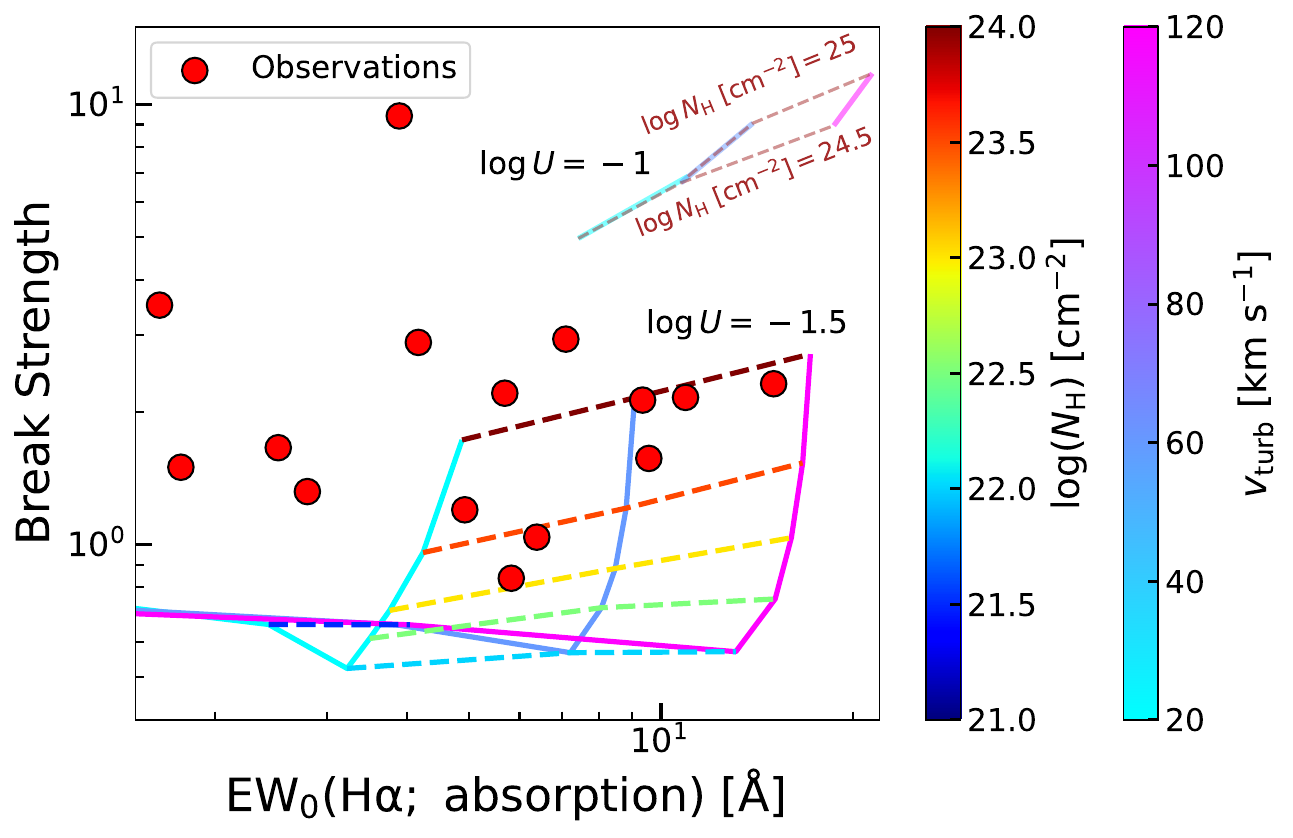}
    \caption{\textbf{Left:} Comparison between Balmer break strengths and $\rm EW_{0}$ of our sample sources and stellar atmosphere models of \protect\cite{Liu2026} showcasing the clear mismatch between our measurements and theoretical predictions of the BH* framework. \textbf{Right:} Showcase of example \texttt{CLOUDY} model grids computed at a fixed $\log{U} = -1.5$ and spanning the full range of $N_H$ and $v_{turb}$ values considered plotted over the same measurements as the left panel. Higher ${U}$ and/or higher $N_H$ would have the effect of moving the whole grid upwards as shown by the smaller model grid, and lower $v_{turb}$ or lower $C_f$ would move the grid to the left \citep[see also][]{Ji2025}, hence covering the entire observed parameter space.}
    \label{fig:EWs}
\end{figure*}

We define the Balmer break strength as the flux ratio $F(4000\mathring{A})/F(3600\mathring{A})$, tracing the optical and UV continua right below and above the break wavelength (3646~\AA), respectively. For comparison with the BH* model predictions, we use the synthetic BH* spectra library from \cite{Liu2026}, who model LRDs as BHs encased in optically thick star-like atmospheres with low surface gravity and $T_{eff} = 4000$--$6000$~~K. As shown in \autoref{fig:EWs}, our sample sources have considerably weaker Balmer breaks and higher $\rm EW_0$ values than predicted by the \cite{Liu2026} models with no apparent correlation between the two quantities. A potential explanation of this observation is that the UV and optical continua have different origins. If the UV emission originates from the host's stellar population or extended nebular emission, the infill would cause an underestimation in Balmer break strength and decorrelate the break and $\rm EW_0$ measurements. Indeed, as shown by \cite{Sun2026}, a large fraction of LRDs resemble pure BH* with extreme Balmer breaks under the assumption that \OIIIs and the UV continuum originate from the host environment.

Nevertheless, models proposing identical origins for the UV and optical continua \citep{Lin2025LRDlocal, Asada2026} can not be discounted either. We investigate the parameter space spanned by single cloud \texttt{CLOUDY} models constructed in a similar way to those used in \cite{Juodzbalis2024b}. Considering that the \texttt{CLOUDY} grids, unlike the BH* models, are not constrained to $C_f = 1$, we find that our measurements overlap well with a grid of high ionization ($\log U > -1.5$), high column density ($\log{N_H/{\rm cm}^2}>22$) and moderate turbulence ($20$~km~s$^{-1}<v_{turb}<120$~km~s$^{-1}$) models (\autoref{fig:EWs}). These results suggest that the $\rm EW_0$ properties of LRD absorption features may be the result of a diverse set of conditions within the absorbing medium rather than radially symmetric stellar atmosphere-like configurations. Considering that the $\Delta v$ distribution shown in \autoref{fig:dv_dist} is skewed towards blueshifted absorbers, this latter interpretation is likely more reasonable as there is little indication that the absorption features reside in stable systems. This interpretation favors the models of \cite{Madau2026_LRD_LBD, Madau2026_LRD_LBD2}, which model the LRD \Has profiles as a combination of absorption and BLR stratification effects. However, multi epoch observations of absorber variability are needed before strong conclusions can be drawn.

\begin{figure}
    \centering
    \includegraphics[width=\linewidth]{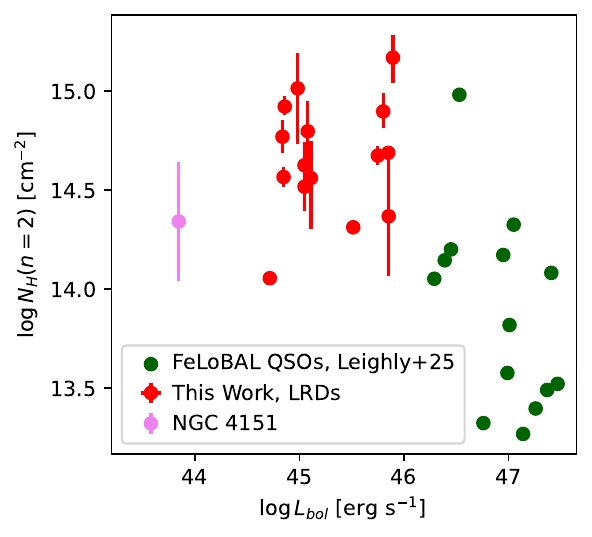}
    \caption{Comparison between our sample LRDs (red points) and FeLoBALs from \protect\cite{Leighly2025} (green points) on the $N_H(n=2)$ -- $L_{\rm bol}$ plane. The violet circle illustrates the position of NGC 4151 - a prototypical local Seyfert 1 AGN with prominent Balmer absorption features \citep{Anderson1969}. As can be seen here, the LRDs show up as an extension of the BAL population, occupying the low luminosity, high column density region.}
    \label{fig:Nh}
\end{figure}

As the $\rm EW_0$ values are well recovered by both R1000 and R2700 data, we can obtain reasonable estimates of the column density of n = 2 hydrogen ($N_H(n=2)$). We measure $N_H(n=2)$ by the apparent optical depth method \citep{Wang2024, SavageSembach1991}:
\begin{equation}
    N_H(n=2) = \frac{m_ec}{\uppi e^2 f_0\lambda_0}\tau,
\end{equation}
where $\tau$ is the integrated line optical depth, $\lambda_0$ is the rest-frame wavelength of the \Has transition and $f_0$ its oscillator strength. Since the above equation integrates \autoref{eq:depth_prof} to get $\tau$, it is unaffected by the degeneracy between $\tau_0$ and $\sigma$ and most of our sample sources have inferred $C_f > 0.5$. Hence, the $N_H(n=2)$ is unlikely to be overestimated by more than 0.5~dex. We find that our LRDs sample spans $\log{N_H(n=2)/{\rm cm^{-2}}}$ from 14 to 15, consistent with the total $N_H$ values of $>10^{22}$~cm$^{-2}$ inferred for our LRDs (\autoref{fig:EWs}) and individual sources \citep{Juodzbalis2024, Wang2024, DEugenio2025b}. Interestingly, LRDs appear as a low luminosity extension of FeLoBALs on the $N_H(n=2)$ -- $L_{\rm bol}$ plane (\autoref{fig:Nh}), with LRDs occupying the low luminosity - high $N_H(n=2)$ quadrant, similar to NGC 4151, a local Seyfert 1 AGN with observed Balmer absorption \citep{Anderson1969}. Notably, however, weak anticorrelation between $N_H(n=2)$ and $L_{\rm bol}$ seen in FeLoBALs is not reflected in LRDs. The relation instead appears to flatten out at low $L_{\rm bol}$. Hence, while both LRD and BAL QSO absorbing environments are likely similar, LRDs appear to maintain higher columns of excited hydrogen, potentially due to smaller spatial scales and higher overall gas densities involved.

\subsection{Sample correlations}
Considering the earlier discussion on the problems of parameter recovery as well as the small sample size, we proceed cautiously when looking for correlations between the absorber parameters. Mainly, we perform jackknife resampling to ensure that any observed correlations are not driven by singular outliers. In addition, we bootstrap the errors on each measured quantity and only keep the correlations for which $p < 0.05$ after the earlier tests. We utilize the Spearman rank correlation as the main metric as it makes the least amount of assumptions about the properties of the underlying data set, only reporting correlations with $|\rho_r| > 0.5$.

\begin{figure*}
    \centering
    \includegraphics[width=0.48\linewidth]{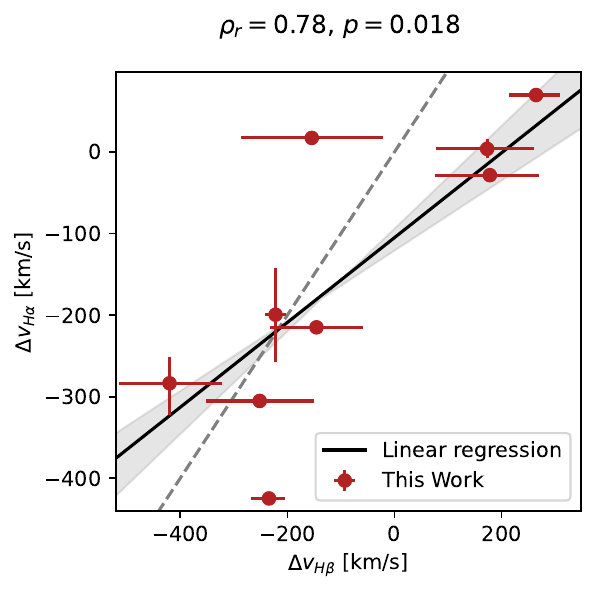}
    \includegraphics[width=0.48\linewidth]{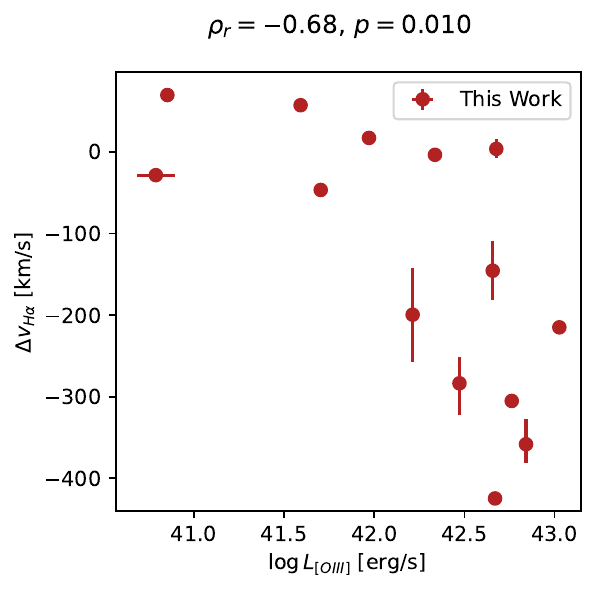}
    \caption{\textbf{Left: }The observed correlation between $\Delta v_{H\beta}$ and $\Delta v_{H\alpha}$. The data is shown in red, the best-fit linear regression with slope $m=0.52^{+0.11}_{-0.09}$ and intercept $b=-105 \pm 15$~km~s$^{-1}$ is shown as a black line with gray shading indicating the uncertainties. The 1:1 relation is indicated by a dashed gray line. \textbf{Right: }Correlation between $\Delta v_{H\alpha}$ and $L_{[OIII]}$. The notation is the same as in the left panel except we do not attempt linear regression as we do not expect the data points to follow a linear relation.}
    \label{fig:correlations_main}
\end{figure*}

Only two non-trivial correlations related to the observed absorption features survive the above procedure - the $\Delta v$ of \Has and \Hbs lines are strongly correlated (\autoref{fig:correlations_main}) and we identify a noticeable anticorrelation between $L_{\rm [OIII]}$ and $\Delta v$ of \Has with more \OIIIs luminous systems hosting more blueshifted absorbers (\autoref{fig:correlations_main}). The correlation between $\Delta v_{\rm H\beta}$ and $\Delta v_{\rm H\alpha}$ is expected as these lines likely originate from the same gas. However, linear regression, performed with bootstrapped errors, shows that the relation between these two quantities significantly deviates from a 1:1 relation, with slope $m=0.52^{+0.11}_{-0.09}$ and intercept $b=-105 \pm 15$~km~s$^{-1}$. This suggests higher order effects present in the data - as discussed before, at least part of the deviation can be attributed to the limitations of JWST spectroscopy. However, the lack of any appreciable correlation between the $EW_0$ of \Has and \Hbs absorbers strongly points towards partial covering playing a significant role in these systems. Partial covering ($C_f < 1$) would enable absorption to approach saturation without reaching zero flux causing apparent optical depths to deviate from the ratios set by relative oscillator strengths \citep{Leighly2019}, erasing the correlation expected from atomic physics alone. This was already demonstrated in observations of individual LRDs \citep{DEugenio2025b} and shown for low redshift FeLoBALs and LRDs alike \citep{Leighly2025, Lin2026}. Aside from partial covering, lack of $\rm EW_0$ correlation is also expected if the absorption arises in the same clouds as the broad line emission. In this case, the \Has and \Hbs transitions probe different cloud depths and are subject to different emission infill as expected in a stratified BLR \citep{Li2017, Rakic2022}.

\begin{figure}
    \centering
    \includegraphics[width=\linewidth]{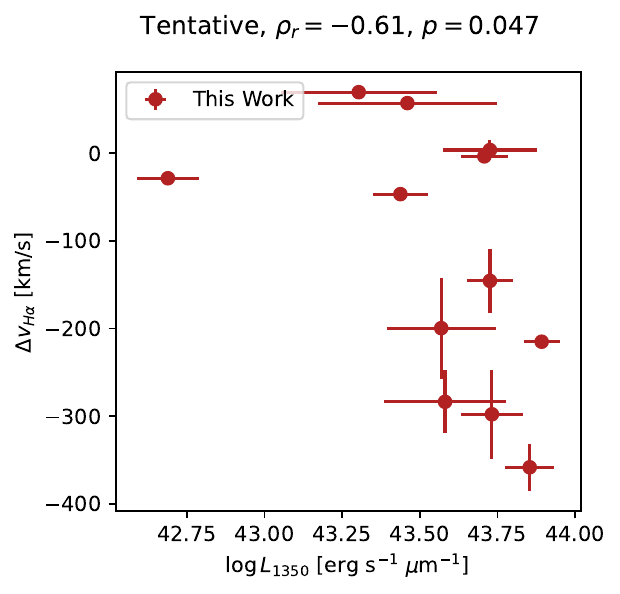}
    \caption{The tentative correlation between $\Delta v_{H\alpha}$ and $L_{1350}$. While the p-value is robust to outliers, the measurement errors give $p = 0.13$.}
    \label{fig:vel_L135_corr}
\end{figure}

The correlations between $\Delta v_{\rm H\alpha}$ and narrow line luminosities are more unexpected - the location of the absorber is generally thought to be located at $<2$~pc from the central ionizing source \citep{Juodzbalis2024, Naidu2025, Matthee2026}, corresponding to the BLR scale, while narrow line emission, whether powered by the host or AGN, is generally found to be extended to scales of $100$--$1000$~pc for both LRDs and regular Seyfert galaxies \citep{Maiolino2026, Bennert2006, Zanchettin2023, Ishikawa2026}. Hence, if the velocity offset and narrow line emission are both driven by energy injection from the central source, we would expect accompanying strong correlations between $\Delta v$ and broad \Has luminosity, which do not appear in our sample. We thus investigate whether either UV or optical continuum luminosity at 1350 and 5100~\AA\ respectively, correlates with $\Delta v_{H\alpha}$. We measure the respective luminosities from the prism spectra of our sample sources - overall 12 targets had the required prism coverage at 1350~\AA\ and 13  at 5100~\AA. We find no significant correlation between $\Delta v_{\rm H\alpha}$ and $L_{5100}$ ($p > 0.05$, $\rho_r = -0.2$). On the other hand, we find a correlation between $\Delta v_{\rm H\alpha}$ and $L_{1350}$ that is robust to outliers with $\rho_r = -0.61$ and $p = 0.046$, \autoref{fig:vel_L135_corr}. However, the relative faintness of LRD UV emission means that $L_{1350}$ measurements are considerably less precise, hence $p = 0.14$ once the measurement errors are taken into account. Therefore, while the correlations between $L_{\rm [OIII]}$, $\Delta v_{\rm H\alpha}$ and $L_{1350}$ are consistent with a common mechanism driving the UV, line emission and the outflow velocity, more precise measurements and larger samples are needed for a robust correlation analysis.

The tentative nature of the current correlation analysis of Balmer absorbers is reflected by the fact that we do not recover the reported tight correlation between $\Delta v_{\rm H\alpha}$ and Balmer break strength reported in \cite{Matthee2026} despite a 50\% sample overlap and the slightly larger statistics of our sample (15 versus 12 absorbers). While it is possible that this discrepancy stems from differing assumptions regarding the BLR profile shapes made by \cite{Matthee2026}, who fit all of their targets with the electron scattering model, we do not recover the \cite{Matthee2026} correlation even when fitting all sources with electron scattering profiles. Notably, \cite{Lin2026} also do not observe correlations between $\Delta v$ and Balmer break strength. However, the currently small sample sizes and resolution-limited nature of the measurements, which are mostly taken at $R < 2000$, imply that any correlations reported on current LRD samples should be considered preliminary and subject to confirmation. Therefore, we do not draw strong conclusions from the correlations observed in our sample.

\section{Discussion}
\label{sec:discussion}
\subsection{Are LRDs faint FeLoBAL QSOs?}
\begin{figure}
    \centering
    \includegraphics[width=\linewidth]{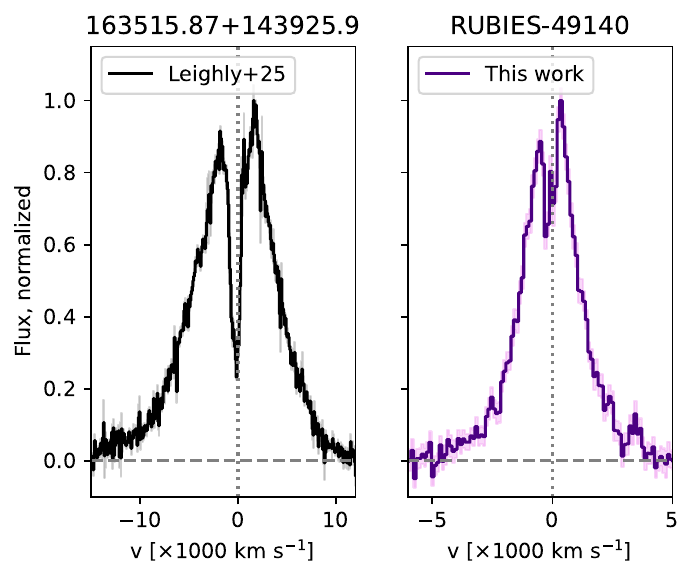}
    \caption{A qualitative comparison of broad \Has profiles of FeLoBAL SDSS 163515.87+143925.9 from \protect\cite{Leighly2025} and one of our LRDs, RUBIES-49140 (also known as 'Irony', \citealp{DEugenio2025b}). The horizontal dashed line denotes the zero flux threshold while the dotted vertical line marks the rest-frame velocity. It can be seen that, despite a factor of two difference in velocity scale, the FeLoBAL and Irony share a near rest-frame absorber and similar \Has profile shapes, including the slight asymmetry apparent at larger velocities.}
    \label{fig:profile_comparison}
\end{figure}

Throughout the previous section, a common trend emerges of LRDs appearing as a low luminosity extension of the FeLoBAL QSO (FeLoBAL) population. This was already noted by \cite{Leighly2025}, who point out that FeLoBALs share the reddened rest-frame optical SED and X-ray weakness of LRDs. In addition, the broad \Has profiles of LRDs and FeLoBALs are qualitatively similar (\autoref{fig:profile_comparison}). Hence, it is tempting to assume a unified LRD - QSO framework at play. However, lower luminosity and higher obscuring column densities are not the only areas where LRDs differ from FeLoBALs. While an in-depth comparison study is beyond the scope of this work, in this section we will discuss the similarities and differences between LRDs and FeLoBALs observed so far.

The most notable difference between LRDs and FeLoBALs is a lack of permitted \ion{Fe}{II} emission and absorption, which is ubiquitous in FeLoBAL systems \citep{Choi2022, Leighly2025}. LRDs instead, appear to display forbidden optical \FeII emission \citep{Ji2025LRDlocal, DEugenio2025b} with only hints of \ion{Fe}{II} absorption and permitted FeII emission \citep{Ji2026QSO} in the UV \citep{DEugenio2025b}; see also the stacks in \cite{PerezGonzalez2026}. Hence, the different iron emission patterns across LRDs and FeLoBALs can not be entirely the result of iron deficiency in the former. Intriguingly, high Balmer continuum optical depths can efficiently destroy \ion{Fe}{II} emission in the UV \citep{Netzer1983, Wills1985}, suggesting that \ion{Fe}{II} emission may be suppressed by the same gas that is producing Balmer absorption. However, the increased collisional excitation in such environments would, in turn, predict strong optical FeII emission \citep{Netzer1983}, which is absent in LRDs and other faint AGN \citep{Trefoloni2025}. This suggests that lower metallicity of the absorbing gas may be a factor suppressing the optical \ion{Fe}{II} bump in LRDs. This is not immediately inconsistent with the detections of forbidden optical and permitted UV FeII emission in LRDs as optical \FeII\ can be efficiently produced by shocks \citep{Knop1996}, while the lines seen in the UV can be produced by Ly$\alpha$ fluorescence \citep{Ji2026QSO}. Hence, significant further theoretical work is required to complete the picture of iron emission patterns in LRDs and establish the driving mechanisms behind the discrepancy with luminous QSOs.

Both LRDs and FeLoBALs display X-ray weakness. However, this weakness is considerably more pronounced in LRDs with only two high redshift `X-ray dots' known so far \citep{Hviding2026, Geris2026} along with one low redshift source \citep{Lin2025LRDlocal, Ji2025LRDlocal}, while the remainder of the population shows no X-ray detections. This results in an $L_{\rm bol}/L_{\rm [2-10keV]}$ ratio in stacked X-ray data exceeding 10000 \citep{Maiolino_xray_weak, Geris2026}. In contrast, FeLoBALs are more consistently X-ray detected and display more modest, yet still significant, X-ray weakness with observed $\langle L_{\rm bol}/L_{\rm [2-10keV]}\rangle \sim 1700$ \citep{Liu2018, Vito2018} and X-ray spectra consistent with obscuration by $N_{\rm H} \sim 10^{23}$~cm$^{-2}$ media \citep{Morabito2011, Rogerson2011}. Considering the $\sim$1~dex higher $N_{\rm H}(n=2)$ densities in LRDs, it is reasonable to suggest that the corresponding higher obscuration drives the greater X-ray weakness in these sources. However, LBDs are similarly X-ray weak while lacking the Balmer absorption features of LRDs. Therefore, the difference in the configuration or accretion properties of the central engines can not be discounted as the root cause of greater X-ray weakness in LRDs when compared to FeLoBALs.

In terms of bolometric luminosities and column densities, however, LRDs appear to strongly resemble the lower luminosity counterparts of FeLoBALs. In particular, they appear as a natural extension of the FeLoBAL population on the $L_{\rm bol}$ -- $N_{\rm H}(n=2)$ plane (\autoref{fig:Nh}). We investigate this resemblance further by comparing the energetics of outflows traced by LRD and FeLoBAL absorbers in the following section.

\subsection{Absorber geometry and energetics}
Motivated by the picture revealed by \autoref{fig:Nh}, we explore deeper the apparent connections between the LRD/FeLoBAL absorber properties and the bolometric luminosities of the sources. We frame our discussion in terms of a simple toy outflow model following the formalism of \cite{Marconi2008} described below.

We consider a single geometrically thin spherical shell (or shell fragment) of hydrogen at a distance $R$ from the accreting BH with a mass $M_{\rm BH}$ and bolometric luminosity $L$. The shell has a number density of $n_H$, thickness of $\Delta R$ (column density $N_H \equiv \Delta R n_H$), a covering factor $C'_f$ and a radial velocity of $v$. We note that the covering factor here represents the covering as seen from the central source, hence it is not necessarily the same as in \autoref{eq:absorption}, which is sensitive to the covering from the view of the observer.

The radiative force acting on a shell is given by:
\begin{equation}
    F_{\rm rad} = \frac{LC'_f}{c}\int_0^{\infty}L_{\nu}(1-e^{\tau_{\nu}})d\nu = \frac{LC'_f}{c}(1-e^{\tau_{eff}}),
\end{equation}
where $L_{\nu}$ and $\tau_{\nu}$ are the frequency dependent monochromatic luminosity and optical depth respectively. The effective optical depth ($\tau_{\rm eff}$) can be approximated as $\tau_{\rm eff} = \sigma_{\rm eff}N_H$, where $\sigma_{\rm eff}$ is the effective luminosity-weighted interaction cross section per hydrogen nucleus. Further assuming that the absorber retains a constant physical area ($A = 4\pi R^2C'_f = const.$) while expanding and does not undergo significant lateral expansion or fragmentation such that $\tau_{\rm eff}$ and $N_H$ are constant. The absorber mass is then constant and given by:
\begin{equation}
    m = AN_H\mu m_p = 4\pi R^2C'_fN_H\mu m_p,
\end{equation}
where $\mu m_p$ is the mean mass per hydrogen nucleus. The final expression for $F_{\rm rad}$ thus becomes:
\begin{equation}
F_{\rm rad}(R)
=\frac{LA}{4\pi cR^2}
\left(1-e^{-\tau_{\rm eff}}\right),
\end{equation}
and therefore has the same \(R^{-2}\) dependence as the gravitational force,
\begin{equation}
F_{\rm grav}(R)
=
\frac{GM_{\rm BH}m}{R^2}.
\end{equation}

The force balance on the shell can thus be expressed as a ratio between $F_{\rm rad}$ and $F_{\rm grav}$:
\begin{equation}
\label{eq:force}
\Gamma_{\rm eff}
\equiv
\frac{F_{\rm rad}}{F_{\rm grav}}
=
\frac{L\left(1-e^{-\tau_{\rm eff}}\right)}{4\pi GM_{\rm BH}cN_H\mu m_p}.
\end{equation}
As both $N_H$ and $\tau_{eff}$ are assumed to remain constant, $\Gamma_{\rm eff}$ is independent of radius. The net energy of the absorber is then given by:
\begin{equation}
\label{eq:energy}
    E=\frac{1}{2}mv^2-\frac{GM_{\rm BH}m}{R}\left(1-\Gamma_{\rm eff}\right).
\end{equation}

The form of \autoref{eq:force} and \autoref{eq:energy} together prevent imposing exact force balance ($\Gamma_{\rm eff} = 1$) together with a parabolic escape trajectory ($E = 0$) for individual configurations. However, we interpret the observed distribution of absorber velocities, clustered around $\Delta v=0$ with a modest excess at $\Delta v<0$ (\autoref{fig:dv_dist}), as qualitatively suggesting that at least part of the absorber population may lie close to the transition between gravitationally bound and radiatively accelerated configurations with $\langle\Gamma_{\rm eff}\rangle = 1$. If the characteristic absorber velocity scales with the gravitational velocity ($v^2\propto {GM_{BH}}/{R}$), the force balance condition implies the following scalings between $L$ and absorber properties depending on whether the shell is optically thick or thin: 
\begin{equation}
\label{eq:Lum_product}
L \propto
\begin{cases}
      N_H v^2 R\ {\rm if\ \tau_{eff} \gg 1}\\
      v^2 R\ {\rm if\ \tau_{eff} \ll 1}
    \end{cases}\
\end{equation}
The above expression has no explicit dependence on the covering factor $C'_f$ and $M_{BH}$.

We can construct the products in \autoref{eq:Lum_product} from our measurements of $N_H(n=2)$ (assuming $N_H(n=2) \propto N_H$) and identifying the $v$ in \autoref{eq:Lum_product} with $\Delta v$. While neither our nor \cite{Leighly2025} measurements can constrain $R$, we proceed under the assumption that LRD and FeLoBAL outflows have similar physical scales, which is consistent with current (albeit extremely uncertain) observational estimates of $R \sim 2$~pc for LRDs \citep{Juodzbalis2024b} and $0.7\pm0.5$~kpc for FeLoBALs \citep{Sharma2025}. In addition, we note that the existence of a scaling relation between radius and luminosity ($R \propto L^{\alpha}$) would have the effect of flattening the slope of unity predicted by \autoref{eq:Lum_product} to $m = \frac{1}{1-\alpha}$. We estimate the $L_{bol}$ of our LRDs using the calibrations from \cite{SternLbol}, which state $L_{bol} = 130L_{H\alpha}$. While some recent results have raised doubts about the applicability of this scaling to LRDs \citep{Greene2026}, we note that their findings have been put in question by more recent results \citep{Sok2026, Geris2026, Ji2026QSO, Tang2026}.  In addition, our results hold even if we use $L_{H\alpha}$ directly, which we checked by converting the reported FeLoBAL $L_{bol}$ to $L_{H\alpha}$ via the \cite{SternLbol} relation. Hence, we use $L_{H\alpha}$ and $L_{bol}$ as proxies for $L$ interchangeably in the following analysis.

Utilizing the mean $N_H(n=2)$ and $|\Delta v|$ from \cite{Leighly2025} FeLoBAL sample and our LRDs (\autoref{tab:BAL_LRD_comparison}), we obtain a ratio of $(N_H\Delta v^2)_{LRD}/(N_H\Delta v^2)_{BAL} = 0.04$. This is in almost exact correspondence to the ratio of the mean bolometric luminosities with $\langle L_{LRD} \rangle/\langle L_{BAL} \rangle = 0.039$. While at first glance this might suggest that the absorbing shells must be optically thick, the results of Section~\ref{sec:vel_dist} show that $\Delta v^2$ appears to likewise scale with $L$ --implying an optically thin regime. Therefore, the current data is insufficient to determine the overall optical depth of the absorbing medium.

\begin{figure}
    \centering
    \includegraphics[width=\linewidth]{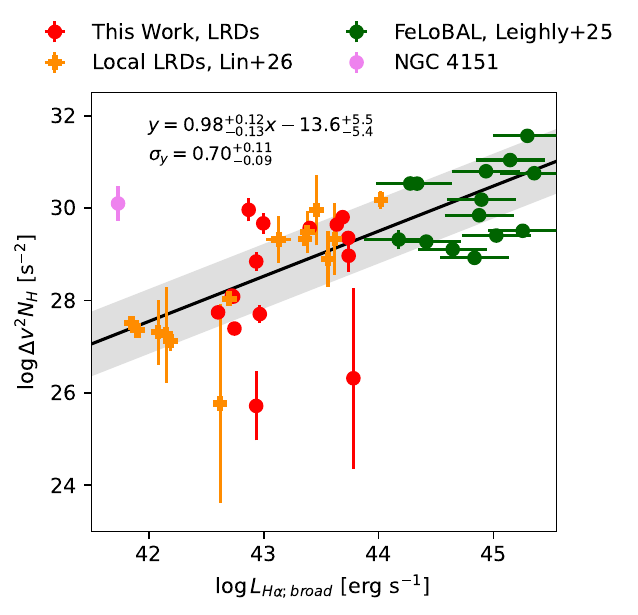}
    \caption{Plot of our sample LRDs (red), \protect\cite{Leighly2025} FeLoBALs (green) and local LRDs from \protect\cite{Lin2026} and Ji in prep. (orange) on the $L_{H\alpha}$ -- outflow energy plane. The black line shows the best fit linear regression which shows a slope consistent with 1 as expected from \autoref{eq:Lum_product}. The 0.7~dex scatter is illustrated in gray. The violet point shows NGC 4151 \protect\citep{Hutchings2002} with the error bar denoting the range of absorber velocity variation. NGC 4151 is a significant outlier with respect to the the LRD -- FeLoBAL relation.}
    \label{fig:LRD_BAL_energy}
\end{figure}

Investigating the scaling at sample level we find that LRDs (both those in our JWST sample and local ones from \citealp{Lin2026}) and FeLoBALs span the same relation between $L_{bol}$ and outflow energy (\autoref{fig:LRD_BAL_energy}) with the only outliers being the $\Delta v = 0$ absorbers from \cite{DEugenio2026} and \cite{DEugenio2025b}. The best-fit linear relation shows a slope $m = 0.98 \pm 0.12$ -- consistent with the expectations from \autoref{eq:Lum_product} although the scatter is large, with $\sigma_K = 0.70 \pm 0.1$~dex. This large scatter is likely driven by a combination of projection effects on $\Delta v$, imperfect correspondence between $N_H(n=2)$ and $N_H$ and variation in spatial scales.

\begin{figure}
    \centering
    \includegraphics[width=\linewidth]{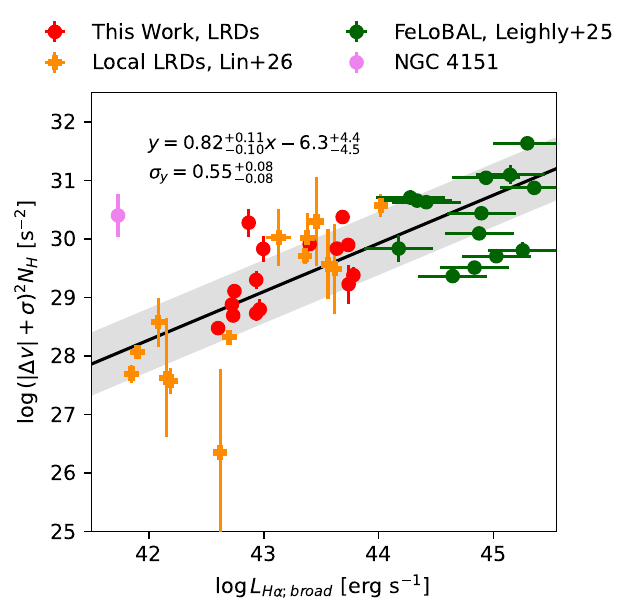}
    \caption{Same as \autoref{fig:LRD_BAL_energy} except with $v \equiv |\Delta v| + \sigma$. As can be seen here, the outlier fraction is considerably lower, corresponding to a lower scatter on the inferred relation. The slope remains consistent with 1, however, the slight flattening may be attributed to temperature effects or fit systematics of the $\sigma$ measurements. As in \autoref{fig:LRD_BAL_energy}, NGC 4151 (violet) remains an outlying point, possibly due to smaller size of its outflowing region.}
    \label{fig:LRD_BAL_energy_sigma}
\end{figure}

\begin{figure}
    \centering
    \includegraphics[width=\linewidth]{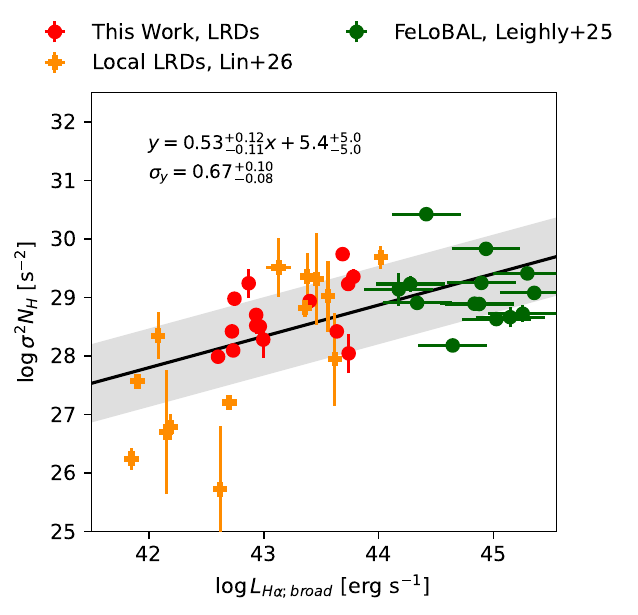}
    \caption{Same as \autoref{fig:LRD_BAL_energy} except with $v \equiv \sigma$. As shown here, the correlation remains, however, the slope of $0.53$ is considerably flatter than \autoref{fig:LRD_BAL_energy}, indicating secondary effects driving the broadening of the absorption features.}
    \label{fig:LRD_BAL_energy_only_sigma}
\end{figure}

If projection effects are a driving factor in the scatter, it may appear that defining $v \equiv |\Delta v| + \sigma$, with $\sigma$ representing the velocity dispersion of the absorption profile, should reduce the scatter. However, this may not necessarily be the case as $\sigma$ could be driven by intrinsic turbulent motion. Indeed, as shown in \autoref{fig:EWs}, the $v_{turb.}$ values inferred for our sample sources are similar in magnitude to the observed $\sigma$. In addition, as discussed in Section 3, $\sigma$ in \autoref{eq:depth_prof} is degenerate with $\tau_0$ and hence is poorly recovered by our fitting. Nevertheless, we repeat the earlier analysis using $v \equiv |\Delta v| + \sigma$ as the outflow velocity and find that the slope decreases slightly to $m = 0.82 \pm 0.11$ (\autoref{fig:LRD_BAL_energy_sigma}). While this is still consistent with unity within 2$\sigma$, this may indicate gas temperature effects contributing to the $\sigma$ value once bulk velocities become comparable to $v_{turb}$. Alternatively, as expected, outflow radius ($R$) may be decreasing for lower luminosity AGN, causing their tail to move up on the $v^2N_H$ axis as $R$ is currently unaccounted for by our measurements. On the other hand, purely instrumental effects, while likely present, can be discounted as the driving force behind the flattening as defining $v \equiv \sigma$ still yields a significant linear relation with $L_{\rm H\alpha;broad}$ (\autoref{fig:LRD_BAL_energy_only_sigma}). However, the slope on the $N_H\sigma^2$--$L_{\rm H\alpha;broad}$ $0.53_{-0.11}^{+0.12}$ is significantly flatter than in \autoref{fig:LRD_BAL_energy} and \autoref{fig:LRD_BAL_energy_sigma} suggesting a weaker correspondence between AGN luminosity and Doppler broadening of the absorber. It is currently unclear whether this is driven by physical or instrumental effects.

Assuming that the flattening of the slope in \autoref{fig:LRD_BAL_energy_sigma} is entirely due to a secondary relation between $R$ and $L$, we infer $\alpha = 0.26 \pm 0.16$ for this putative relation. This is consistent with slopes of $0.22$--$0.29$ inferred for ionized QSO outflows by \cite{Kim2023}. However, the large uncertainties on our estimate mean that a BLR-like $R-L$ relation with $\alpha = 0.5$ cannot be ruled out. Interestingly, the local Seyfert 1 AGN, NGC 4151, appears as a significant outlier in both figures, lying $\sim 3$~dex above the LRD -- FeLoBAL relation. The most likely explanation for this, given the current data, is that the absorbing region in NGC 4151 is considerably smaller than that of LRDs and FeLoBALs, resulting in a higher than expected value of $v^2N_H$ if $R$ is not taken into account. This is tentatively supported by the rapid variability of the NGC 4151 absorber over 3 month time scales found by \cite{Hutchings2002}, who also estimate the absorber to be located at 0.001~pc from the nucleus; ie $>$3~dex smaller than LRDs and FeLoBALs. However, as LRDs currently lack comparable monitoring of their absorption features, strong conclusions can not be drawn.

The $0.55_{-0.08}^{+0.08}$~dex scatter on the relation in \autoref{fig:LRD_BAL_energy_sigma} is considerably lower than the $0.7$~dex inferred in \autoref{fig:LRD_BAL_energy}. Assuming that the earlier scatter originates from projection effects and other sources added in quadrature, we obtain $\sigma_{proj.} = 0.43 \pm 0.06$~dex. Considering that projection angle $i$ is expected to be uniformly random on a sample level and assuming a conical outflow geometry, we can attempt to infer the opening angle of this cone through the relation:
\begin{equation}
    \Delta v = \Delta v_{int}\sin{\left(i + \frac{\theta}{2}\right)}
\end{equation}
where $\Delta v$ is the observed absorber velocity offset as defined by \autoref{eq:depth_prof} while $\Delta v_{int}$ is the intrinsic, deprojected one. Uniformly sampling $i \in [0, 90^{\circ}]$ we can iteratively solve the above equation for $\theta$, obtaining that $\theta = 54 \pm 9^{\circ}$ would give the required scatter. This is consistent with the opening angles of the dusty torus inferred for classical AGN, which range from 20$^{\circ}$ to 60$^{\circ}$ \citep{Arshakian2005}. This may indirectly point to a scenario in which the poles of the torus in LRDs are occupied by a dense reservoir of gas rather than the more tenuous medium found in the general AGN population. Alternatively, under the assumption that LRDs are LBDs viewed edge-on \citep{Madau2026_LRD_LBD, Geris2026}, our inferred $\theta$ is then the angle covered by the torus itself. However, considering the simplicity of the model discussed here, detailed interpretation is premature.

Regardless of the exact interpretation, LRD and FeLoBAL absorbers following the same luminosity scaling suggests that absorption in both is produced by radiative pressure driven outflows. In this scenario, LRD absorbers appear more common than their more luminous counterparts due to lower luminosities of their central engines struggling to drive the dense outflows out into the ISM, causing them to `loiter' around the nucleus and build up higher column densities. However, a more in-depth comparison across larger samples of both LRDs and FeLoBALs is required to further test this scenario. In particular, extensive R2700 observations are required to decouple $\tau$ from $\sigma$ of the absorption features and accurately capture the outflow energetics in LRDs.

\begin{table*}
    \centering
    \begin{tabular}{ccc}
    \hline
        Factor & LRDs & FeLoBALs \\
    \hline
         FeII & Forbidden \FeII in the optical, permitted in UV & Ubiquitous permitted FeII emission and absorption\\
        $\langle L_{\rm bol.}\rangle$ [erg~s$^{-1}$] & $10^{45.2}$ & $ 10^{46.9}$\\
        $\langle N_H(n=2) \rangle$ [cm$^{-2}$]& $10^{14.7}$ & $ 10^{13.9}$ \\
        $\langle |\Delta v_{H\alpha}| \rangle $ [km~s$^{-1}$]& $164$ & $1975$ \\
        $\langle L_{bol}/L_{2-10 keV} \rangle$& $>10^4$ &  $1700$\\
        \hline
    \end{tabular}
    \caption{Summary of the comparisons and mean properties of LRDs and FeLoBALs from \protect\cite{Leighly2025} discussed in the text.}
    \label{tab:BAL_LRD_comparison}
\end{table*}

\subsection{Parallels with emission line stars - the case for radiatively driven outflows}
AGN are not the only systems displaying outflowing \Has absorption. Indeed, luminous blue variable (LBV) stars were among the first systems in which such absorption has been seen \citep{Leitherer1987}. In LBV systems, the absorption is generated by dense shells of gas ejected during eruption events by radiation-driven winds \citep{Genderen2001} and residing at $\sim 10$~AU scales \citep{Balan2010}. Considering that P Cygni profiles were successfully used in modeling the \Has profiles of LRDs \citep{Matthee2026, Rusakov2025} and LRD physics have increasingly been linked with stellar processes \citep{Naidu2025, deGraaff2025, Liu2026, Naidu2026}, we further compare the energies of LRD and FeLoBAL outflows and winds of LBV stars.

Assembling a reasonable comparison sample of LBVs is not a straightforward task as these systems are known to strongly vary on timescales of months and years owing to their small sizes and unstable nature \citep{Kalari2018}. In addition, some of the parameters of the stellar wind models used to fit P Cygni profiles do not have clear correspondence with parameters of \autoref{eq:Lum_product}. Hence, we utilize the measured velocities of the P Cygni absorption minima as the $v$ parameter and construct our comparison sample from the time averaged $\Delta v$ and L$_{H\alpha}$ values of P Cygni, $\eta$ Car and NGC 2363-V1 \citep{Richardson2010, Richardson2011,Drissen2001} as well as four additional extragalactic LBVs reported by \cite{Guseva2022, Guseva2024}.

\begin{figure}
    \centering
    \includegraphics[width=\linewidth]{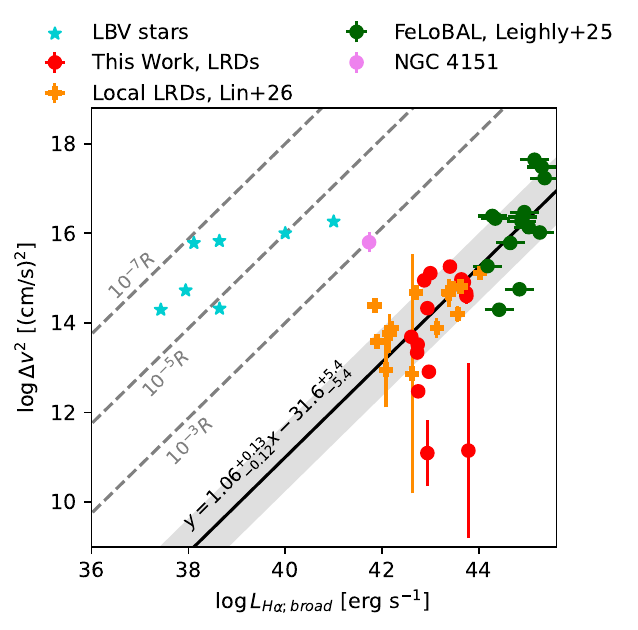}
    \caption{Comparison of the absorber velocity -- broad H$\alpha$ luminosity for our LRDs (red points), local LRDs (orange pluses), FeLoBALs (green points), NGC 4151 (violet point) and LBV stars (blue stars). The solid black line shows the best-fit relation for the AGN sequence with grey shading illustrating the $0.75_{-0.09}^{+0.12}$~dex scatter. The dashed gray lines illustrate the same relation shifted by different physical scales relative to the $R$ for LRDs and FeLoBALs.}
    \label{fig:black_hole_star}
\end{figure}

The comparison of the LBV stars, our LRDs and \cite{Leighly2025} FeLoBALs is presented in \autoref{fig:black_hole_star} and shows that AGN and LBVs are consistent with the $L \propto \Delta v^2$ relation expected by the simplified model of \autoref{eq:Lum_product}, yet LBVs are offset from the LRD -- FeLoBAL sequence by approximately 5~dex with the local Seyfert 1 AGN, NGC 4151, located in between. This is most readily explained by a difference in spatial scales across these systems -- the LBV shells have been resolved to be extended on $\sim$10~AU scales \citep{Balan2010} and the current estimates for NGC 4151 place its absorber at $200$~AU from its center \citep{Hutchings2002}. On the other hand, LRD and FeLoBAL absorbers are expected to lie at 1-2~pc or even greater distances from the central engine \citep{Juodzbalis2024b, Sharma2025}. Following \autoref{eq:Lum_product}, these factors of $10^{5}$--$10^{3}$ difference in spatial scales, caused by different lifetimes of AGN episodes and LBV star eruptions, would naturally account for the offset seen on the $L_{H\alpha}$ -- $\Delta v^2$ plane. On the other hand, a similarly extreme variation in $N_H$ is unlikely, given that both AGN and stars show similar \Has optical depths and hence likely have similar $N_H(n=2)$ values.

The above analysis, however, is still preliminary as our LBV comparison sample is not homogeneous. In addition, at least some LBV eruptions are likely catalyzed by the presence of a companion \citep{Mehner2010,Mehner2021}. Therefore, at least a fraction of the energy in the outflowing shell of some LBVs may be injected from tidal interaction with a companion rather than purely radiatively. The correspondence between these sources breaks down further when considering that the continuum emission of LBVs, by definition, is extremely blue, while LRDs are optically red. Thus, while there are hints that LBV outflows and AGN outflows may be launched through a similar physical mechanism -- radiation pressure from a point-like luminous source, a more comprehensive comparison study is needed to conclusively verify the connection between the physics of AGN and stellar absorbers.

\section{Conclusions}
\label{sec:conclusions}
In conclusion, the search for Balmer absorbers in JWST AGN is affected by significant incompleteness, particularly for absorption features observed at rest-frame ($\Delta v = 0$), due to narrow line infill. However, the current data sets are sufficient to show that strong \Has absorbers are a feature of LRDs rather than LBDs with LRDs approaching incidence rates of 40 -- 70\%, while we constrain an upper limit of 25\% incidence of \Has absorption in LBDs. While this is consistent with LRDs showing Balmer breaks in their continua  \citep{Setton2025, Naidu2025}, which are absent in LBDs \citep{Brazzini2026}, both types of faint AGN are similarly X-ray weak \citep{Maiolino_xray_weak, Brazzini2026, Geris2026}. This suggests that there may be more to the X-ray weakness of faint high redshift JWST AGN than simple obscuration by high hydrogen column densities. While recent theoretical efforts by \cite{Sneppen2026} have shown that apparently unabsorbed broad line profiles can be produced by electron scattering in a high column density environment, we find, consistently with \cite{Scholtz2026}, that exponential broad line profiles are not universally preferred for our sample sources while electron scattering is not the unique interpretation of their shapes \citep{Scholtz2026, Madau2026_LRD_LBD, Madau2026_LRD_LBD2}. Hence, it appears that scenarios invoking radiatively inefficient super-Eddington accretion \citep{Madau2024A, Madau2026} or a missing/weak hot corona \citep{Maiolino_xray_weak} in combination with absorption are the more likely explanations for the near universal X-ray weakness of high redshift AGN.

We also find that the current JWST spectroscopic data sets, which are mostly comprised of R1000 observations, are strongly resolution limited when it comes to accurately assessing the finer properties of \Hbs and \Has absorbers found in LRDs with only velocity offsets and equivalent widths reliably recoverable. The higher resolution R2700 gratings mitigate this issue, yet still struggle with recovering weaker rest-frame absorbers with narrow \Has superimposed. This implies that the possibility of all LRDs having \Has absorption can not be discounted by the current data. However, R2700 performs substantially better than R1000 in terms of disentangling Doppler broadening from the effects of optical depth and covering fraction. Hence, large samples of LRDs with R2700 data capturing absorption in multiple hydrogen transitions are required to fully disentangle the $\tau$ -- $\sigma$ -- $C_f$ degeneracy and constrain the effects of partial covering and emission line infill on the absorber properties. Additionally, in light of the discovery of multiple LRDs at $z < 0.3$ \citep{Ji2025LRDlocal, Lin2025LRDlocal}, ground based high resolution ($R \sim 10000$) follow up of these sources offers another highly promising avenue of constraining the physics of the absorbing medium.

In light of the above findings, the statistical power of our sample is rather limited in terms of in-depth constraints on the absorber properties. However, we find that $\rm EW_0$ and $\Delta v$ are reasonably well recovered in the current data, with similar performance between R1000 and R2700 data. The demographics of our sample across these two parameters reveal that, while absorption is predominantly within $|\Delta v| < 100$~km~s$^{-1}$, once the incompleteness is accounted for, the absorber distribution is skewed towards blueshifted outflows of $\Delta v \approx -300$~km~s$^{-1}$ with no absorbers at similar 300~km~s$^{-1}$ redshifts, \autoref{fig:dv_dist}. This suggests that the absorption does not originate in a quasi static atmosphere as proposed by the BH* model \citep{Naidu2025, Liu2026}, but instead resembles the `Loitering outflows' in FeLoBAL QSOs \citep{Leighly2025}. In addition, we find that $\rm EW_0$ of \Has absorption is  uncorrelated with Balmer break strength and considerably larger than stellar atmosphere models predict. This implies either different origins of the UV and optical continua or a diversity in the physical conditions and covering fractions of the absorbing medium -- in line with the models of \cite{Madau2026_LRD_LBD, Madau2026_LRD_LBD2}.

The small sample size and intrinsic uncertainties in the fitting mean that extensive discussion of correlations of the absorber properties is largely premature at this stage, with sample sizes larger by a factor of 2-3 required to search for robust relations between the observed properties. Nevertheless, we identify a clear, albeit expected, correlation between $\Delta v_{\rm H\alpha}$ and $\Delta v_{\rm H\beta}$ - consistent with \Has and \Hb\ absorbers tracing the same gas. The significant deviation of the relation between these velocities together with a lack of correlation between $\rm EW_0$ of the two lines indicates that partial covering and stratification of the absorbing medium are at play - consistent with the earlier findings that the sample absorbers span a spectrum of gas conditions and likely represent clumpy and non-uniform media rather than quasi-static shells of gas. Additionally, we identify a significant correlation between $\Delta v_{\rm H\alpha}$ and $L_{\rm [OIII]}$, potentially implying a common mechanism driving narrow line emission and outflow energetics. If confirmed, this would be direct evidence of the radiation from the BLR escaping into the surrounding medium. Yet, we see only a tentative correlation between $L_{1350}$ and $\Delta v_{\rm H\alpha}$ while no relation between $L_{\rm [OIII]}$ and broad \Has luminosity or $L_{5100}$ is seen. Hence, further investigation is needed to test whether the observed $\Delta v_{\rm H\alpha}$ -- $L_{\rm [OIII]}$ correlation is robust. In addition, we do not confirm the correlations claimed in the literature \citep[e.g.][]{Matthee2026} further highlighting the tentative nature of current statistical analyses.

In terms of absorber energetics, LRDs appear to closely resemble lower luminosity counterparts of FeLoBAL QSOs, occupying the low luminosity -- high $N_H(n=2)$ region of the parameter space (\autoref{fig:Nh}). The energetics of the absorbing media of both types of sources appear to be consistent with radiation driven outflows (\autoref{fig:LRD_BAL_energy}). In particular, the relation between \Has luminosities and absorber kinetic energies in LRDs and FeLoBALs mimics that of luminous blue variable (LBV) stars (\autoref{fig:black_hole_star}), with the offset accountable for by the currently loosely constrained outflow radii ($R$). This finding offers a potential avenue towards a unified LRD-QSO framework in which the prevalence and column densities of Balmer absorbers in LRDs can be accounted for through a combination of lower luminosities and lower metallicities producing an abundance of loitering outflows. Such high density, low metallicity environments may also be able to explain the lack of standard FeII emission in LRDs through suppression of the UV and optical FeII emission by Balmer continuum absorption and reduced presence of iron respectively. However, a more careful analysis of the properties of these two AGN populations is needed to fully establish whether a unified framework is at play. In particular, high resolution (R2700) JWST observations of LRDs are required in order to decouple absorber kinematics from radiative transfer effects, while constraints on FeII emission properties in LRDs will generally require deep low and medium resolution spectra.

Lastly, the parallels between nuclear AGN outflows and those driven by LBV stars offer an intriguing window into how radiatively driven outflows operate across different spatial scales. However, the current comparisons are limited as only $v$ and $N_{\rm H}(n=2)$ can be constrained by the current data. Therefore, observations of local LRDs, luminous FeLoBALs and LBV stars with state of the art optical interferometers, like GRAVITY+ \citep{Gravity22, Gravity26}, aimed at resolving the \Has absorbers are needed to establish whether AGN and stellar outflows truly share the same scaling relation. 

\section*{Acknowledgements}
IJ acknowledges support by the Huo Family Foundation through a P.C. Ho PhD Studentship. XJ, FDE, JS and RM acknowledge support by the Science and Technology Facilities Council (STFC), by the ERC through Advanced Grant 695671 “QUENCH”, and by the UKRI Frontier Research grant RISEandFALL. RM also acknowledges funding from a research professorship from the Royal Society. EB acknowledges funding from the Bando Ricerca Fondamentale INAF 2024 (GO grant 
 ``A JWST/MIRI MIRACLE: Mid-IR Activity of Circumnuclear Line Emission'' and MiniGrant RSN1 1.05.24.07.01). AJB acknowledges funding from the "FirstGalaxies" Advanced Grant from the European Research Council (ERC) under the European Union’s Horizon 2020 research and innovation programme (Grant agreement No. 789056). SC and GV acknowledge support by European Union’s HE ERC Starting Grant No. 101040227 - WINGS. GC acknowledges support from the INAF GO grant 2024 “A JWST/MIRI MIRACLE: MidIR Activity of Circumnuclear Line Emission”. ECL acknowledges support of an STFC Webb Fellowship (ST/W001438/1). BER, ZM and JZ acknowledge support from the NIRCam Science Team contract to the University of Arizona, NAS5-02105, and JWST Program 3215. H\"U acknowledges support by the Max Planck Society through the Lise Meitner Excellence Program. H\"U acknowledges funding by the European Union (ERC APEX, 101164796). Views and opinions expressed are however those of the authors only and do not necessarily reflect those of the European Union or the European Research Council Executive Agency. Neither the European Union nor the granting authority can be held responsible for them. GV acknowledges financial support from the Italian National Institute for Astrophysics (INAF) under the IAF - Astrophysics Fellowships in Italy grant CUP C59J21034720001 (AD MAJORA).
\section*{Data Availability}
All data used in this study is publicly available as part of JADES DR4 \citep{Curtis-Lake2026} as well as in MAST \url{https://mast.stsci.edu/portal/Mashup/Clients/Mast/Portal.html} and the Dawn JWST archive -- \url{https://dawn-cph.github.io/dja/}.



\bibliographystyle{mnras}
\bibliography{example} 

@article{oke+gunn1983,
 adsurl = {https://ui.adsabs.harvard.edu/abs/1983ApJ...266..713O},
 author = {{Oke}, J.~B. and {Gunn}, J.~E.},
 doi = {10.1086/160817},
 journal = {\apj},
 month = {March},
 pages = {713-717},
 title = {{Secondary standard stars for absolute spectrophotometry.}},
 volume = {266},
 year = {1983}
}

@ARTICLE{Gravity26,
       author = {{GRAVITY+ Collaboration} and {Abuter}, R. and {Allouche}, F. and {Bailet}, C. and {Benisty}, M. and {Berdeu}, A. and {Berger}, J.-P. and {Berio}, P. and {Bigioli}, A. and {Blanchard}, C. and {Boebion}, O. and {Bonnet}, H. and {Bourdarot}, G. and {Bourget}, P. and {Brandner}, W. and {Brul{\'e}}, J. and {Burgos}, P. and {Carbillet}, M. and {Correia}, C. and {Courtney-Barrer}, B. and {Curaba}, S. and {Davies}, R. and {Defr{\`e}re}, D. and {Delboulb{\'e}}, A. and {Delplancke}, F. and {Dembet}, R. and {Drescher}, A. and {Dubost}, N. and {Eckart}, A. and {{\'E}douard}, C. and {Eisenhauer}, F. and {Esteras Otal}, L. and {Fabricius}, M. and {Feuchtgruber}, H. and {F{\'e}dou}, P. and {Finger}, G. and {Schreiber}, N.~M. F{\"o}rster and {Frahm}, R. and {Garcia}, E. and {Garcia}, P. and {Lopez}, R. Garcia and {Genzel}, R. and {Gil}, J.~P. and {Gillessen}, S. and {Gomes}, T. and {Gont{\'e}}, F. and {Gopinath}, V. and {Gouvret}, C. and {Graf}, J. and {Guajardo}, P. and {Guieu}, S. and {Hackenberg}, W. and {Hartl}, M. and {Haubois}, X. and {Hau{\ss}mann}, F. and {Henning}, T. and {Hibon}, P. and {H{\"o}nig}, S. and {Horrobin}, M. and {Houll{\'e}}, M. and {Hubin}, N. and {Taieb}, I. Ibn and {Jochum}, L. and {Jocou}, L. and {Jost}, A. and {Kammerer}, J. and {Karl}, L. and {Kaufer}, A. and {Kern}, P. and {Kervella}, P. and {Kolb}, J. and {Korhonen}, H. and {Kreidberg}, L. and {Krempl}, P. and {Lacour}, S. and {Lagarde}, S. and {Lai}, O. and {Lapeyr{\`e}re}, V. and {Laugier}, R. and {Leal}, V. and {Le Bouquin}, J.-B. and {Leftley}, J. and {L{\'e}na}, P. and {Lopez}, B. and {Lutz}, D. and {Magnard}, Y. and {Mang}, F. and {Marcotto}, A. and {Maurel}, D. and {M{\'e}rand}, A. and {Millour}, F. and {Montarges}, M. and {More}, N. and {Moruj{\~a}o}, N. and {Moulin}, T. and {Nowacki}, H. and {Nowak}, M. and {Oberti}, S. and {Ott}, T. and {Pallanca}, L. and {Patru}, F. and {Paumard}, T. and {Perraut}, K. and {Perrin}, G. and {Petrucci}, P.~O. and {Petrov}, R. and {Pfuhl}, O. and {Pourr{\'e}}, N. and {Rabien}, S. and {Rau}, C. and {Riquelme}, M. and {Robbe-Dubois}, S. and {Rochat}, S. and {Salman}, M. and {S{\'a}nchez-Berm{\'u}dez}, J. and {Schubert}, J. and {Scigliuto}, J. and {Shchekaturov}, P. and {Schuhler}, N. and {Shangguan}, J. and {Shimizu}, T. and {Scheithauer}, S. and {Soenke}, C. and {Soulez}, F. and {Stadler}, E. and {Stadler}, J. and {Straubmeier}, C. and {Sturm}, E. and {Subroweit}, M. and {Sykes}, C. and {Tacconi}, L.~J. and {Tristram}, K.~R.~W. and {Uysal}, S. and {von Fellenberg}, S. and {Widmann}, F. and {Wieprecht}, E. and {Wiezorrek}, E. and {Woillez}, J. and {Yazici}, S. and {Zins}, G.},
        title = "{First light for the GRAVITY+ Adaptive Optics: Extreme adaptive optics for the Very Large Telescope Interferometer}",
      journal = {\aap},
         year = 2026,
        month = mar,
       volume = {707},
          eid = {A115},
        pages = {A115},
          doi = {10.1051/0004-6361/202555666},
archivePrefix = {arXiv},
       eprint = {2509.21431},
 primaryClass = {astro-ph.IM},
       adsurl = {https://ui.adsabs.harvard.edu/abs/2026A&A...707A.115G}
}

@ARTICLE{Gravity22,
       author = {{GRAVITY+ Collaboration} and {Abuter}, R. and {Allouche}, F. and {Amorim}, A. and {Bailet}, C. and {Baub{\"o}ck}, M. and {Berger}, J.-P. and {Berio}, P. and {Bigioli}, A. and {Boebion}, O. and {Bolzer}, M.~L. and {Bonnet}, H. and {Bourdarot}, G. and {Bourget}, P. and {Brandner}, W. and {Cl{\'e}net}, Y. and {Courtney-Barrer}, B. and {Dallilar}, Y. and {Davies}, R. and {Defr{\`e}re}, D. and {Delboulb{\'e}}, A. and {Delplancke}, F. and {Dembet}, R. and {de Zeeuw}, P.~T. and {Drescher}, A. and {Eckart}, A. and {{\'E}douard}, C. and {Eisenhauer}, F. and {Fabricius}, M. and {Feuchtgruber}, H. and {Finger}, G. and {F{\"o}rster Schreiber}, N.~M. and {Garcia}, E. and {Garcia}, P. and {Gao}, F. and {Gendron}, E. and {Genzel}, R. and {Gil}, J.~P. and {Gillessen}, S. and {Gomes}, T. and {Gont{\'e}}, F. and {Gouvret}, C. and {Guajardo}, P. and {Guieu}, S. and {Hartl}, M. and {Haubois}, X. and {Hau{\ss}mann}, F. and {Hei{\ss}el}, G. and {Henning}, Th. and {Hippler}, S. and {H{\"o}nig}, S. and {Horrobin}, M. and {Hubin}, N. and {Jacqmart}, E. and {Jochum}, L. and {Jocou}, L. and {Kaufer}, A. and {Kervella}, P. and {Korhonen}, H. and {Kreidberg}, L. and {Lacour}, S. and {Lagarde}, S. and {Lai}, O. and {Lapeyr{\`e}re}, V. and {Laugier}, R. and {Le Bouquin}, J.-B. and {Leftley}, J. and {L{\'e}na}, P. and {Lutz}, D. and {Mang}, F. and {Marcotto}, A. and {Maurel}, D. and {M{\'e}rand}, A. and {Millour}, F. and {More}, N. and {Nowacki}, H. and {Nowak}, M. and {Oberti}, S. and {Ott}, T. and {Pallanca}, L. and {Pasquini}, L. and {Paumard}, T. and {Perraut}, K. and {Perrin}, G. and {Petrov}, R. and {Pfuhl}, O. and {Pourr{\'e}}, N. and {Rabien}, S. and {Rau}, C. and {Robbe-Dubois}, S. and {Rochat}, S. and {Salman}, M. and {Sch{\"o}ller}, M. and {Schubert}, J. and {Schuhler}, N. and {Shangguan}, J. and {Shimizu}, T. and {Scheithauer}, S. and {Sevin}, A. and {Soulez}, F. and {Spang}, A. and {Stadler}, E. and {Stadler}, J. and {Straubmeier}, C. and {Sturm}, E. and {Tacconi}, L.~J. and {Tristram}, K.~R.~W. and {Vincent}, F. and {von Fellenberg}, S. and {Uysal}, S. and {Widmann}, F. and {Wieprecht}, E. and {Wiezorrek}, E. and {Woillez}, J. and {Yazici}, S. and {Young}, A. and {Zins}, G.},
        title = "{First light for GRAVITY Wide. Large separation fringe tracking for the Very Large Telescope Interferometer}",
      journal = {\aap},
         year = 2022,
        month = sep,
       volume = {665},
          eid = {A75},
        pages = {A75},
          doi = {10.1051/0004-6361/202243941},
archivePrefix = {arXiv},
       eprint = {2206.00684},
 primaryClass = {astro-ph.IM},
       adsurl = {https://ui.adsabs.harvard.edu/abs/2022A&A...665A..75G}
}

@ARTICLE{Maiolino24_GN-z11,
       author = {{Maiolino}, Roberto and {Scholtz}, Jan and {Witstok}, Joris and {Carniani}, Stefano and {D'Eugenio}, Francesco and {de Graaff}, Anna and {{\"U}bler}, Hannah and {Tacchella}, Sandro and {Curtis-Lake}, Emma and {Arribas}, Santiago and {Bunker}, Andrew and {Charlot}, St{\'e}phane and {Chevallard}, Jacopo and {Curti}, Mirko and {Looser}, Tobias J. and {Maseda}, Michael V. and {Rawle}, Timothy D. and {Rodr{\'\i}guez del Pino}, Bruno and {Willott}, Chris J. and {Egami}, Eiichi and {Eisenstein}, Daniel J. and {Hainline}, Kevin N. and {Robertson}, Brant and {Williams}, Christina C. and {Willmer}, Christopher N.~A. and {Baker}, William M. and {Boyett}, Kristan and {DeCoursey}, Christa and {Fabian}, Andrew C. and {Helton}, Jakob M. and {Ji}, Zhiyuan and {Jones}, Gareth C. and {Kumari}, Nimisha and {Laporte}, Nicolas and {Nelson}, Erica J. and {Perna}, Michele and {Sandles}, Lester and {Shivaei}, Irene and {Sun}, Fengwu},
        title = "{A small and vigorous black hole in the early Universe}",
      journal = {\nat},
         year = 2024,
        month = mar,
       volume = {627},
       number = {8002},
        pages = {59-63},
          doi = {10.1038/s41586-024-07052-5},
archivePrefix = {arXiv},
       eprint = {2305.12492},
 primaryClass = {astro-ph.GA},
       adsurl = {https://ui.adsabs.harvard.edu/abs/2024Natur.627...59M}
}

@ARTICLE{Maiolino_AGN,
       author = {{Maiolino}, Roberto and {Scholtz}, Jan and {Curtis-Lake}, Emma and {Carniani}, Stefano and {Baker}, William and {de Graaff}, Anna and {Tacchella}, Sandro and {{\"U}bler}, Hannah and {D'Eugenio}, Francesco and {Witstok}, Joris and {Curti}, Mirko and {Arribas}, Santiago and {Bunker}, Andrew J. and {Charlot}, St{\'e}phane and {Chevallard}, Jacopo and {Eisenstein}, Daniel J. and {Egami}, Eiichi and {Ji}, Zhiyuan and {Jones}, Gareth C. and {Lyu}, Jianwei and {Rawle}, Tim and {Robertson}, Brant and {Rujopakarn}, Wiphu and {Perna}, Michele and {Sun}, Fengwu and {Venturi}, Giacomo and {Williams}, Christina C. and {Willott}, Chris},
        title = "{JADES: The diverse population of infant black holes at 4 < z < 11: Merging, tiny, poor, but mighty}",
      journal = {\aap},
         year = 2024,
        month = nov,
       volume = {691},
          eid = {A145},
        pages = {A145},
          doi = {10.1051/0004-6361/202347640},
archivePrefix = {arXiv},
       eprint = {2308.01230},
 primaryClass = {astro-ph.GA},
       adsurl = {https://ui.adsabs.harvard.edu/abs/2024A&A...691A.145M}
}

@ARTICLE{Kocevski_AGN,
       author = {{Kocevski}, Dale D. and {Onoue}, Masafusa and {Inayoshi}, Kohei and {Trump}, Jonathan R. and {Arrabal Haro}, Pablo and {Grazian}, Andrea and {Dickinson}, Mark and {Finkelstein}, Steven L. and {Kartaltepe}, Jeyhan S. and {Hirschmann}, Michaela and {Aird}, James and {Holwerda}, Benne W. and {Fujimoto}, Seiji and {Juneau}, St{\'e}phanie and {Amor{\'\i}n}, Ricardo O. and {Backhaus}, Bren E. and {Bagley}, Micaela B. and {Barro}, Guillermo and {Bell}, Eric F. and {Bisigello}, Laura and {Calabr{\`o}}, Antonello and {Cleri}, Nikko J. and {Cooper}, M.~C. and {Ding}, Xuheng and {Grogin}, Norman A. and {Ho}, Luis C. and {Hutchison}, Taylor A. and {Inoue}, Akio K. and {Jiang}, Linhua and {Jones}, Brenda and {Koekemoer}, Anton M. and {Li}, Wenxiu and {Li}, Zhengrong and {McGrath}, Elizabeth J. and {Molina}, Juan and {Papovich}, Casey and {P{\'e}rez-Gonz{\'a}lez}, Pablo G. and {Pirzkal}, Nor and {Wilkins}, Stephen M. and {Yang}, Guang and {Yung}, L.~Y. Aaron},
        title = "{Hidden Little Monsters: Spectroscopic Identification of Low-mass, Broad-line AGNs at z > 5 with CEERS}",
      journal = {\apjl},
         year = 2023,
        month = sep,
       volume = {954},
       number = {1},
          eid = {L4},
        pages = {L4},
          doi = {10.3847/2041-8213/ace5a0},
archivePrefix = {arXiv},
       eprint = {2302.00012},
 primaryClass = {astro-ph.GA},
       adsurl = {https://ui.adsabs.harvard.edu/abs/2023ApJ...954L...4K}
}

@ARTICLE{Furtak2023_AGN,
       author = {{Furtak}, Lukas J. and {Labb{\'e}}, Ivo and {Zitrin}, Adi and {Greene}, Jenny E. and {Dayal}, Pratika and {Chemerynska}, Iryna and {Kokorev}, Vasily and {Miller}, Tim B. and {Goulding}, Andy D. and {de Graaff}, Anna and {Bezanson}, Rachel and {Brammer}, Gabriel B. and {Cutler}, Sam E. and {Leja}, Joel and {Pan}, Richard and {Price}, Sedona H. and {Wang}, Bingjie and {Weaver}, John R. and {Whitaker}, Katherine E. and {Atek}, Hakim and {Bogd{\'a}n}, {\'A}kos and {Charlot}, St{\'e}phane and {Curtis-Lake}, Emma and {van Dokkum}, Pieter and {Endsley}, Ryan and {Feldmann}, Robert and {Fudamoto}, Yoshinobu and {Fujimoto}, Seiji and {Glazebrook}, Karl and {Juneau}, St{\'e}phanie and {Marchesini}, Danilo and {Maseda}, Micheal V. and {Nelson}, Erica and {Oesch}, Pascal A. and {Plat}, Ad{\`e}le and {Setton}, David J. and {Stark}, Daniel P. and {Williams}, Christina C.},
        title = "{A high black-hole-to-host mass ratio in a lensed AGN in the early Universe}",
      journal = {\nat},
         year = 2024,
        month = apr,
       volume = {628},
       number = {8006},
        pages = {57-61},
          doi = {10.1038/s41586-024-07184-8},
archivePrefix = {arXiv},
       eprint = {2308.05735},
 primaryClass = {astro-ph.GA},
       adsurl = {https://ui.adsabs.harvard.edu/abs/2024Natur.628...57F}
}

@ARTICLE{Bunker2023_nirspec,
       author = {{Bunker}, Andrew J. and {Cameron}, Alex J. and {Curtis-Lake}, Emma and {Jakobsen}, Peter and {Carniani}, Stefano and {Curti}, Mirko and {Witstok}, Joris and {Maiolino}, Roberto and {D'Eugenio}, Francesco and {Looser}, Tobias J. and {Willott}, Chris and {Bonaventura}, Nina and {Hainline}, Kevin and {Uebler}, Hannah and {Willmer}, Christopher N.~A. and {Saxena}, Aayush and {Smit}, Renske and {Alberts}, Stacey and {Arribas}, Santiago and {Baker}, William M. and {Baum}, Stefi and {Bhatawdekar}, Rachana and {Bowler}, Rebecca A.~A. and {Boyett}, Kristan and {Charlot}, Stephane and {Chen}, Zuyi and {Chevallard}, Jacopo and {Circosta}, Chiara and {DeCoursey}, Christa and {de Graaff}, Anna and {Egami}, Eiichi and {Eisenstein}, Daniel J. and {Endsley}, Ryan and {Ferruit}, Pierre and {Giardino}, Giovanna and {Hausen}, Ryan and {Helton}, Jakob M. and {Hviding}, Raphael E. and {Ji}, Zhiyuan and {Johnson}, Benjamin D. and {Jones}, Gareth C. and {Kumari}, Nimisha and {Laseter}, Isaac and {Luetzgendorf}, Nora and {Maseda}, Michael V. and {Nelson}, Erica and {Parlanti}, Eleonora and {Perna}, Michele and {Rawle}, Tim and {Rix}, Hans-Walter and {Rieke}, Marcia and {Robertson}, Brant and {Rodriguez Del Pino}, Bruno and {Sandles}, Lester and {Scholtz}, Jan and {Sharpe}, Katherine and {Skarbinski}, Maya and {Stark}, Daniel P. and {Sun}, Fengwu and {Tacchella}, Sandro and {Topping}, Michael W. and {Villanueva}, Natalia C. and {Wallace}, Imaan E.~B. and {Williams}, Christina C. and {Woodrum}, Charity},
        title = "{JADES NIRSpec Initial Data Release for the Hubble Ultra Deep Field: Redshifts and Line Fluxes of Distant Galaxies from the Deepest JWST Cycle 1 NIRSpec Multi-Object Spectroscopy}",
      journal = {arXiv e-prints},
         year = 2023,
        month = jun,
          eid = {arXiv:2306.02467},
        pages = {arXiv:2306.02467},
          doi = {10.48550/arXiv.2306.02467},
archivePrefix = {arXiv},
       eprint = {2306.02467},
 primaryClass = {astro-ph.GA},
       adsurl = {https://ui.adsabs.harvard.edu/abs/2023arXiv230602467B}
}

@ARTICLE{Juodzbalis2024,
       author = {{Juod{\v{z}}balis}, Ignas and {Maiolino}, Roberto and {Baker}, William M. and {Tacchella}, Sandro and {Scholtz}, Jan and {D'Eugenio}, Francesco and {Witstok}, Joris and {Schneider}, Raffaella and {Trinca}, Alessandro and {Valiante}, Rosa and {DeCoursey}, Christa and {Curti}, Mirko and {Carniani}, Stefano and {Chevallard}, Jacopo and {de Graaff}, Anna and {Arribas}, Santiago and {Bennett}, Jake S. and {Bourne}, Martin A. and {Bunker}, Andrew J. and {Charlot}, St{\'e}phane and {Jiang}, Brian and {Koudmani}, Sophie and {Perna}, Michele and {Robertson}, Brant and {Sijacki}, Debora and {{\"U}bler}, Hannah and {Williams}, Christina C. and {Willott}, Chris},
        title = "{A dormant overmassive black hole in the early Universe}",
      journal = {\nat},
         year = 2024,
        month = dec,
       volume = {636},
       number = {8043},
        pages = {594-597},
          doi = {10.1038/s41586-024-08210-5},
archivePrefix = {arXiv},
       eprint = {2403.03872},
 primaryClass = {astro-ph.GA},
       adsurl = {https://ui.adsabs.harvard.edu/abs/2024Natur.636..594J}
}

@ARTICLE{cloudy,
       author = {{Ferland}, G.~J. and {Chatzikos}, M. and {Guzm{\'a}n}, F. and {Lykins}, M.~L. and {van Hoof}, P.~A.~M. and {Williams}, R.~J.~R. and {Abel}, N.~P. and {Badnell}, N.~R. and {Keenan}, F.~P. and {Porter}, R.~L. and {Stancil}, P.~C.},
        title = "{The 2017 Release Cloudy}",
      journal = {\rmxaa},
         year = 2017,
        month = oct,
       volume = {53},
        pages = {385-438},
          doi = {10.48550/arXiv.1705.10877},
archivePrefix = {arXiv},
       eprint = {1705.10877},
 primaryClass = {astro-ph.GA},
       adsurl = {https://ui.adsabs.harvard.edu/abs/2017RMxAA..53..385F}
}

@ARTICLE{emcee,
       author = {{Foreman-Mackey}, Daniel and {Hogg}, David W. and {Lang}, Dustin and {Goodman}, Jonathan},
        title = "{emcee: The MCMC Hammer}",
      journal = {\pasp},
         year = 2013,
        month = mar,
       volume = {125},
       number = {925},
        pages = {306},
          doi = {10.1086/670067},
archivePrefix = {arXiv},
       eprint = {1202.3665},
 primaryClass = {astro-ph.IM},
       adsurl = {https://ui.adsabs.harvard.edu/abs/2013PASP..125..306F}
}

@ARTICLE{Wang2024,
       author = {{Wang}, Bingjie and {de Graaff}, Anna and {Davies}, Rebecca L. and {Greene}, Jenny E. and {Leja}, Joel and {Goulding}, Andy D. and {Williams}, Christina C. and {Brammer}, Gabriel B. and {Suess}, Katherine A. and {Weibel}, Andrea and {Bezanson}, Rachel and {Boogaard}, Leindert A. and {Cleri}, Nikko J. and {Hirschmann}, Michaela and {Katz}, Harley and {Labbe}, Ivo and {Maseda}, Michael V. and {Matthee}, Jorryt and {McConachie}, Ian and {Naidu}, Rohan P. and {Oesch}, Pascal A. and {Rix}, Hans-Walter and {Setton}, David J. and {Whitaker}, Katherine E.},
        title = "{RUBIES: JWST/NIRSpec Confirmation of an Infrared-luminous, Broad-line Little Red Dot with an Ionized Outflow}",
      journal = {arXiv e-prints},
         year = 2024,
        month = mar,
          eid = {arXiv:2403.02304},
        pages = {arXiv:2403.02304},
          doi = {10.48550/arXiv.2403.02304},
archivePrefix = {arXiv},
       eprint = {2403.02304},
 primaryClass = {astro-ph.GA},
       adsurl = {https://ui.adsabs.harvard.edu/abs/2024arXiv240302304W}
}

@ARTICLE{SavageSembach1991,
       author = {{Savage}, Blair D. and {Sembach}, Kenneth R.},
        title = "{The Analysis of Apparent Optical Depth Profiles for Interstellar Absorption Lines}",
      journal = {\apj},
         year = 1991,
        month = sep,
       volume = {379},
        pages = {245},
          doi = {10.1086/170498},
       adsurl = {https://ui.adsabs.harvard.edu/abs/1991ApJ...379..245S}
}

@ARTICLE{SternLbol,
       author = {{Stern}, Jonathan and {Laor}, Ari},
        title = "{Type 1 AGN at low z- I. Emission properties}",
      journal = {\mnras},
         year = 2012,
        month = jun,
       volume = {423},
       number = {1},
        pages = {600-631},
          doi = {10.1111/j.1365-2966.2012.20901.x},
archivePrefix = {arXiv},
       eprint = {1203.3158},
 primaryClass = {astro-ph.CO},
       adsurl = {https://ui.adsabs.harvard.edu/abs/2012MNRAS.423..600S}
}

@ARTICLE{JADES_desc,
       author = {{Eisenstein}, Daniel J. and {Willott}, Chris and {Alberts}, Stacey and {Arribas}, Santiago and {Bonaventura}, Nina and {Bunker}, Andrew J. and {Cameron}, Alex J. and {Carniani}, Stefano and {Charlot}, Stephane and {Curtis-Lake}, Emma and {D'Eugenio}, Francesco and {Endsley}, Ryan and {Ferruit}, Pierre and {Giardino}, Giovanna and {Hainline}, Kevin and {Hausen}, Ryan and {Jakobsen}, Peter and {Johnson}, Benjamin D. and {Maiolino}, Roberto and {Rieke}, Marcia and {Rieke}, George and {Rix}, Hans-Walter and {Robertson}, Brant and {Stark}, Daniel P. and {Tacchella}, Sandro and {Williams}, Christina C. and {Willmer}, Christopher N.~A. and {Baker}, William M. and {Baum}, Stefi and {Bhatawdekar}, Rachana and {Boyett}, Kristan and {Chen}, Zuyi and {Chevallard}, Jacopo and {Circosta}, Chiara and {Curti}, Mirko and {Danhaive}, A. Lola and {DeCoursey}, Christa and {de Graaff}, Anna and {Dressler}, Alan and {Egami}, Eiichi and {Helton}, Jakob M. and {Hviding}, Raphael E. and {Ji}, Zhiyuan and {Jones}, Gareth C. and {Kumari}, Nimisha and {L{\"u}tzgendorf}, Nora and {Laseter}, Isaac and {Looser}, Tobias J. and {Lyu}, Jianwei and {Maseda}, Michael V. and {Nelson}, Erica and {Parlanti}, Eleonora and {Perna}, Michele and {Pusk{\'a}s}, D{\'a}vid and {Rawle}, Tim and {Rodr{\'\i}guez Del Pino}, Bruno and {Sandles}, Lester and {Saxena}, Aayush and {Scholtz}, Jan and {Sharpe}, Katherine and {Shivaei}, Irene and {Silcock}, Maddie S. and {Simmonds}, Charlotte and {Skarbinski}, Maya and {Smit}, Renske and {Stone}, Meredith and {Suess}, Katherine A. and {Sun}, Fengwu and {Tang}, Mengtao and {Topping}, Michael W. and {{\"U}bler}, Hannah and {Villanueva}, Natalia C. and {Wallace}, Imaan E.~B. and {Whitler}, Lily and {Witstok}, Joris and {Woodrum}, Charity},
        title = "{Overview of the JWST Advanced Deep Extragalactic Survey (JADES)}",
      journal = {arXiv e-prints},
         year = 2023,
        month = jun,
          eid = {arXiv:2306.02465},
        pages = {arXiv:2306.02465},
          doi = {10.48550/arXiv.2306.02465},
archivePrefix = {arXiv},
       eprint = {2306.02465},
 primaryClass = {astro-ph.GA},
       adsurl = {https://ui.adsabs.harvard.edu/abs/2023arXiv230602465E}
}

@ARTICLE{Maiolino_xray_weak,
       author = {{Maiolino}, Roberto and {Risaliti}, Guido and {Signorini}, Matilde and {Trefoloni}, Bartolomeo and {Juod{\v{z}}balis}, Ignas and {Scholtz}, Jan and {{\"U}bler}, Hannah and {D'Eugenio}, Francesco and {Carniani}, Stefano and {Fabian}, Andy and {Ji}, Xihan and {Mazzolari}, Giovanni and {Bertola}, Elena and {Brusa}, Marcella and {Bunker}, Andrew J. and {Charlot}, Stephane and {Comastri}, Andrea and {Cresci}, Giovanni and {DeCoursey}, Christa Noel and {Egami}, Eiichi and {Fiore}, Fabrizio and {Gilli}, Roberto and {Perna}, Michele and {Tacchella}, Sandro and {Venturi}, Giacomo},
        title = "{JWST meets Chandra: a large population of Compton thick, feedback-free, and intrinsically X-ray weak AGN, with a sprinkle of SNe}",
      journal = {\mnras},
         year = 2025,
        month = apr,
       volume = {538},
       number = {3},
        pages = {1921-1943},
          doi = {10.1093/mnras/staf359},
archivePrefix = {arXiv},
       eprint = {2405.00504},
 primaryClass = {astro-ph.GA},
       adsurl = {https://ui.adsabs.harvard.edu/abs/2025MNRAS.538.1921M}
}

@ARTICLE{Matthee2024,
       author = {{Matthee}, Jorryt and {Naidu}, Rohan P. and {Brammer}, Gabriel and {Chisholm}, John and {Eilers}, Anna-Christina and {Goulding}, Andy and {Greene}, Jenny and {Kashino}, Daichi and {Labbe}, Ivo and {Lilly}, Simon J. and {Mackenzie}, Ruari and {Oesch}, Pascal A. and {Weibel}, Andrea and {Wuyts}, Stijn and {Xiao}, Mengyuan and {Bordoloi}, Rongmon and {Bouwens}, Rychard and {van Dokkum}, Pieter and {Illingworth}, Garth and {Kramarenko}, Ivan and {Maseda}, Michael V. and {Mason}, Charlotte and {Meyer}, Romain A. and {Nelson}, Erica J. and {Reddy}, Naveen A. and {Shivaei}, Irene and {Simcoe}, Robert A. and {Yue}, Minghao},
        title = "{Little Red Dots: An Abundant Population of Faint Active Galactic Nuclei at z {\ensuremath{\sim}} 5 Revealed by the EIGER and FRESCO JWST Surveys}",
      journal = {\apj},
         year = 2024,
        month = mar,
       volume = {963},
       number = {2},
          eid = {129},
        pages = {129},
          doi = {10.3847/1538-4357/ad2345},
archivePrefix = {arXiv},
       eprint = {2306.05448},
 primaryClass = {astro-ph.GA},
       adsurl = {https://ui.adsabs.harvard.edu/abs/2024ApJ...963..129M}
}

@ARTICLE{Harikane2023,
       author = {{Harikane}, Yuichi and {Zhang}, Yechi and {Nakajima}, Kimihiko and {Ouchi}, Masami and {Isobe}, Yuki and {Ono}, Yoshiaki and {Hatano}, Shun and {Xu}, Yi and {Umeda}, Hiroya},
        title = "{A JWST/NIRSpec First Census of Broad-line AGNs at z = 4-7: Detection of 10 Faint AGNs with M $_{BH}$ {}10$^{6}$-{}10$^{8}$ M $_{{\ensuremath{\odot}}}$ and Their Host Galaxy Properties}",
      journal = {\apj},
         year = 2023,
        month = dec,
       volume = {959},
       number = {1},
          eid = {39},
        pages = {39},
          doi = {10.3847/1538-4357/ad029e},
archivePrefix = {arXiv},
       eprint = {2303.11946},
 primaryClass = {astro-ph.GA},
       adsurl = {https://ui.adsabs.harvard.edu/abs/2023ApJ...959...39H}
}

@ARTICLE{Juodzbalis2024b,
       author = {{Juod{\v{z}}balis}, Ignas and {Ji}, Xihan and {Maiolino}, Roberto and {D'Eugenio}, Francesco and {Scholtz}, Jan and {Risaliti}, Guido and {Fabian}, Andrew C. and {Mazzolari}, Giovanni and {Gilli}, Roberto and {Prandoni}, Isabella and {Arribas}, Santiago and {Bunker}, Andrew J. and {Carniani}, Stefano and {Charlot}, St{\'e}phane and {Curtis-Lake}, Emma and {de Graaff}, Anna and {Hainline}, Kevin and {Parlanti}, Eleonora and {Perna}, Michele and {P{\'e}rez-Gonz{\'a}lez}, Pablo G. and {Robertson}, Brant and {Tacchella}, Sandro and {{\"U}bler}, Hannah and {Williams}, Christina C. and {Willott}, Chris and {Witstok}, Joris},
        title = "{JADES - the Rosetta stone of JWST-discovered AGN: deciphering the intriguing nature of early AGN}",
      journal = {\mnras},
         year = 2024,
        month = nov,
       volume = {535},
       number = {1},
        pages = {853-873},
          doi = {10.1093/mnras/stae2367},
archivePrefix = {arXiv},
       eprint = {2407.08643},
 primaryClass = {astro-ph.GA},
       adsurl = {https://ui.adsabs.harvard.edu/abs/2024MNRAS.535..853J}
}

@ARTICLE{Yue2024,
       author = {{Yue}, Minghao and {Eilers}, Anna-Christina and {Ananna}, Tonima Tasnim and {Panagiotou}, Christos and {Kara}, Erin and {Miyaji}, Takamitsu},
        title = "{Stacking X-Ray Observations of ``Little Red Dots'': Implications for Their Active Galactic Nucleus Properties}",
      journal = {\apjl},
         year = 2024,
        month = oct,
       volume = {974},
       number = {2},
          eid = {L26},
        pages = {L26},
          doi = {10.3847/2041-8213/ad7eba},
archivePrefix = {arXiv},
       eprint = {2404.13290},
 primaryClass = {astro-ph.GA},
       adsurl = {https://ui.adsabs.harvard.edu/abs/2024ApJ...974L..26Y}
}

@ARTICLE{DEugenio2024,
       author = {{D'Eugenio}, Francesco and {Cameron}, Alex J. and {Scholtz}, Jan and {Carniani}, Stefano and {Willott}, Chris J. and {Curtis-Lake}, Emma and {Bunker}, Andrew J. and {Parlanti}, Eleonora and {Maiolino}, Roberto and {Willmer}, Christopher N.~A. and {Jakobsen}, Peter and {Robertson}, Brant E. and {Johnson}, Benjamin D. and {Tacchella}, Sandro and {Cargile}, Phillip A. and {Rawle}, Tim and {Arribas}, Santiago and {Chevallard}, Jacopo and {Curti}, Mirko and {Egami}, Eiichi and {Eisenstein}, Daniel J. and {Kumari}, Nimisha and {Looser}, Tobias J. and {Rieke}, Marcia J. and {Rodr{\'\i}guez Del Pino}, Bruno and {Saxena}, Aayush and {{\"U}bler}, Hannah and {Venturi}, Giacomo and {Witstok}, Joris and {Baker}, William M. and {Bhatawdekar}, Rachana and {Bonaventura}, Nina and {Boyett}, Kristan and {Charlot}, St{\'e}phane and {Danhaive}, A. Lola and {Hainline}, Kevin N. and {Hausen}, Ryan and {Helton}, Jakob M. and {Ji}, Xihan and {Ji}, Zhiyuan and {Jones}, Gareth C. and {Joud{\v{z}}balis}, Ignas and {Maseda}, Michael V. and {P{\'e}rez-Gonz{\'a}lez}, Pablo G. and {Perna}, Michele and {Pusk{\'a}s}, D{\'a}vid and {Shivaei}, Irene and {Silcock}, Maddie S. and {Simmonds}, Charlotte and {Smit}, Renske and {Sun}, Fengwu and {Villanueva}, Natalia C. and {Williams}, Christina C. and {Zhu}, Yongda},
        title = "{JADES Data Release 3 -- NIRSpec/MSA spectroscopy for 4,000 galaxies in the GOODS fields}",
      journal = {arXiv e-prints},
         year = 2024,
        month = apr,
          eid = {arXiv:2404.06531},
        pages = {arXiv:2404.06531},
          doi = {10.48550/arXiv.2404.06531},
archivePrefix = {arXiv},
       eprint = {2404.06531},
 primaryClass = {astro-ph.GA},
       adsurl = {https://ui.adsabs.harvard.edu/abs/2024arXiv240406531D}
}

@ARTICLE{Inayoshi2024,
       author = {{Inayoshi}, Kohei and {Maiolino}, Roberto},
        title = "{Extremely Dense Gas around Little Red Dots and High-redshift AGNs: A Non-stellar Origin of the Balmer Break and Absorption Features}",
      journal = {arXiv e-prints},
         year = 2024,
        month = sep,
          eid = {arXiv:2409.07805},
        pages = {arXiv:2409.07805},
          doi = {10.48550/arXiv.2409.07805},
archivePrefix = {arXiv},
       eprint = {2409.07805},
 primaryClass = {astro-ph.GA},
       adsurl = {https://ui.adsabs.harvard.edu/abs/2024arXiv240907805I}
}

@ARTICLE{Ji2025,
       author = {{Ji}, Xihan and {Maiolino}, Roberto and {{\"U}bler}, Hannah and {Scholtz}, Jan and {D'Eugenio}, Francesco and {Sun}, Fengwu and {Perna}, Michele and {Turner}, Hannah and {Arribas}, Santiago and {Bennett}, Jake S. and {Bunker}, Andrew and {Carniani}, Stefano and {Charlot}, St{\'e}phane and {Cresci}, Giovanni and {Curti}, Mirko and {Egami}, Eiichi and {Fabian}, Andy and {Inayoshi}, Kohei and {Isobe}, Yuki and {Jones}, Gareth and {Juod{\v{z}}balis}, Ignas and {Kumari}, Nimisha and {Lyu}, Jianwei and {Mazzolari}, Giovanni and {Parlanti}, Eleonora and {Robertson}, Brant and {Rodr{\'\i}guez Del Pino}, Bruno and {Schneider}, Raffaella and {Sijacki}, Debora and {Tacchella}, Sandro and {Trinca}, Alessandro and {Valiante}, Rosa and {Venturi}, Giacomo and {Volonteri}, Marta and {Willott}, Chris and {Witten}, Callum and {Witstok}, Joris},
        title = "{BlackTHUNDER -- A non-stellar Balmer break in a black hole-dominated little red dot at $z=7.04$}",
      journal = {arXiv e-prints},
         year = 2025,
        month = jan,
          eid = {arXiv:2501.13082},
        pages = {arXiv:2501.13082},
          doi = {10.48550/arXiv.2501.13082},
archivePrefix = {arXiv},
       eprint = {2501.13082},
 primaryClass = {astro-ph.GA},
       adsurl = {https://ui.adsabs.harvard.edu/abs/2025arXiv250113082J}
}

@ARTICLE{Ananna2024,
       author = {{Ananna}, Tonima Tasnim and {Bogd{\'a}n}, {\'A}kos and {Kov{\'a}cs}, Orsolya E. and {Natarajan}, Priyamvada and {Hickox}, Ryan C.},
        title = "{X-Ray View of Little Red Dots: Do They Host Supermassive Black Holes?}",
      journal = {\apjl},
         year = 2024,
        month = jul,
       volume = {969},
       number = {1},
          eid = {L18},
        pages = {L18},
          doi = {10.3847/2041-8213/ad5669},
archivePrefix = {arXiv},
       eprint = {2404.19010},
 primaryClass = {astro-ph.GA},
       adsurl = {https://ui.adsabs.harvard.edu/abs/2024ApJ...969L..18A}
}

@ARTICLE{Planck2020,
       author = {{Planck Collaboration} and {Aghanim}, N. and {Akrami}, Y. and {Ashdown}, M. and {Aumont}, J. and {Baccigalupi}, C. and {Ballardini}, M. and {Banday}, A.~J. and {Barreiro}, R.~B. and {Bartolo}, N. and {Basak}, S. and {Battye}, R. and {Benabed}, K. and {Bernard}, J. -P. and {Bersanelli}, M. and {Bielewicz}, P. and {Bock}, J.~J. and {Bond}, J.~R. and {Borrill}, J. and {Bouchet}, F.~R. and {Boulanger}, F. and {Bucher}, M. and {Burigana}, C. and {Butler}, R.~C. and {Calabrese}, E. and {Cardoso}, J. -F. and {Carron}, J. and {Challinor}, A. and {Chiang}, H.~C. and {Chluba}, J. and {Colombo}, L.~P.~L. and {Combet}, C. and {Contreras}, D. and {Crill}, B.~P. and {Cuttaia}, F. and {de Bernardis}, P. and {de Zotti}, G. and {Delabrouille}, J. and {Delouis}, J. -M. and {Di Valentino}, E. and {Diego}, J.~M. and {Dor{\'e}}, O. and {Douspis}, M. and {Ducout}, A. and {Dupac}, X. and {Dusini}, S. and {Efstathiou}, G. and {Elsner}, F. and {En{\ss}lin}, T.~A. and {Eriksen}, H.~K. and {Fantaye}, Y. and {Farhang}, M. and {Fergusson}, J. and {Fernandez-Cobos}, R. and {Finelli}, F. and {Forastieri}, F. and {Frailis}, M. and {Fraisse}, A.~A. and {Franceschi}, E. and {Frolov}, A. and {Galeotta}, S. and {Galli}, S. and {Ganga}, K. and {G{\'e}nova-Santos}, R.~T. and {Gerbino}, M. and {Ghosh}, T. and {Gonz{\'a}lez-Nuevo}, J. and {G{\'o}rski}, K.~M. and {Gratton}, S. and {Gruppuso}, A. and {Gudmundsson}, J.~E. and {Hamann}, J. and {Handley}, W. and {Hansen}, F.~K. and {Herranz}, D. and {Hildebrandt}, S.~R. and {Hivon}, E. and {Huang}, Z. and {Jaffe}, A.~H. and {Jones}, W.~C. and {Karakci}, A. and {Keih{\"a}nen}, E. and {Keskitalo}, R. and {Kiiveri}, K. and {Kim}, J. and {Kisner}, T.~S. and {Knox}, L. and {Krachmalnicoff}, N. and {Kunz}, M. and {Kurki-Suonio}, H. and {Lagache}, G. and {Lamarre}, J. -M. and {Lasenby}, A. and {Lattanzi}, M. and {Lawrence}, C.~R. and {Le Jeune}, M. and {Lemos}, P. and {Lesgourgues}, J. and {Levrier}, F. and {Lewis}, A. and {Liguori}, M. and {Lilje}, P.~B. and {Lilley}, M. and {Lindholm}, V. and {L{\'o}pez-Caniego}, M. and {Lubin}, P.~M. and {Ma}, Y. -Z. and {Mac{\'\i}as-P{\'e}rez}, J.~F. and {Maggio}, G. and {Maino}, D. and {Mandolesi}, N. and {Mangilli}, A. and {Marcos-Caballero}, A. and {Maris}, M. and {Martin}, P.~G. and {Martinelli}, M. and {Mart{\'\i}nez-Gonz{\'a}lez}, E. and {Matarrese}, S. and {Mauri}, N. and {McEwen}, J.~D. and {Meinhold}, P.~R. and {Melchiorri}, A. and {Mennella}, A. and {Migliaccio}, M. and {Millea}, M. and {Mitra}, S. and {Miville-Desch{\^e}nes}, M. -A. and {Molinari}, D. and {Montier}, L. and {Morgante}, G. and {Moss}, A. and {Natoli}, P. and {N{\o}rgaard-Nielsen}, H.~U. and {Pagano}, L. and {Paoletti}, D. and {Partridge}, B. and {Patanchon}, G. and {Peiris}, H.~V. and {Perrotta}, F. and {Pettorino}, V. and {Piacentini}, F. and {Polastri}, L. and {Polenta}, G. and {Puget}, J. -L. and {Rachen}, J.~P. and {Reinecke}, M. and {Remazeilles}, M. and {Renzi}, A. and {Rocha}, G. and {Rosset}, C. and {Roudier}, G. and {Rubi{\~n}o-Mart{\'\i}n}, J.~A. and {Ruiz-Granados}, B. and {Salvati}, L. and {Sandri}, M. and {Savelainen}, M. and {Scott}, D. and {Shellard}, E.~P.~S. and {Sirignano}, C. and {Sirri}, G. and {Spencer}, L.~D. and {Sunyaev}, R. and {Suur-Uski}, A. -S. and {Tauber}, J.~A. and {Tavagnacco}, D. and {Tenti}, M. and {Toffolatti}, L. and {Tomasi}, M. and {Trombetti}, T. and {Valenziano}, L. and {Valiviita}, J. and {Van Tent}, B. and {Vibert}, L. and {Vielva}, P. and {Villa}, F. and {Vittorio}, N. and {Wandelt}, B.~D. and {Wehus}, I.~K. and {White}, M. and {White}, S.~D.~M. and {Zacchei}, A. and {Zonca}, A.},
        title = "{Planck 2018 results. VI. Cosmological parameters}",
      journal = {\aap},
         year = 2020,
        month = sep,
       volume = {641},
          eid = {A6},
        pages = {A6},
          doi = {10.1051/0004-6361/201833910},
archivePrefix = {arXiv},
       eprint = {1807.06209},
 primaryClass = {astro-ph.CO},
       adsurl = {https://ui.adsabs.harvard.edu/abs/2020A&A...641A...6P}
}

@ARTICLE{Rusakov2025,
       author = {{Rusakov}, V. and {Watson}, D. and {Nikopoulos}, G.~P. and {Brammer}, G. and {Gottumukkala}, R. and {Harvey}, T. and {Heintz}, K.~E. and {Nielsen}, R.~D. and {Sim}, S.~A. and {Sneppen}, A. and {Vijayan}, A.~P. and {Adams}, N. and {Austin}, D. and {Conselice}, C.~J. and {Goolsby}, C.~M. and {Toft}, S.},
        title = "{JWST's little red dots: an emerging population of young, low-mass AGN cocooned in dense ionized gas}",
      journal = {arXiv e-prints},
         year = 2025,
        month = mar,
          eid = {arXiv:2503.16595},
        pages = {arXiv:2503.16595},
          doi = {10.48550/arXiv.2503.16595},
archivePrefix = {arXiv},
       eprint = {2503.16595},
 primaryClass = {astro-ph.GA},
       adsurl = {https://ui.adsabs.harvard.edu/abs/2025arXiv250316595R}
}

@ARTICLE{Laor2006,
       author = {{Laor}, Ari},
        title = "{Evidence for Line Broadening by Electron Scattering in the Broad-Line Region of NGC 4395}",
      journal = {\apj},
         year = 2006,
        month = may,
       volume = {643},
       number = {1},
        pages = {112-119},
          doi = {10.1086/502798},
archivePrefix = {arXiv},
       eprint = {astro-ph/0601688},
 primaryClass = {astro-ph},
       adsurl = {https://ui.adsabs.harvard.edu/abs/2006ApJ...643..112L}
}

@ARTICLE{Naidu2025,
       author = {{Naidu}, Rohan P. and {Matthee}, Jorryt and {Katz}, Harley and {de Graaff}, Anna and {Oesch}, Pascal and {Smith}, Aaron and {Greene}, Jenny E. and {Brammer}, Gabriel and {Weibel}, Andrea and {Hviding}, Raphael and {Chisholm}, John and {Labb\textbackslash'e}, Ivo and {Simcoe}, Robert A. and {Witten}, Callum and {Atek}, Hakim and {Baggen}, Josephine F.~W. and {Belli}, Sirio and {Bezanson}, Rachel and {Boogaard}, Leindert A. and {Bose}, Sownak and {Covelo-Paz}, Alba and {Dayal}, Pratika and {Fudamoto}, Yoshinobu and {Furtak}, Lukas J. and {Giovinazzo}, Emma and {Goulding}, Andy and {Gronke}, Max and {Heintz}, Kasper E. and {Hirschmann}, Michaela and {Illingworth}, Garth and {Inoue}, Akio K. and {Johnson}, Benjamin D. and {Leja}, Joel and {Leonova}, Ecaterina and {McConachie}, Ian and {Maseda}, Michael V. and {Natarajan}, Priyamvada and {Nelson}, Erica and {Setton}, David J. and {Shivaei}, Irene and {Sobral}, David and {Stefanon}, Mauro and {Tacchella}, Sandro and {Toft}, Sune and {Torralba}, Alberto and {van Dokkum}, Pieter and {van der Wel}, Arjen and {Volonteri}, Marta and {Walter}, Fabian and {Wang}, Bingjie and {Watson}, Darach},
        title = "{A ``Black Hole Star'' Reveals the Remarkable Gas-Enshrouded Hearts of the Little Red Dots}",
      journal = {arXiv e-prints},
         year = 2025,
        month = mar,
          eid = {arXiv:2503.16596},
        pages = {arXiv:2503.16596},
          doi = {10.48550/arXiv.2503.16596},
archivePrefix = {arXiv},
       eprint = {2503.16596},
 primaryClass = {astro-ph.GA},
       adsurl = {https://ui.adsabs.harvard.edu/abs/2025arXiv250316596N}
}

@ARTICLE{Hainline2025,
       author = {{Hainline}, Kevin N. and {Maiolino}, Roberto and {Juod{\v{z}}balis}, Ignas and {Scholtz}, Jan and {{\"U}bler}, Hannah and {D'Eugenio}, Francesco and {Helton}, Jakob M. and {Sun}, Yang and {Sun}, Fengwu and {Robertson}, Brant and {Tacchella}, Sandro and {Bunker}, Andrew J. and {Carniani}, Stefano and {Charlot}, Stephane and {Curtis-Lake}, Emma and {Egami}, Eiichi and {Johnson}, Benjamin D. and {Lin}, Xiaojing and {Lyu}, Jianwei and {P{\'e}rez-Gonz{\'a}lez}, Pablo G. and {Rinaldi}, Pierluigi and {Silcock}, Maddie S. and {Venturi}, Giacomo and {Williams}, Christina C. and {Willmer}, Christopher N.~A. and {Willott}, Chris and {Zhang}, Junyu and {Zhu}, Yongda},
        title = "{An Investigation into the Selection and Colors of Little Red Dots and Active Galactic Nuclei}",
      journal = {\apj},
         year = 2025,
        month = feb,
       volume = {979},
       number = {2},
          eid = {138},
        pages = {138},
          doi = {10.3847/1538-4357/ad9920},
archivePrefix = {arXiv},
       eprint = {2410.00100},
 primaryClass = {astro-ph.GA},
       adsurl = {https://ui.adsabs.harvard.edu/abs/2025ApJ...979..138H}
}

@ARTICLE{Juodzbalis2026,
       author = {{Juod{\v{z}}balis}, Ignas and {Maiolino}, Roberto and {Baker}, William M. and {Lake}, Emma Curtis and {Scholtz}, Jan and {D'Eugenio}, Francesco and {Trefoloni}, Bartolomeo and {Isobe}, Yuki and {Tacchella}, Sandro and {Bunker}, Andrew J. and {Carniani}, Stefano and {Charlot}, St{\'e}phane and {Jones}, Gareth C. and {Parlanti}, Eleonora and {Perna}, Michele and {Rinaldi}, Pierluigi and {Robertson}, Brant and {{\"U}bler}, Hannah and {Venturi}, Giacomo and {Willott}, Chris},
        title = "{JADES: comprehensive census of broad-line AGN from reionization to cosmic noon revealed by JWST}",
      journal = {\mnras},
         year = 2026,
        month = mar,
       volume = {546},
       number = {3},
          eid = {stag086},
        pages = {stag086},
          doi = {10.1093/mnras/stag086},
archivePrefix = {arXiv},
       eprint = {2504.03551},
 primaryClass = {astro-ph.GA},
       adsurl = {https://ui.adsabs.harvard.edu/abs/2026MNRAS.546ag086J}
}

@ARTICLE{Scholtz2026,
       author = {{Scholtz}, J. and {D'Eugenio}, F. and {Maiolino}, R. and {Brazzini}, M. and {{\"U}bler}, H. and {Ji}, X. and {Perna}, M. and {Sun}, F. and {Brocchi}, G. and {Carniani}, S. and {Cresci}, G. and {Ivey}, L.~R. and {Juod{\v{z}}balis}, I. and {Marconi}, A. and {Mazzolari}, G. and {Risaliti}, G. and {Trefoloni}, B.},
        title = "{Little Red and Blue Dots: simply stratified Broad Line Regions}",
      journal = {arXiv e-prints},
         year = 2026,
        month = mar,
          eid = {arXiv:2603.22277},
        pages = {arXiv:2603.22277},
          doi = {10.48550/arXiv.2603.22277},
archivePrefix = {arXiv},
       eprint = {2603.22277},
 primaryClass = {astro-ph.GA},
       adsurl = {https://ui.adsabs.harvard.edu/abs/2026arXiv260322277S}
}

@ARTICLE{DEugenio2026,
       author = {{D'Eugenio}, Francesco and {Juod{\v{z}}balis}, Ignas and {Ji}, Xihan and {Scholtz}, Jan and {Maiolino}, Roberto and {Carniani}, Stefano and {Perna}, Michele and {Mazzolari}, Giovanni and {{\"U}bler}, Hannah and {Arribas}, Santiago and {Bhatawdekar}, Rachana and {Bunker}, Andrew J. and {Cresci}, Giovanni and {Curtis-Lake}, Emma and {Hainline}, Kevin and {Inayoshi}, Kohei and {Isobe}, Yuki and {Ji}, Zhiyuan and {Johnson}, Benjamin D. and {Jones}, Gareth C. and {Looser}, Tobias J. and {Nelson}, Erica J. and {Parlanti}, Eleonora and {Pusk{\'a}s}, D{\'a}vid and {Rinaldi}, Pierluigi and {Robertson}, Brant and {Rodr{\'\i}guez Del Pino}, Bruno and {Shivaei}, Irene and {Sun}, Fengwu and {Tacchella}, Sandro and {Venturi}, Giacomo and {Volonteri}, Marta and {Williams}, Christina C. and {Willmer}, Christopher N.~A. and {Willott}, Chris and {Witstok}, Joris},
        title = "{JADES and BlackTHUNDER: rest-frame Balmer-line absorption and the local environment in a Little Red Dot at z = 5}",
      journal = {\mnras},
         year = 2026,
        month = jan,
       volume = {545},
       number = {3},
          eid = {staf2117},
        pages = {staf2117},
          doi = {10.1093/mnras/staf2117},
archivePrefix = {arXiv},
       eprint = {2506.14870},
 primaryClass = {astro-ph.GA},
       adsurl = {https://ui.adsabs.harvard.edu/abs/2026MNRAS.545f2117D}
}

@ARTICLE{Matthee2026,
       author = {{Matthee}, Jorryt and {Torralba}, Alberto and {Pezzulli}, Gabriele and {Naidu}, Rohan P. and {Chisholm}, John and {Mascia}, Sara and {Greene}, Jenny E. and {Ishikawa}, Yuzo and {Gronke}, Max and {Wuyts}, Stijn and {Bordoloi}, Rongmon and {Brammer}, Gabriel and {Chang}, Seok-Jun and {Eilers}, Anna-Christina and {de Graaff}, Anna and {Hviding}, Raphael E. and {Iani}, Edoardo and {Illingworth}, Garth and {Kashino}, Daichi and {Labbe}, Ivo and {Ma}, Yilun and {Maseda}, Michael V. and {Meyer}, Romain and {Nelson}, Erica and {Oesch}, Pascal and {Xiao}, Mengyuan},
        title = "{The Engine and its Flows: Little Red Dot spectra are shaped by the column densities of their gas envelopes}",
      journal = {arXiv e-prints},
         year = 2026,
        month = mar,
          eid = {arXiv:2603.17667},
        pages = {arXiv:2603.17667},
          doi = {10.48550/arXiv.2603.17667},
archivePrefix = {arXiv},
       eprint = {2603.17667},
 primaryClass = {astro-ph.GA},
       adsurl = {https://ui.adsabs.harvard.edu/abs/2026arXiv260317667M}
}

@ARTICLE{Leighly2025,
       author = {{Leighly}, Karen M. and {Gallagher}, Sarah C. and {Choi}, Hyunseop and {Terndrup}, Donald M. and {Voelker}, Julianna R. and {Richards}, Gordon T. and {Morabito}, Leah K.},
        title = "{Balmer Absorption in Iron Low-ionization Broad Absorption Line Quasars}",
      journal = {\apj},
         year = 2025,
        month = nov,
       volume = {993},
       number = {1},
          eid = {129},
        pages = {129},
          doi = {10.3847/1538-4357/ae04df},
archivePrefix = {arXiv},
       eprint = {2509.07611},
 primaryClass = {astro-ph.GA},
       adsurl = {https://ui.adsabs.harvard.edu/abs/2025ApJ...993..129L}
}

@ARTICLE{Liu2026,
       author = {{Liu}, Hanpu and {Jiang}, Yan-Fei and {Quataert}, Eliot and {Greene}, Jenny E. and {Ma}, Yilun and {Lin}, Xiaojing},
        title = "{Synthetic Spectral Library of Optically Thick Atmospheres for Little Red Dots}",
      journal = {arXiv e-prints},
         year = 2026,
        month = mar,
          eid = {arXiv:2603.02317},
        pages = {arXiv:2603.02317},
          doi = {10.48550/arXiv.2603.02317},
archivePrefix = {arXiv},
       eprint = {2603.02317},
 primaryClass = {astro-ph.GA},
       adsurl = {https://ui.adsabs.harvard.edu/abs/2026arXiv260302317L}
}

@ARTICLE{Sun2026,
       author = {{Sun}, Wendy Q. and {Naidu}, Rohan P. and {Matthee}, Jorryt and {de Graaff}, Anna and {Chisholm}, John and {Greene}, Jenny E. and {Oesch}, Pascal A. and {Torralba}, Alberto and {Hviding}, Raphael E. and {Brammer}, Gabriel and {Simcoe}, Robert A. and {Bose}, Sownak and {Bouwens}, Rychard and {Dayal}, Pratika and {Eilers}, Anna-Christina and {Fei}, Qinyue and {Furtak}, Lukas J. and {Gottumukkala}, Rashmi and {Goulding}, Andy and {Heintz}, Kasper E. and {Hirschmann}, Michaela and {Kokorev}, Vasily and {Leja}, Joel and {Liu}, Zhaoran and {Natarajan}, Priyamvada and {Santarelli}, Andrew D. and {Setton}, David J. and {Smith}, Aaron and {Tacchella}, Sandro and {Volonteri}, Marta and {Walter}, Fabian and {Weibel}, Andrea and {Williams}, Christina C.},
        title = "{Little Red Dot $-$ Host Galaxy $=$ Black Hole Star: A Gas-Enshrouded Heart at the Center of Every Little Red Dot}",
      journal = {arXiv e-prints},
         year = 2026,
        month = jan,
          eid = {arXiv:2601.20929},
        pages = {arXiv:2601.20929},
          doi = {10.48550/arXiv.2601.20929},
archivePrefix = {arXiv},
       eprint = {2601.20929},
 primaryClass = {astro-ph.GA},
       adsurl = {https://ui.adsabs.harvard.edu/abs/2026arXiv260120929S}
}

@ARTICLE{DEugenio2025b,
       author = {{D'Eugenio}, Francesco and {Nelson}, Erica and {Ji}, Xihan and {Baggen}, Josephine and {Greene}, Jenny and {Labb{\'e}}, Ivo and {Pezzulli}, Gabriele and {Brown}, Vanessa and {Maiolino}, Roberto and {Matthee}, Jorryt and {Terlevich}, Elena and {Terlevich}, Roberto and {Torralba}, Alberto and {Carniani}, Stefano},
        title = "{Irony at z=6.68: a bright AGN with forbidden Fe emission and multi-component Balmer absorption}",
      journal = {arXiv e-prints},
         year = 2025,
        month = sep,
          eid = {arXiv:2510.00101},
        pages = {arXiv:2510.00101},
          doi = {10.48550/arXiv.2510.00101},
archivePrefix = {arXiv},
       eprint = {2510.00101},
 primaryClass = {astro-ph.GA},
       adsurl = {https://ui.adsabs.harvard.edu/abs/2025arXiv251000101D}
}

@ARTICLE{Maiolino2026,
       author = {{Maiolino}, Roberto and {{\"U}bler}, Hannah and {D'Eugenio}, Francesco and {Scholtz}, Jan and {Juod{\v{z}}balis}, Ignas and {Ji}, Xihan and {Perna}, Michele and {Bromm}, Volker and {Dayal}, Pratika and {Koudmani}, Sophie and {Liu}, Boyuan and {Schneider}, Raffaella and {Sijacki}, Debora and {Valiante}, Rosa and {Trinca}, Alessandro and {Zhang}, Saiyang and {Volonteri}, Marta and {Inayoshi}, Kohei and {Carniani}, Stefano and {Nakajima}, Kimihiko and {Isobe}, Yuki and {Witstok}, Joris and {Jones}, Gareth C. and {Tacchella}, Sandro and {Arribas}, Santiago and {Bunker}, Andrew and {Cataldi}, Elisa and {Charlot}, Stephane and {Curti}, Giovanni Cresci Mirko and {Fabian}, Andrew C. and {Katz}, Harley and {Kumari}, Nimisha and {Laporte}, Nicolas and {Mazzolari}, Giovanni and {Robertson}, Brant and {Sun}, Fengwu and {Rodriguez Del Pino}, Bruno and {Venturi}, Giacomo},
        title = "{A black hole in a near pristine galaxy 700 Myr after the big bang}",
      journal = {\mnras},
         year = 2026,
        month = may,
       volume = {548},
       number = {1},
          eid = {staf2109},
        pages = {staf2109},
          doi = {10.1093/mnras/staf2109},
archivePrefix = {arXiv},
       eprint = {2505.22567},
 primaryClass = {astro-ph.GA},
       adsurl = {https://ui.adsabs.harvard.edu/abs/2026MNRAS.548f2109M}
}

@ARTICLE{Bennert2006,
       author = {{Bennert}, N. and {Jungwiert}, B. and {Komossa}, S. and {Haas}, M. and {Chini}, R.},
        title = "{Size and properties of the narrow-line region in Seyfert-2 galaxies from spatially-resolved optical spectroscopy}",
      journal = {\aap},
         year = 2006,
        month = sep,
       volume = {456},
       number = {3},
        pages = {953-966},
          doi = {10.1051/0004-6361:20065319},
archivePrefix = {arXiv},
       eprint = {astro-ph/0607636},
 primaryClass = {astro-ph},
       adsurl = {https://ui.adsabs.harvard.edu/abs/2006A&A...456..953B}
}

@ARTICLE{Brazzini2026,
       author = {{Brazzini}, M. and {D'Eugenio}, F. and {Maiolino}, R. and {Lyu}, J. and {DeCoursey}, C. and {{\"U}bler}, H. and {Ji}, X. and {Juod{\v{z}}balis}, I. and {Scholtz}, J. and {Jones}, G.~C. and {Hainline}, K. and {Dalla Bont{\`a}}, E. and {{\'e}rez-Gonz{\'a}lez}, P.~G. P and {Geris}, S. and {Harshan}, A. and {Feruglio}, C. and {Bischetti}, M. and {Mazzolari}, G. and {Rieke}, G. and {Alberts}, S. and {Trefoloni}, B. and {Carniani}, S. and {Parlanti}, E. and {Marconi}, A. and {Risaliti}, G. and {Ramos Almeida}, C. and {Rinaldi}, P. and {Perna}, M. and {Zamora}, S. and {Lamperti}, I. and {Venturi}, G. and {Cresci}, G. and {Bunker}, Andrew J. and {Ivey}, L.~R.},
        title = "{The Little Blue and Red Dots Rosetta Stones: Non-Gaussian broad lines, hot dust, and X-ray weakness}",
      journal = {arXiv e-prints},
         year = 2026,
        month = jan,
          eid = {arXiv:2601.22214},
        pages = {arXiv:2601.22214},
          doi = {10.48550/arXiv.2601.22214},
archivePrefix = {arXiv},
       eprint = {2601.22214},
 primaryClass = {astro-ph.GA},
       adsurl = {https://ui.adsabs.harvard.edu/abs/2026arXiv260122214B}
}

@ARTICLE{Setton2025,
       author = {{Setton}, David J. and {Greene}, Jenny E. and {de Graaff}, Anna and {Ma}, Yilun and {Leja}, Joel and {Matthee}, Jorryt and {Bezanson}, Rachel and {Boogaard}, Leindert A. and {Cleri}, Nikko J. and {Katz}, Harley and {Labbe}, Ivo and {Maseda}, Michael V. and {McConachie}, Ian and {Miller}, Tim B. and {Price}, Sedona H. and {Suess}, Katherine A. and {van Dokkum}, Pieter and {Wang}, Bingjie and {Weibel}, Andrea and {Whitaker}, Katherine E. and {Williams}, Christina C.},
        title = "{Little Red Dots at an Inflection Point: Ubiquitous V-shaped Turnover Consistently Occurs at the Balmer Limit}",
      journal = {\apj},
         year = 2025,
        month = dec,
       volume = {995},
       number = {1},
          eid = {118},
        pages = {118},
          doi = {10.3847/1538-4357/ae1500},
archivePrefix = {arXiv},
       eprint = {2411.03424},
 primaryClass = {astro-ph.GA},
       adsurl = {https://ui.adsabs.harvard.edu/abs/2025ApJ...995..118S}
}

@ARTICLE{Ji2025LRDlocal,
       author = {{Ji}, Xihan and {D'Eugenio}, Francesco and {Juod{\v{z}}balis}, Ignas and {Walton}, Dominic J. and {Fabian}, Andrew C. and {Maiolino}, Roberto and {Ramos Almeida}, Cristina and {Acosta Pulido}, Jose A. and {Belokurov}, Vasily A. and {Isobe}, Yuki and {Jones}, Gareth and {Maraston}, Claudia and {Scholtz}, Jan and {Simmonds}, Charlotte and {Tacchella}, Sandro and {Terlevich}, Elena and {Terlevich}, Roberto},
        title = "{Lord of LRDs: Insights into a ``Little Red Dot'' with a low-ionization spectrum at z = 0.1}",
      journal = {arXiv e-prints},
         year = 2025,
        month = jul,
          eid = {arXiv:2507.23774},
        pages = {arXiv:2507.23774},
          doi = {10.48550/arXiv.2507.23774},
archivePrefix = {arXiv},
       eprint = {2507.23774},
 primaryClass = {astro-ph.GA},
       adsurl = {https://ui.adsabs.harvard.edu/abs/2025arXiv250723774J}
}

@ARTICLE{Lin2025LRDlocal,
       author = {{Lin}, Xiaojing and {Fan}, Xiaohui and {Cai}, Zheng and {Bian}, Fuyan and {Liu}, Hanpu and {Sun}, Fengwu and {Ma}, Yilun and {Greene}, Jenny E. and {Strauss}, Michael A. and {Green}, Richard and {Lyu}, Jianwei and {Champagne}, Jaclyn B. and {Goulding}, Andy D. and {Inayoshi}, Kohei and {Jin}, Xiangyu and {Leung}, Gene C.~K. and {Li}, Mingyu and {Liu}, Yichen and {Mao}, Junjie and {Pudoka}, Maria Anne and {Tee}, Wei Leong and {Wang}, Ben and {Wang}, Feige and {Wu}, Yunjing and {Yang}, Jinyi and {Zhang}, Haowen and {Zhu}, Yongda},
        title = "{The Discovery of Little Red Dots in the Local Universe: Signatures of Cool Gas Envelopes}",
      journal = {arXiv e-prints},
         year = 2025,
        month = jul,
          eid = {arXiv:2507.10659},
        pages = {arXiv:2507.10659},
          doi = {10.48550/arXiv.2507.10659},
archivePrefix = {arXiv},
       eprint = {2507.10659},
 primaryClass = {astro-ph.GA},
       adsurl = {https://ui.adsabs.harvard.edu/abs/2025arXiv250710659L}
}

@ARTICLE{Sneppen2026,
       author = {{Sneppen}, A. and {Watson}, D. and {Matthews}, J.~H. and {Nikopoulos}, G. and {Allen}, N. and {Brammer}, G. and {Damgaard}, R. and {Heintz}, K.~E. and {Knigge}, C. and {Long}, K.~S. and {Rusakov}, V. and {Sim}, S.~A. and {Witstok}, J.},
        title = "{Inside the cocoon: a comprehensive explanation of the spectra of Little Red Dots}",
      journal = {arXiv e-prints},
         year = 2026,
        month = jan,
          eid = {arXiv:2601.18864},
        pages = {arXiv:2601.18864},
          doi = {10.48550/arXiv.2601.18864},
archivePrefix = {arXiv},
       eprint = {2601.18864},
 primaryClass = {astro-ph.GA},
       adsurl = {https://ui.adsabs.harvard.edu/abs/2026arXiv260118864S}
}

@ARTICLE{Madau2024A,
       author = {{Madau}, Piero and {Haardt}, Francesco},
        title = "{X-Ray Weak Active Galactic Nuclei from Super-Eddington Accretion onto Infant Black Holes}",
      journal = {\apjl},
         year = 2024,
        month = dec,
       volume = {976},
       number = {2},
          eid = {L24},
        pages = {L24},
          doi = {10.3847/2041-8213/ad90e1},
archivePrefix = {arXiv},
       eprint = {2410.00417},
 primaryClass = {astro-ph.GA},
       adsurl = {https://ui.adsabs.harvard.edu/abs/2024ApJ...976L..24M}
}

@ARTICLE{Madau2026,
       author = {{Madau}, Piero},
        title = "{Chasing the light: Shadowing, collimation, and the super-Eddington growth of infant black holes in JWST broad-line AGNs}",
      journal = {\aap},
         year = 2026,
        month = apr,
       volume = {708},
          eid = {A116},
        pages = {A116},
          doi = {10.1051/0004-6361/202659244},
archivePrefix = {arXiv},
       eprint = {2501.09854},
 primaryClass = {astro-ph.HE},
       adsurl = {https://ui.adsabs.harvard.edu/abs/2026A&A...708A.116M}
}

@ARTICLE{Anderson1969,
       author = {{Anderson}, Kurt S. and {Kraft}, Robert P.},
        title = "{Evidence for the Ejection of Matter from the Nucleus of the Seyfert Galaxy NGC 4151}",
      journal = {\apj},
         year = 1969,
        month = dec,
       volume = {158},
        pages = {859},
          doi = {10.1086/150246},
       adsurl = {https://ui.adsabs.harvard.edu/abs/1969ApJ...158..859A}
}

@ARTICLE{Lin2026,
       author = {{Lin}, Xiaojing and {Fan}, Xiaohui and {Cai}, Zheng and {Liu}, Yichen and {Sun}, Fengwu and {Bian}, Fuyan and {Li}, Mingyu and {Mao}, Junjie and {Greene}, Jenny E. and {Liu}, Hanpu and {Li}, Jiaxuan and {Liu}, Weizhe and {Ma}, Yilun and {Sun}, Zechang and {Zhang}, Zijian},
        title = "{(LRDs)$^2$: The Low-ReDshift Little Red Dots Survey. II. DESI DR1 Sample}",
      journal = {arXiv e-prints},
         year = 2026,
        month = may,
          eid = {arXiv:2605.21574},
        pages = {arXiv:2605.21574},
          doi = {10.48550/arXiv.2605.21574},
archivePrefix = {arXiv},
       eprint = {2605.21574},
 primaryClass = {astro-ph.GA},
       adsurl = {https://ui.adsabs.harvard.edu/abs/2026arXiv260521574L}
}

@ARTICLE{Choi2022,
       author = {{Choi}, Hyunseop and {Leighly}, Karen M. and {Terndrup}, Donald M. and {Dabbieri}, Collin and {Gallagher}, Sarah C. and {Richards}, Gordon T.},
        title = "{The Physical Properties of Low-redshift FeLoBAL Quasars. I. Spectral-synthesis Analysis of the Broad Absorption-line (BAL) Outflows Using SimBAL}",
      journal = {\apj},
         year = 2022,
        month = oct,
       volume = {937},
       number = {2},
          eid = {74},
        pages = {74},
          doi = {10.3847/1538-4357/ac61d9},
archivePrefix = {arXiv},
       eprint = {2203.11964},
 primaryClass = {astro-ph.GA},
       adsurl = {https://ui.adsabs.harvard.edu/abs/2022ApJ...937...74C}
}

@ARTICLE{Wills1985,
       author = {{Wills}, B.~J. and {Netzer}, H. and {Wills}, D.},
        title = "{Broad emission features in QSOs and active galactic nuclei. II. New observations and theory of Fe II and HI emission.}",
      journal = {\apj},
         year = 1985,
        month = jan,
       volume = {288},
        pages = {94-116},
          doi = {10.1086/162767},
       adsurl = {https://ui.adsabs.harvard.edu/abs/1985ApJ...288...94W}
}

@ARTICLE{Netzer1983,
       author = {{Netzer}, H. and {Wills}, B.~J.},
        title = "{Broad emission features in QSOs and active galactic nuclei. I. New calculations of Fe II line strengths.}",
      journal = {\apj},
         year = 1983,
        month = dec,
       volume = {275},
        pages = {445-460},
          doi = {10.1086/161545},
       adsurl = {https://ui.adsabs.harvard.edu/abs/1983ApJ...275..445N}
}

@ARTICLE{Trefoloni2025,
       author = {{Trefoloni}, Bartolomeo and {Ji}, Xihan and {Maiolino}, Roberto and {D'Eugenio}, Francesco and {{\"U}bler}, Hannah and {Scholtz}, Jan and {Marconi}, Alessandro and {Marconcini}, Cosimo and {Mazzolari}, Giovanni},
        title = "{The missing Fe II bump in faint JWST active galactic nuclei: Possible evidence of metal-poor broad-line regions at early cosmic times}",
      journal = {\aap},
         year = 2025,
        month = aug,
       volume = {700},
          eid = {A203},
        pages = {A203},
          doi = {10.1051/0004-6361/202452795},
archivePrefix = {arXiv},
       eprint = {2410.21867},
 primaryClass = {astro-ph.GA},
       adsurl = {https://ui.adsabs.harvard.edu/abs/2025A&A...700A.203T}
}

@ARTICLE{Ji2026QSO,
       author = {{Ji}, Xihan and {Pezzulli}, Gabriele and {D'Eugenio}, Francesco and {Maiolino}, Roberto and {Carniani}, Stefano and {Tacchella}, Sandro and {Jones}, Gareth and {Smith}, Aaron and {Witstok}, Joris and {Fabian}, Andrew C. and {Geris}, Sophia and {Harshan}, Anishya and {Isobe}, Yuki and {Ivey}, Lucy R. and {Juod{\v{z}}balis}, Ignas and {Pascalau}, Robert and {Scholtz}, Jan and {Witten}, Callum},
        title = "{Holes in the BH$^\star$? AGN signatures in the FUV spectrum of a black-hole dominated Little Red Dot at $z=7.04$}",
      journal = {arXiv e-prints},
         year = 2026,
        month = apr,
          eid = {arXiv:2604.03370},
        pages = {arXiv:2604.03370},
          doi = {10.48550/arXiv.2604.03370},
archivePrefix = {arXiv},
       eprint = {2604.03370},
 primaryClass = {astro-ph.GA},
       adsurl = {https://ui.adsabs.harvard.edu/abs/2026arXiv260403370J}
}

@ARTICLE{Knop1996,
       author = {{Knop}, R.~A. and {Armus}, L. and {Larkin}, J.~E. and {Mathews}, K. and {Shupe}, D.~L. and {Soifer}, B.~T.},
        title = "{Infrared Spectroscopy of Pa(Beta) and [Fe II] Emission in NGC 4151}",
      journal = {\aj},
         year = 1996,
        month = jul,
       volume = {112},
        pages = {81},
          doi = {10.1086/117990},
       adsurl = {https://ui.adsabs.harvard.edu/abs/1996AJ....112...81K}
}

@ARTICLE{Sharma2025,
       author = {{Sharma}, Mayank and {Arav}, Nahum and {Korista}, Kirk T. and {Bautista}, Manuel and {Dehghanian}, Maryam and {Byun}, Doyee and {Walker}, Gwen and {Mintz}, Sasha},
        title = "{Physical characterization of the FeLoBAL outflow in SDSS J0932+0840: Analysis of VLT/UVES observations}",
      journal = {\aap},
         year = 2025,
        month = jan,
       volume = {693},
          eid = {A254},
        pages = {A254},
          doi = {10.1051/0004-6361/202452735},
archivePrefix = {arXiv},
       eprint = {2412.06929},
 primaryClass = {astro-ph.GA},
       adsurl = {https://ui.adsabs.harvard.edu/abs/2025A&A...693A.254S}
}

@ARTICLE{Hviding2026,
       author = {{Hviding}, Raphael E. and {de Graaff}, Anna and {Liu}, Hanpu and {Goulding}, Andy D. and {Ma}, Yilun and {Greene}, Jenny E. and {Boogaard}, Leindert A. and {Bunker}, Andrew J. and {Cleri}, Nikko J. and {Franx}, Marijn and {Hirschmann}, Michaela and {Leja}, Joel and {Matthee}, Jorryt and {Naidu}, Rohan P. and {Setton}, David J. and {{\"U}bler}, Hannah and {Venturi}, Giacomo and {Wang}, Bingjie},
        title = "{The X-Ray Dot: Exotic Dust or a Late-stage Little Red Dot?}",
      journal = {\apjl},
         year = 2026,
        month = mar,
       volume = {1000},
       number = {1},
          eid = {L18},
        pages = {L18},
          doi = {10.3847/2041-8213/ae4c88},
archivePrefix = {arXiv},
       eprint = {2601.09778},
 primaryClass = {astro-ph.GA},
       adsurl = {https://ui.adsabs.harvard.edu/abs/2026ApJ..1000L..18H}
}

@ARTICLE{Greene2026,
       author = {{Greene}, Jenny E. and {Setton}, David J. and {Furtak}, Lukas J. and {Naidu}, Rohan P. and {Volonteri}, Marta and {Dayal}, Pratika and {Labbe}, Ivo and {van Dokkum}, Pieter and {Bezanson}, Rachel and {Brammer}, Gabriel and {Cutler}, Sam E. and {Glazebrook}, Karl and {de Graaff}, Anna and {Hirschmann}, Michaela and {Hviding}, Raphael E. and {Kokorev}, Vasily and {Leja}, Joel and {Liu}, Hanpu and {Ma}, Yilun and {Matthee}, Jorryt and {Nanayakkara}, Themiya and {Oesch}, Pascal A. and {Pan}, Richard and {Price}, Sedona H. and {Spilker}, Justin S. and {Wang}, Bingjie and {Weaver}, John R. and {Whitaker}, Katherine E. and {Williams}, Christina C. and {Zitrin}, Adi},
        title = "{What You See Is What You Get: Empirically Measured Bolometric Luminosities of Little Red Dots}",
      journal = {\apj},
         year = 2026,
        month = jan,
       volume = {996},
       number = {2},
          eid = {129},
        pages = {129},
          doi = {10.3847/1538-4357/ae1836},
archivePrefix = {arXiv},
       eprint = {2509.05434},
 primaryClass = {astro-ph.GA},
       adsurl = {https://ui.adsabs.harvard.edu/abs/2026ApJ...996..129G}
}

@ARTICLE{Casey2024,
       author = {{Casey}, Caitlin M. and {Akins}, Hollis B. and {Kokorev}, Vasily and {McKinney}, Jed and {Cooper}, Olivia R. and {Long}, Arianna S. and {Franco}, Maximilien and {Manning}, Sinclaire M.},
        title = "{Dust in Little Red Dots}",
      journal = {\apjl},
         year = 2024,
        month = nov,
       volume = {975},
       number = {1},
          eid = {L4},
        pages = {L4},
          doi = {10.3847/2041-8213/ad7ba7},
archivePrefix = {arXiv},
       eprint = {2407.05094},
 primaryClass = {astro-ph.GA},
       adsurl = {https://ui.adsabs.harvard.edu/abs/2024ApJ...975L...4C}
}

@ARTICLE{Labbe2024,
       author = {{Labb\'e}, Ivo and {Greene}, Jenny E. and {Matthee}, Jorryt and {Treiber}, Helena and {Kokorev}, Vasily and {Miller}, Tim B. and {Kramarenko}, Ivan and {Setton}, David J. and {Ma}, Yilun and {Goulding}, Andy D. and {Bezanson}, Rachel and {Naidu}, Rohan P. and {Williams}, Christina C. and {Atek}, Hakim and {Brammer}, Gabriel and {Cutler}, Sam E. and {Chemerynska}, Iryna and {Cloonan}, Aidan P. and {Dayal}, Pratika and {de Graaff}, Anna and {Fudamoto}, Yoshinobu and {Fujimoto}, Seiji and {Furtak}, Lukas J. and {Glazebrook}, Karl and {Heintz}, Kasper E. and {Leja}, Joel and {Marchesini}, Danilo and {Nanayakkara}, Themiya and {Nelson}, Erica J. and {Oesch}, Pascal A. and {Pan}, Richard and {Price}, Sedona H. and {Shivaei}, Irene and {Sobral}, David and {Suess}, Katherine A. and {van Dokkum}, Pieter and {Wang}, Bingjie and {Weaver}, John R. and {Whitaker}, Katherine E. and {Zitrin}, Adi},
        title = "{An unambiguous AGN and a Balmer break in an Ultraluminous Little Red Dot at z=4.47 from Ultradeep UNCOVER and All the Little Things Spectroscopy}",
      journal = {arXiv e-prints},
         year = 2024,
        month = dec,
          eid = {arXiv:2412.04557},
        pages = {arXiv:2412.04557},
          doi = {10.48550/arXiv.2412.04557},
archivePrefix = {arXiv},
       eprint = {2412.04557},
 primaryClass = {astro-ph.GA},
       adsurl = {https://ui.adsabs.harvard.edu/abs/2024arXiv241204557L}
}

@ARTICLE{deGraaff2025,
       author = {{de Graaff}, Anna and {Rix}, Hans-Walter and {Naidu}, Rohan P. and {Labb{\'e}}, Ivo and {Wang}, Bingjie and {Leja}, Joel and {Matthee}, Jorryt and {Katz}, Harley and {Greene}, Jenny E. and {Hviding}, Raphael E. and {Baggen}, Josephine and {Bezanson}, Rachel and {Boogaard}, Leindert A. and {Brammer}, Gabriel and {Dayal}, Pratika and {van Dokkum}, Pieter and {Goulding}, Andy D. and {Hirschmann}, Michaela and {Maseda}, Michael V. and {McConachie}, Ian and {Miller}, Tim B. and {Nelson}, Erica and {Oesch}, Pascal A. and {Setton}, David J. and {Shivaei}, Irene and {Weibel}, Andrea and {Whitaker}, Katherine E. and {Williams}, Christina C.},
        title = "{A remarkable ruby: Absorption in dense gas, rather than evolved stars, drives the extreme Balmer break of a little red dot at z = 3.5}",
      journal = {\aap},
         year = 2025,
        month = sep,
       volume = {701},
          eid = {A168},
        pages = {A168},
          doi = {10.1051/0004-6361/202554681},
archivePrefix = {arXiv},
       eprint = {2503.16600},
 primaryClass = {astro-ph.GA},
       adsurl = {https://ui.adsabs.harvard.edu/abs/2025A&A...701A.168D}
}

@ARTICLE{Juodzbalis2026qso1,
       author = {{Juod{\v{z}}balis}, Ignas and {Marconcini}, Cosimo and {D'Eugenio}, Francesco and {Maiolino}, Roberto and {Marconi}, Alessandro and {{\"U}bler}, Hannah and {Scholtz}, Jan and {Ji}, Xihan and {Jones}, Gareth C. and {Perna}, Michele and {Arribas}, Santiago and {Bennett}, Jake S. and {Bromm}, Volker and {Bunker}, Andrew J. and {Carniani}, Stefano and {Charlot}, St{\'e}phane and {Cresci}, Giovanni and {Dayal}, Pratika and {Egami}, Eiichi and {Fabian}, Andrew and {Inayoshi}, Kohei and {Isobe}, Yuki and {Ivey}, Lucy R. and {Koudmani}, Sophie and {Laporte}, Nicolas and {Liu}, Boyuan and {Lyu}, Jianwei and {Mazzolari}, Giovanni and {Monty}, Stephanie and {Parlanti}, Eleonora and {P{\'e}rez-Gonz{\'a}lez}, Pablo G. and {Robertson}, Brant and {Schneider}, Raffaella and {Sijacki}, Debora and {Tacchella}, Sandro and {Trinca}, Alessandro and {Valiante}, Rosa and {Volonteri}, Marta and {Witstok}, Joris and {Zhang}, Saiyang},
        title = "{A direct black-hole mass measurement in a little red dot at high redshift}",
      journal = {\nat},
         year = 2026,
        month = may,
       volume = {653},
       number = {8116},
        pages = {1017-1021},
          doi = {10.1038/s41586-026-10579-4},
       adsurl = {https://ui.adsabs.harvard.edu/abs/2026Natur.653.1017J}
}

@ARTICLE{Leitherer1987,
       author = {{Leitherer}, C. and {Zickgraf}, F.-J.},
        title = "{The detection of a circumstellarr shell around P Cygni by direct CCD imaging.}",
      journal = {\aap},
         year = 1987,
        month = mar,
       volume = {174},
        pages = {103-106},
       adsurl = {https://ui.adsabs.harvard.edu/abs/1987A&A...174..103L}
}

@ARTICLE{Balan2010,
       author = {{Balan}, Aurelian and {Tycner}, C. and {Zavala}, R.~T. and {Benson}, J.~A. and {Hutter}, D.~J. and {Templeton}, M.},
        title = "{The Spatially Resolved H{\ensuremath{\alpha}}-emitting Wind Structure of P Cygni}",
      journal = {\aj},
         year = 2010,
        month = jun,
       volume = {139},
       number = {6},
        pages = {2269-2278},
          doi = {10.1088/0004-6256/139/6/2269},
archivePrefix = {arXiv},
       eprint = {1004.0376},
 primaryClass = {astro-ph.SR},
       adsurl = {https://ui.adsabs.harvard.edu/abs/2010AJ....139.2269B}
}

@ARTICLE{Genderen2001,
       author = {{van Genderen}, A.~M.},
        title = "{S Doradus variables in the Galaxy and the Magellanic Clouds}",
      journal = {\aap},
         year = 2001,
        month = feb,
       volume = {366},
        pages = {508-531},
          doi = {10.1051/0004-6361:20000022},
       adsurl = {https://ui.adsabs.harvard.edu/abs/2001A&A...366..508V}
}

@ARTICLE{Kalari2018,
       author = {{Kalari}, V.~M. and {Vink}, J.~S. and {Dufton}, P.~L. and {Fraser}, M.},
        title = "{How common is LBV S Doradus variability at low metallicity?}",
      journal = {\aap},
         year = 2018,
        month = oct,
       volume = {618},
          eid = {A17},
        pages = {A17},
          doi = {10.1051/0004-6361/201833484},
archivePrefix = {arXiv},
       eprint = {1807.01309},
 primaryClass = {astro-ph.SR},
       adsurl = {https://ui.adsabs.harvard.edu/abs/2018A&A...618A..17K}
}

@ARTICLE{Drissen2001,
       author = {{Drissen}, Laurent and {Crowther}, Paul A. and {Smith}, Linda J. and {Robert}, Carmelle and {Roy}, Jean-Ren{\'e} and {Hillier}, D. John},
        title = "{Physical Parameters of Erupting Luminous Blue Variables: NGC 2363-V1 Caught in the Act}",
      journal = {\apj},
         year = 2001,
        month = jan,
       volume = {546},
       number = {1},
        pages = {484-495},
          doi = {10.1086/318264},
archivePrefix = {arXiv},
       eprint = {astro-ph/0008221},
 primaryClass = {astro-ph},
       adsurl = {https://ui.adsabs.harvard.edu/abs/2001ApJ...546..484D}
}

@ARTICLE{Guseva2022,
       author = {{Guseva}, N.~G. and {Thuan}, T.~X. and {Izotov}, Y.~I.},
        title = "{Decade-long time-monitoring of candidate luminous blue variable stars in the two very metal-deficient star-forming galaxies DDO 68 and PHL 293B}",
      journal = {\mnras},
         year = 2022,
        month = may,
       volume = {512},
       number = {3},
        pages = {4298-4307},
          doi = {10.1093/mnras/stac820},
archivePrefix = {arXiv},
       eprint = {2203.11630},
 primaryClass = {astro-ph.GA},
       adsurl = {https://ui.adsabs.harvard.edu/abs/2022MNRAS.512.4298G}
}

@ARTICLE{Richardson2010,
       author = {{Richardson}, N.~D. and {Gies}, D.~R. and {Henry}, T.~J. and {Fern{\'a}ndez-Laj{\'u}s}, E. and {Okazaki}, A.~T.},
        title = "{The H{\ensuremath{\alpha}} Variations of {\ensuremath{\eta}} Carinae During the 2009.0 Spectroscopic Event}",
      journal = {\aj},
         year = 2010,
        month = apr,
       volume = {139},
       number = {4},
        pages = {1534-1541},
          doi = {10.1088/0004-6256/139/4/1534},
archivePrefix = {arXiv},
       eprint = {1001.3414},
 primaryClass = {astro-ph.SR},
       adsurl = {https://ui.adsabs.harvard.edu/abs/2010AJ....139.1534R}
}

@ARTICLE{Richardson2011,
       author = {{Richardson}, N.~D. and {Morrison}, N.~D. and {Gies}, D.~R. and {Markova}, N. and {Hesselbach}, E.~N. and {Percy}, J.~R.},
        title = "{The H{\ensuremath{\alpha}} Variations of the Luminous Blue Variable P Cygni: Discrete Absorption Components and the Short S Doradus-phase}",
      journal = {\aj},
         year = 2011,
        month = apr,
       volume = {141},
       number = {4},
          eid = {120},
        pages = {120},
          doi = {10.1088/0004-6256/141/4/120},
archivePrefix = {arXiv},
       eprint = {1101.4319},
 primaryClass = {astro-ph.SR},
       adsurl = {https://ui.adsabs.harvard.edu/abs/2011AJ....141..120R}
}

@ARTICLE{Guseva2024,
       author = {{Guseva}, N.~G. and {Thuan}, T.~X. and {Izotov}, Y.~I.},
        title = "{Monitoring broad emission-line components in spectra of the two low-metallicity dwarf compact star-forming galaxies SBS 1420+540 and J1444+4840}",
      journal = {\mnras},
         year = 2024,
        month = jan,
       volume = {527},
       number = {2},
        pages = {3932-3944},
          doi = {10.1093/mnras/stad3485},
archivePrefix = {arXiv},
       eprint = {2311.08286},
 primaryClass = {astro-ph.GA},
       adsurl = {https://ui.adsabs.harvard.edu/abs/2024MNRAS.527.3932G}
}

@ARTICLE{Mehner2021,
       author = {{Mehner}, A. and {Janssens}, S. and {Agliozzo}, C. and {de Wit}, W.-J. and {Boffin}, H.~M.~J. and {Baade}, D. and {Bodensteiner}, J. and {Groh}, J.~H. and {Mahy}, L. and {Vogt}, F.~P.~A.},
        title = "{LBV phenomenon and binarity: The environment of HR Car}",
      journal = {\aap},
         year = 2021,
        month = nov,
       volume = {655},
          eid = {A33},
        pages = {A33},
          doi = {10.1051/0004-6361/202141473},
archivePrefix = {arXiv},
       eprint = {2109.13416},
 primaryClass = {astro-ph.SR},
       adsurl = {https://ui.adsabs.harvard.edu/abs/2021A&A...655A..33M}
}

@ARTICLE{Mehner2010,
       author = {{Mehner}, Andrea and {Davidson}, Kris and {Ferland}, Gary J. and {Humphreys}, Roberta M.},
        title = "{High-excitation Emission Lines near Eta Carinae, and Its Likely Companion Star}",
      journal = {\apj},
         year = 2010,
        month = feb,
       volume = {710},
       number = {1},
        pages = {729-742},
          doi = {10.1088/0004-637X/710/1/729},
archivePrefix = {arXiv},
       eprint = {0912.1067},
 primaryClass = {astro-ph.SR},
       adsurl = {https://ui.adsabs.harvard.edu/abs/2010ApJ...710..729M}
}

@ARTICLE{Liu2018,
       author = {{Liu}, Hezhen and {Luo}, B. and {Brandt}, W.~N. and {Gallagher}, S.~C. and {Garmire}, G.~P.},
        title = "{The Frequency of Intrinsic X-Ray Weakness among Broad Absorption Line Quasars}",
      journal = {\apj},
         year = 2018,
        month = jun,
       volume = {859},
       number = {2},
          eid = {113},
        pages = {113},
          doi = {10.3847/1538-4357/aabe8d},
archivePrefix = {arXiv},
       eprint = {1804.05074},
 primaryClass = {astro-ph.HE},
       adsurl = {https://ui.adsabs.harvard.edu/abs/2018ApJ...859..113L}
}

@ARTICLE{Vito2018,
       author = {{Vito}, F. and {Brandt}, W.~N. and {Luo}, B. and {Shemmer}, O. and {Vignali}, C. and {Gilli}, R.},
        title = "{No evidence for an Eddington-ratio dependence of X-ray weakness in BALQSOs}",
      journal = {\mnras},
         year = 2018,
        month = oct,
       volume = {479},
       number = {4},
        pages = {5335-5342},
          doi = {10.1093/mnras/sty1765},
archivePrefix = {arXiv},
       eprint = {1807.03868},
 primaryClass = {astro-ph.GA},
       adsurl = {https://ui.adsabs.harvard.edu/abs/2018MNRAS.479.5335V}
}

@ARTICLE{Morabito2011,
       author = {{Morabito}, Leah K. and {Dai}, Xinyu and {Leighly}, Karen M. and {Sivakoff}, Gregory R. and {Shankar}, Francesco},
        title = "{Suzaku Observations of Three FeLoBAL Quasi-stellar Objects: SDSS J0943+5417, J1352+4239, and J1723+5553}",
      journal = {\apj},
         year = 2011,
        month = aug,
       volume = {737},
       number = {1},
          eid = {46},
        pages = {46},
          doi = {10.1088/0004-637X/737/1/46},
archivePrefix = {arXiv},
       eprint = {1011.4327},
 primaryClass = {astro-ph.CO},
       adsurl = {https://ui.adsabs.harvard.edu/abs/2011ApJ...737...46M}
}

@ARTICLE{Scholtz2026_JADES,
       author = {{Scholtz}, J. and {Carniani}, S. and {Parlanti}, E. and {D'Eugenio}, F. and {Curtis-Lake}, E. and {Jakobsen}, P. and {Bunker}, A.~J. and {Cameron}, A.~J. and {Arribas}, S. and {Baker}, W.~M. and {Charlot}, S. and {Chevellard}, J. and {Circosta}, C. and {Curti}, M. and {Duan}, Q. and {Eisenstein}, D.~J. and {Hainline}, K. and {Ji}, Z. and {Johnson}, B.~D. and {Jones}, G.~C. and {Kumari}, N. and {Maiolino}, R. and {Maseda}, M.~V. and {Perna}, M. and {P{\'e}rez-Gonz{\'a}lez}, P.~G. and {Rawle}, T. and {Rieke}, M. and {Rinaldi}, P. and {Robertson}, B. and {Saxena}, A. and {Shivaei}, I. and {Silcock}, M.~S. and {Sun}, Y. and {Rodr{\'\i}guez Del Pino}, B. and {Tacchella}, S. and {{\"U}bler}, H. and {Venturi}, G. and {Williams}, C.~C. and {Willmer}, C.~N.~A. and {Willott}, C. and {Witstok}, J.},
        title = "{JADES Data Release 4 - Paper II: Data reduction, analysis and emission-line fluxes of the complete spectroscopic sample}",
      journal = {\mnras},
         year = 2026,
        month = may,
          doi = {10.1093/mnras/stag939},
archivePrefix = {arXiv},
       eprint = {2510.01034},
 primaryClass = {astro-ph.GA},
       adsurl = {https://ui.adsabs.harvard.edu/abs/2026MNRAS.tmp..886S}
}

@ARTICLE{Arshakian2005,
       author = {{Arshakian}, T.~G.},
        title = "{Direct evidence of the receding ``torus'' around central nuclei of powerful radio sources}",
      journal = {\aap},
         year = 2005,
        month = jun,
       volume = {436},
       number = {3},
        pages = {817-824},
          doi = {10.1051/0004-6361:20042341},
archivePrefix = {arXiv},
       eprint = {astro-ph/0411636},
 primaryClass = {astro-ph},
       adsurl = {https://ui.adsabs.harvard.edu/abs/2005A&A...436..817A}
}

@ARTICLE{Geris2026,
       author = {{Geris}, Sophia and {Maiolino}, Roberto and {Ji}, Xihan and {Risaliti}, Guido and {Lanzuisi}, Giorgio and {D'Eugenio}, Francesco and {Isobe}, Yuki and {Jones}, Gareth and {Harshan}, Anishya and {Brazzini}, Matilde and {Juod{\v{z}}balis}, Ignas and {Scholtz}, Jan and {Rinaldi}, Pierluigi and {{\"U}bler}, Hannah and {Baker}, William and {Bunker}, Andrew J. and {Brusa}, Marcella and {Carniani}, Stefano and {Charlot}, Stephane and {Curti}, Mirko and {Comastri}, Andrea and {Lake}, Emma Curtis and {Gilli}, Roberto and {Hainline}, Kevin and {Madau}, Piero and {Marchesi}, Stefano and {Mazzolari}, Giovanni and {Napolitano}, Lorenzo and {Parlanti}, Eleonora and {Pentericci}, Laura and {Ramos Almeida}, Cristina and {Robertson}, Brant and {Silcock}, Maddie S. and {Tripodi}, Roberta and {Venturi}, Giacomo and {Vignali}, Cristian and {Vito}, Fabio and {Zhu}, Yongda},
        title = "{Little Red and Blue Dots: AGN-excited narrow lines, Lyman-$α$ emission, and resemblance to standard quasars}",
      journal = {arXiv e-prints},
         year = 2026,
        month = jun,
          eid = {arXiv:2606.21614},
        pages = {arXiv:2606.21614},
archivePrefix = {arXiv},
       eprint = {2606.21614},
 primaryClass = {astro-ph.GA},
       adsurl = {https://ui.adsabs.harvard.edu/abs/2026arXiv260621614G}
}

@ARTICLE{Sacchi2025,
       author = {{Sacchi}, Andrea and {Bogd{\'a}n}, {\'A}kos},
        title = "{Chandra Rules Out Super-Eddington Accretion Models for Little Red Dots}",
      journal = {\apjl},
         year = 2025,
        month = aug,
       volume = {989},
       number = {2},
          eid = {L30},
        pages = {L30},
          doi = {10.3847/2041-8213/adf5c8},
archivePrefix = {arXiv},
       eprint = {2505.09669},
 primaryClass = {astro-ph.GA},
       adsurl = {https://ui.adsabs.harvard.edu/abs/2025ApJ...989L..30S}
}

@ARTICLE{Comastri2026,
       author = {{Comastri}, A. and {Lanzuisi}, G. and {Vito}, F. and {Marchesi}, S. and {Brusa}, M. and {Gilli}, R. and {Juod{\v{z}}balis}, I. and {Maiolino}, R. and {Mazzolari}, G. and {Risaliti}, G. and {Scholtz}, J. and {Vignali}, C.},
        title = "{JWST-discovered AGN: Evidence of heavy obscuration in the type 2 sample from the first stacked X-ray detection}",
      journal = {\aap},
         year = 2026,
        month = feb,
       volume = {706},
          eid = {A302},
        pages = {A302},
          doi = {10.1051/0004-6361/202557017},
archivePrefix = {arXiv},
       eprint = {2510.00112},
 primaryClass = {astro-ph.GA},
       adsurl = {https://ui.adsabs.harvard.edu/abs/2026A&A...706A.302C}
}

@ARTICLE{Aoki2006,
       author = {{Aoki}, Kentaro and {Iwata}, Ikuru and {Ohta}, Kouji and {Ando}, Masataka and {Akiyama}, Masayuki and {Tamura}, Naoyuki},
        title = "{Discovery of H{\ensuremath{\alpha}} Absorption in the Unusual Broad Absorption Line Quasar SDSS J083942.11+380526.3}",
      journal = {\apj},
         year = 2006,
        month = nov,
       volume = {651},
       number = {1},
        pages = {84-92},
          doi = {10.1086/507438},
archivePrefix = {arXiv},
       eprint = {astro-ph/0607036},
 primaryClass = {astro-ph},
       adsurl = {https://ui.adsabs.harvard.edu/abs/2006ApJ...651...84A}
}

@ARTICLE{Finkelstein2023,
       author = {{Finkelstein}, Steven L. and {Bagley}, Micaela B. and {Ferguson}, Henry C. and {Wilkins}, Stephen M. and {Kartaltepe}, Jeyhan S. and {Papovich}, Casey and {Yung}, L.~Y. Aaron and {Arrabal Haro}, Pablo and {Behroozi}, Peter and {Dickinson}, Mark and {Kocevski}, Dale D. and {Koekemoer}, Anton M. and {Larson}, Rebecca L. and {Le Bail}, Aur{\'e}lien and {Morales}, Alexa M. and {P{\'e}rez-Gonz{\'a}lez}, Pablo G. and {Burgarella}, Denis and {Dav{\'e}}, Romeel and {Hirschmann}, Michaela and {Somerville}, Rachel S. and {Wuyts}, Stijn and {Bromm}, Volker and {Casey}, Caitlin M. and {Fontana}, Adriano and {Fujimoto}, Seiji and {Gardner}, Jonathan P. and {Giavalisco}, Mauro and {Grazian}, Andrea and {Grogin}, Norman A. and {Hathi}, Nimish P. and {Hutchison}, Taylor A. and {Jha}, Saurabh W. and {Jogee}, Shardha and {Kewley}, Lisa J. and {Kirkpatrick}, Allison and {Long}, Arianna S. and {Lotz}, Jennifer M. and {Pentericci}, Laura and {Pierel}, Justin D.~R. and {Pirzkal}, Nor and {Ravindranath}, Swara and {Ryan}, Russell E. and {Trump}, Jonathan R. and {Yang}, Guang and {Bhatawdekar}, Rachana and {Bisigello}, Laura and {Buat}, V{\'e}ronique and {Calabr{\`o}}, Antonello and {Castellano}, Marco and {Cleri}, Nikko J. and {Cooper}, M.~C. and {Croton}, Darren and {Daddi}, Emanuele and {Dekel}, Avishai and {Elbaz}, David and {Franco}, Maximilien and {Gawiser}, Eric and {Holwerda}, Benne W. and {Huertas-Company}, Marc and {Jaskot}, Anne E. and {Leung}, Gene C.~K. and {Lucas}, Ray A. and {Mobasher}, Bahram and {Pandya}, Viraj and {Tacchella}, Sandro and {Weiner}, Benjamin J. and {Zavala}, Jorge A.},
        title = "{CEERS Key Paper. I. An Early Look into the First 500 Myr of Galaxy Formation with JWST}",
      journal = {\apjl},
         year = 2023,
        month = mar,
       volume = {946},
       number = {1},
          eid = {L13},
        pages = {L13},
          doi = {10.3847/2041-8213/acade4},
archivePrefix = {arXiv},
       eprint = {2211.05792},
 primaryClass = {astro-ph.GA},
       adsurl = {https://ui.adsabs.harvard.edu/abs/2023ApJ...946L..13F}
}

@ARTICLE{deGraaf2025,
       author = {{de Graaff}, Anna and {Brammer}, Gabriel and {Weibel}, Andrea and {Lewis}, Zach and {Maseda}, Michael V. and {Oesch}, Pascal A. and {Bezanson}, Rachel and {Boogaard}, Leindert A. and {Cleri}, Nikko J. and {Cooper}, Olivia R. and {Gottumukkala}, Rashmi and {Greene}, Jenny E. and {Hirschmann}, Michaela and {Hviding}, Raphael E. and {Katz}, Harley and {Labb{\'e}}, Ivo and {Leja}, Joel and {Matthee}, Jorryt and {McConachie}, Ian and {Miller}, Tim B. and {Naidu}, Rohan P. and {Price}, Sedona H. and {Rix}, Hans-Walter and {Setton}, David J. and {Suess}, Katherine A. and {Wang}, Bingjie and {Whitaker}, Katherine E. and {Williams}, Christina C.},
        title = "{RUBIES: A complete census of the bright and red distant Universe with JWST/NIRSpec}",
      journal = {\aap},
         year = 2025,
        month = may,
       volume = {697},
          eid = {A189},
        pages = {A189},
          doi = {10.1051/0004-6361/202452186},
archivePrefix = {arXiv},
       eprint = {2409.05948},
 primaryClass = {astro-ph.GA},
       adsurl = {https://ui.adsabs.harvard.edu/abs/2025A&A...697A.189D}
}

@ARTICLE{Hviding2025,
       author = {{Hviding}, Raphael E. and {de Graaff}, Anna and {Miller}, Tim B. and {Setton}, David J. and {Greene}, Jenny E. and {Labb{\'e}}, Ivo and {Brammer}, Gabriel and {Bezanson}, Rachel and {Boogaard}, Leindert A. and {Cleri}, Nikko J. and {Leja}, Joel and {Maseda}, Michael V. and {McConachie}, Ian and {Matthee}, Jorryt and {Naidu}, Rohan P. and {Oesch}, Pascal A. and {Wang}, Bingjie and {Whitaker}, Katherine E. and {Williams}, Christina C.},
        title = "{RUBIES: A spectroscopic census of little red dots: All point sources with v-shaped continua have broad lines}",
      journal = {\aap},
         year = 2025,
        month = oct,
       volume = {702},
          eid = {A57},
        pages = {A57},
          doi = {10.1051/0004-6361/202555816},
archivePrefix = {arXiv},
       eprint = {2506.05459},
 primaryClass = {astro-ph.GA},
       adsurl = {https://ui.adsabs.harvard.edu/abs/2025A&A...702A..57H}
}

@ARTICLE{Rinaldi2026,
       author = {{Rinaldi}, Pierluigi and {Hainline}, Kevin and {D'Eugenio}, Francesco and {P{\'e}rez-Gonz{\'a}lez}, Pablo G. and {Eisenstein}, Daniel J. and {Willmer}, Christopher N.~A. and {Carreira}, Courtney and {Robertson}, Brant and {Johnson}, Benjamin D. and {Alberts}, Stacey and {Baker}, William M. and {Bunker}, Andrew J. and {Carniani}, Stefano and {Egami}, Eiichi and {Helton}, Jakob M. and {Ji}, Zhiyuan and {Juod{\v{z}}balis}, Ignas and {Lin}, Xiaojing and {Lyu}, Jianwei and {Ma}, Zheng and {Maiolino}, Roberto and {Parlanti}, Eleonora and {Scholtz}, Jan and {Sun}, Yang and {Tacchella}, Sandro and {Venturi}, Giacomo and {Williams}, Christina C. and {Willott}, Chris and {Witstok}, Joris and {Wu}, Zihao},
        title = "{The Way We Tally Becomes the Tale: the Impact of Selection Strategies on the Inferred Evolution of Little Red Dots Across Cosmic Time}",
      journal = {arXiv e-prints},
         year = 2026,
        month = apr,
          eid = {arXiv:2604.07138},
        pages = {arXiv:2604.07138},
          doi = {10.48550/arXiv.2604.07138},
archivePrefix = {arXiv},
       eprint = {2604.07138},
 primaryClass = {astro-ph.GA},
       adsurl = {https://ui.adsabs.harvard.edu/abs/2026arXiv260407138R}
}

@ARTICLE{Ivey2026,
       author = {{Ivey}, L.~R. and {D'Eugenio}, F. and {Maiolino}, R. and {Isobe}, Y. and {Juod{\v{z}}balis}, I. and {Koudmani}, S. and {Perna}, M. and {Zhang}, S. and {Bromm}, V. and {Bunker}, A.~J. and {Carniani}, S. and {Fabian}, A.~C. and {Inayoshi}, K. and {Ji}, X. and {Jones}, G.~C. and {Liu}, B. and {Pascalau}, R. and {Rinaldi}, P. and {Robertson}, B. and {Scholtz}, J. and {Tacchella}, S.},
        title = "{The Cliff: A Metal-Poor Little Red Dot Hosting an Overmassive Black Hole at z = 3.55}",
      journal = {\mnras},
         year = 2026,
        month = jun,
          doi = {10.1093/mnras/stag1220},
archivePrefix = {arXiv},
       eprint = {2604.09177},
 primaryClass = {astro-ph.GA},
       adsurl = {https://ui.adsabs.harvard.edu/abs/2026MNRAS.tmp.1141I}
}

@ARTICLE{Escott2025,
       author = {{Escott}, Emmy L. and {Morabito}, Leah K. and {Scholtz}, Jan and {Hickox}, Ryan C. and {Harrison}, Chris M. and {Alexander}, David M. and {Arnaudova}, Marina I. and {Smith}, Daniel J.~B. and {Duncan}, Kenneth J. and {Petley}, James and {Kondapally}, Rohit and {Calistro Rivera}, Gabriela and {Kolwa}, Sthabile},
        title = "{Unveiling AGN outflows: [O III] outflow detection rates and correlation with low-frequency radio emission}",
      journal = {\mnras},
         year = 2025,
        month = jan,
       volume = {536},
       number = {2},
        pages = {1166-1179},
          doi = {10.1093/mnras/stae2645},
archivePrefix = {arXiv},
       eprint = {2411.19326},
 primaryClass = {astro-ph.GA},
       adsurl = {https://ui.adsabs.harvard.edu/abs/2025MNRAS.536.1166E}
}

@ARTICLE{Hutchings2002,
       author = {{Hutchings}, J.~B. and {Crenshaw}, D.~M. and {Kraemer}, S.~B. and {Gabel}, J.~R. and {Kaiser}, M.~E. and {Weistrop}, D. and {Gull}, T.~R.},
        title = "{Balmer and He I Absorption in the Nuclear Spectrum of NGC 4151}",
      journal = {\aj},
         year = 2002,
        month = nov,
       volume = {124},
       number = {5},
        pages = {2543-2547},
          doi = {10.1086/344080},
archivePrefix = {arXiv},
       eprint = {astro-ph/0208262},
 primaryClass = {astro-ph},
       adsurl = {https://ui.adsabs.harvard.edu/abs/2002AJ....124.2543H}
}

@ARTICLE{Leighly2019,
       author = {{Leighly}, Karen M. and {Terndrup}, Donald M. and {Lucy}, Adrian B. and {Choi}, Hyunseop and {Gallagher}, Sarah C. and {Richards}, Gordon T. and {Dietrich}, Matthias and {Raney}, Catie},
        title = "{The z = 0.54 LoBAL Quasar SDSS J085053.12+445122.5. II. The Nature of Partial Covering in the Broad-absorption-line Outflow}",
      journal = {\apj},
         year = 2019,
        month = jul,
       volume = {879},
       number = {1},
          eid = {27},
        pages = {27},
          doi = {10.3847/1538-4357/ab212a},
archivePrefix = {arXiv},
       eprint = {1811.04174},
 primaryClass = {astro-ph.GA},
       adsurl = {https://ui.adsabs.harvard.edu/abs/2019ApJ...879...27L}
}

@ARTICLE{Zanchettin2023,
       author = {{Zanchettin}, M.~V. and {Feruglio}, C. and {Massardi}, M. and {Lapi}, A. and {Bischetti}, M. and {Cantalupo}, S. and {Fiore}, F. and {Bongiorno}, A. and {Malizia}, A. and {Marinucci}, A. and {Molina}, M. and {Piconcelli}, E. and {Tombesi}, F. and {Travascio}, A. and {Tozzi}, G. and {Tripodi}, R.},
        title = "{NGC 2992: Interplay between the multiphase disc, wind, and radio bubbles}",
      journal = {\aap},
         year = 2023,
        month = nov,
       volume = {679},
          eid = {A88},
        pages = {A88},
          doi = {10.1051/0004-6361/202245729},
archivePrefix = {arXiv},
       eprint = {2308.04108},
 primaryClass = {astro-ph.GA},
       adsurl = {https://ui.adsabs.harvard.edu/abs/2023A&A...679A..88Z}
}

@ARTICLE{Ishikawa2026,
       author = {{Ishikawa}, Yuzo and {Eilers}, Anna-Christina and {Naidu}, Rohan P. and {Matthee}, Jorryt and {Bordoloi}, Rongmon and {Chisholm}, John and {Greene}, Jenny E. and {Ma}, Yilun and {Oesch}, Pascal A. and {Sun}, Wendy Q. and {Torralba}, Alberto and {Weaver}, John R. and {Wuyts}, Stijn and {Xiao}, Mengyuan},
        title = "{Spatial decomposition of Little Red Dots with JWST/NIRSpec IFU into broad-line red cores and narrow-line blue host galaxies}",
      journal = {arXiv e-prints},
         year = 2026,
        month = jul,
          eid = {arXiv:2607.09647},
        pages = {arXiv:2607.09647},
          doi = {10.48550/arXiv.2607.09647},
archivePrefix = {arXiv},
       eprint = {2607.09647},
 primaryClass = {astro-ph.GA},
       adsurl = {https://ui.adsabs.harvard.edu/abs/2026arXiv260709647I}
}

@ARTICLE{PerezGonzalez2026,
       author = {{P{\'e}rez-Gonz{\'a}lez}, Pablo G. and {Barro}, Guillermo and {Carniani}, Stefano and {D'Eugenio}, Francesco and {Rieke}, George H. and {Tripodi}, Roberta and {Bunker}, Andrew J. and {Ji}, Xihan and {Marques-Chaves}, Rui and {Schaerer}, Daniel and {Venturi}, Giacomo and {Ar{\'e}valo-Gonz{\'a}lez}, Flor and {Arribas}, Santiago and {Rinaldi}, Pierluigi and {Rodr{\'\i}guez Del Pino}, Bruno and {Witstok}, Joris and {Bhatawdekar}, Rachana and {Boogaard}, Leindert A. and {Charlot}, Stephane and {Chevallard}, Jacopo and {Costantin}, Luca and {Curti}, Mirko and {Curtis-Lake}, Emma and {Daddi}, Emanuele and {Davis}, Kelcey and {Dickinson}, Mark and {Donnan}, Callum T. and {Donnan}, Fergus R. and {Dunlop}, James S. and {Eisenstein}, Daniel J. and {Ferguson}, Henry C. and {Fern{\'a}ndez Aranda}, Rom{\'a}n and {Finkelstein}, Steven L. and {Fujimoto}, Seiji and {Gandolfi}, Giovanni and {Giavalisco}, Mauro and {Grogin}, Norman A. and {Hamed}, Mahmoud and {Hirschmann}, Michaela and {Kartaltepe}, Jeyhan S. and {Kocevski}, Dale D. and {Koekemoer}, Anton M. and {Leung}, Gene C.~K. and {Lofaro}, Cristina M. and {Lucas}, Ray A. and {McLeod}, Derek J. and {Melinder}, Jens and {{\"O}stlin}, Goran and {Papovich}, Casey and {Pentericci}, Laura and {P{\'e}rez-D{\'\i}az}, Borja and {Rieke}, Marcia and {Scholtz}, Jan and {Somerville}, Rachel S. and {Stanton}, Thomas M. and {Stevenson}, Struan D. and {Shivaei}, Irene and {Tacchella}, Sandro and {Trump}, Jonathan R. and {{\"U}bler}, Hannah and {Wang}, Xin and {Williams}, Christina C. and {Willmer}, Christopher N.~A. and {Yung}, L.~Y. Aaron and {Zhu}, Yongda},
        title = "{Little Red Dots: One Photometric Tag Concealing Diverse Spectroscopic Flavors of Massive Star Formation and Black Hole Activity}",
      journal = {arXiv e-prints},
         year = 2026,
        month = feb,
          eid = {arXiv:2602.20247},
        pages = {arXiv:2602.20247},
          doi = {10.48550/arXiv.2602.20247},
archivePrefix = {arXiv},
       eprint = {2602.20247},
 primaryClass = {astro-ph.GA},
       adsurl = {https://ui.adsabs.harvard.edu/abs/2026arXiv260220247P}
}

@ARTICLE{Kim2023,
       author = {{Kim}, Changseok and {Woo}, Jong-Hak and {Luo}, Rongxin and {Chung}, Aeree and {Baek}, Junhyun and {Le}, Huynh Anh N. and {Son}, Donghoon},
        title = "{Unraveling the Complex Structure of AGN-driven Outflows. VI. Strong Ionized Outflows in Type 1 AGNs and the Outflow Size-Luminosity Relation}",
      journal = {\apj},
         year = 2023,
        month = dec,
       volume = {958},
       number = {2},
          eid = {145},
        pages = {145},
          doi = {10.3847/1538-4357/acf92b},
archivePrefix = {arXiv},
       eprint = {2310.06928},
 primaryClass = {astro-ph.GA},
       adsurl = {https://ui.adsabs.harvard.edu/abs/2023ApJ...958..145K}
}

@ARTICLE{Naidu2026,
       author = {{Naidu}, Rohan P. and {Matthee}, Jorryt and {de Graaff}, Anna and {Torralba}, Alberto and {Ashall}, Chris and {Katz}, Harley and {Chisholm}, John and {Brammer}, Gabriel and {Dessart}, Luc and {Eilers}, Anna-Christina and {Hviding}, Raphael E. and {Jones}, David O. and {Kokorev}, Vasily and {Leja}, Joel and {Liu}, Hanpu and {Liu}, Zhaoran and {Nandal}, Devesh and {Oesch}, Pascal A. and {Ransome}, Conor L. and {Simcoe}, Robert A. and {Sun}, Wendy Q. and {Weibel}, Andrea and {Xiao}, Mengyuan},
        title = "{Little Red Dots as Intermediate Mass, Super-Eddington Engines: Insights from Type IIn Supernovae and The 1837-1856 Great Eruption of $η$ Carinae}",
      journal = {arXiv e-prints},
         year = 2026,
        month = jun,
          eid = {arXiv:2606.30711},
        pages = {arXiv:2606.30711},
          doi = {10.48550/arXiv.2606.30711},
archivePrefix = {arXiv},
       eprint = {2606.30711},
 primaryClass = {astro-ph.GA},
       adsurl = {https://ui.adsabs.harvard.edu/abs/2026arXiv260630711N}
}

@ARTICLE{Rogerson2011,
       author = {{Rogerson}, Jesse A. and {Hall}, Patrick B. and {Snedden}, Stephanie A. and {Brotherton}, Michael S. and {Anderson}, Scott F.},
        title = "{Chandra X-ray observations of two unusual BAL quasars}",
      journal = {\na},
         year = 2011,
        month = feb,
       volume = {16},
       number = {2},
        pages = {128-137},
          doi = {10.1016/j.newast.2010.07.002},
archivePrefix = {arXiv},
       eprint = {1007.3464},
 primaryClass = {astro-ph.CO},
       adsurl = {https://ui.adsabs.harvard.edu/abs/2011NewA...16..128R}
}

@ARTICLE{Marconi2008,
       author = {{Marconi}, Alessandro and {Axon}, David J. and {Maiolino}, Roberto and {Nagao}, Tohru and {Pastorini}, Guia and {Pietrini}, Paola and {Robinson}, Andrew and {Torricelli}, Guidetta},
        title = "{The Effect of Radiation Pressure on Virial Black Hole Mass Estimates and the Case of Narrow-Line Seyfert 1 Galaxies}",
      journal = {\apj},
         year = 2008,
        month = may,
       volume = {678},
       number = {2},
        pages = {693-700},
          doi = {10.1086/529360},
archivePrefix = {arXiv},
       eprint = {0802.2021},
 primaryClass = {astro-ph},
       adsurl = {https://ui.adsabs.harvard.edu/abs/2008ApJ...678..693M}
}

@ARTICLE{Asada2026,
       author = {{Asada}, Yoshihisa and {Inayoshi}, Kohei and {Fei}, Qinyue and {Fujimoto}, Seiji and {Willott}, Chris},
        title = "{Origins of the UV continuum and Balmer emission lines in Little Red Dots: observational validation of dense gas envelope models enshrouding the AGN}",
      journal = {arXiv e-prints},
         year = 2026,
        month = jan,
          eid = {arXiv:2601.10573},
        pages = {arXiv:2601.10573},
          doi = {10.48550/arXiv.2601.10573},
archivePrefix = {arXiv},
       eprint = {2601.10573},
 primaryClass = {astro-ph.GA},
       adsurl = {https://ui.adsabs.harvard.edu/abs/2026arXiv260110573A}
}

@ARTICLE{Madau2026_wings,
       author = {{Madau}, Piero and {Maiolino}, Roberto and {Scholtz}, Jan and {D'Eugenio}, Francesco},
        title = "{Wings of little dots: Exponential broad lines from a stratified BLR}",
      journal = {arXiv e-prints},
         year = 2026,
        month = apr,
          eid = {arXiv:2604.04216},
        pages = {arXiv:2604.04216},
          doi = {10.48550/arXiv.2604.04216},
archivePrefix = {arXiv},
       eprint = {2604.04216},
 primaryClass = {astro-ph.GA},
       adsurl = {https://ui.adsabs.harvard.edu/abs/2026arXiv260404216M}
}

@ARTICLE{Madau2026_LRD_LBD,
       author = {{Madau}, Piero and {Maiolino}, Roberto},
        title = "{Little Red Dots as Obscured Little Blue Dots: A Super-Eddington Unification Model}",
      journal = {arXiv e-prints},
         year = 2026,
        month = feb,
          eid = {arXiv:2602.22386},
        pages = {arXiv:2602.22386},
          doi = {10.48550/arXiv.2602.22386},
archivePrefix = {arXiv},
       eprint = {2602.22386},
 primaryClass = {astro-ph.GA},
       adsurl = {https://ui.adsabs.harvard.edu/abs/2026arXiv260222386M}
}

@ARTICLE{Madau2026_LRD_LBD2,
       author = {{Madau}, Piero and {Maiolino}, Roberto},
        title = "{Little red dots as obscured little blue dots: relative abundances, luminosities, and black-hole masses}",
      journal = {arXiv e-prints},
         year = 2026,
        month = may,
          eid = {arXiv:2605.05074},
        pages = {arXiv:2605.05074},
          doi = {10.48550/arXiv.2605.05074},
archivePrefix = {arXiv},
       eprint = {2605.05074},
 primaryClass = {astro-ph.GA},
       adsurl = {https://ui.adsabs.harvard.edu/abs/2026arXiv260505074M}
}

@ARTICLE{Brazzini2025M,
       author = {{Brazzini}, Matilde and {D'Eugenio}, Francesco and {Maiolino}, Roberto and {Juod{\v{z}}balis}, Ignas and {Ji}, Xihan and {Scholtz}, Jan and {Chang}, Seok-Jun},
        title = "{Ruling out dominant electron scattering in Little Red Dots' Rosetta Stone using multiple hydrogen lines}",
      journal = {\mnras},
         year = 2025,
        month = nov,
       volume = {544},
       number = {1},
        pages = {L167-L173},
          doi = {10.1093/mnrasl/slaf116},
archivePrefix = {arXiv},
       eprint = {2507.08929},
 primaryClass = {astro-ph.GA},
       adsurl = {https://ui.adsabs.harvard.edu/abs/2025MNRAS.544L.167B}
}

@ARTICLE{Bischetti2023,
       author = {{Bischetti}, Manuela and {Fiore}, Fabrizio and {Feruglio}, Chiara and {D'Odorico}, Valentina and {Arav}, Nahum and {Costa}, Tiago and {Zubovas}, Kastytis and {Becker}, George and {Bosman}, Sarah E.~I. and {Cupani}, Guido and {Davies}, Rebecca and {Eilers}, Anna-Christina and {Farina}, Emanuele Paolo and {Ferrara}, Andrea and {Gaspari}, Massimo and {Mazzucchelli}, Chiara and {Onoue}, Masafusa and {Piconcelli}, Enrico and {Zanchettin}, Maria Vittoria and {Zhu}, Yongda},
        title = "{The Fraction and Kinematics of Broad Absorption Line Quasars across Cosmic Time}",
      journal = {\apj},
         year = 2023,
        month = jul,
       volume = {952},
       number = {1},
          eid = {44},
        pages = {44},
          doi = {10.3847/1538-4357/accea4},
archivePrefix = {arXiv},
       eprint = {2301.09731},
}

@ARTICLE{Li2017,
       author = {{Li}, Jennifer and {Shen}, Yue and {Horne}, Keith and {Brandt}, W.~N. and {Greene}, Jenny E. and {Grier}, C.~J. and {Ho}, Luis C. and {Kochanek}, Chris and {Schneider}, Donald P. and {Trump}, Jonathan R. and {Dawson}, Kyle S. and {Pan}, Kaike and {Bizyaev}, Dmitry and {Oravetz}, Daniel and {Simmons}, Audrey and {Malanushenko}, Elena},
        title = "{The Sloan Digital Sky Survey Reverberation Mapping Project: Composite Lags at z {\ensuremath{\leq}} 1}",
      journal = {\apj},
         year = 2017,
        month = sep,
       volume = {846},
       number = {1},
          eid = {79},
        pages = {79},
          doi = {10.3847/1538-4357/aa845d},
archivePrefix = {arXiv},
       eprint = {1712.02366},
 primaryClass = {astro-ph.GA},
       adsurl = {https://ui.adsabs.harvard.edu/abs/2017ApJ...846...79L}
}

@ARTICLE{Rakic2022,
       author = {{Raki{\'c}}, N.},
        title = "{Kinematics of the H {\ensuremath{\alpha}} and H {\ensuremath{\beta}} broad-line region in an SDSS sample of type-1 AGNs}",
      journal = {\mnras},
         year = 2022,
        month = oct,
       volume = {516},
       number = {2},
        pages = {1624-1634},
          doi = {10.1093/mnras/stac2259},
archivePrefix = {arXiv},
       eprint = {2208.04359},
 primaryClass = {astro-ph.GA},
       adsurl = {https://ui.adsabs.harvard.edu/abs/2022MNRAS.516.1624R}
}

@ARTICLE{Blustin2005,
       author = {{Blustin}, A.~J. and {Page}, M.~J. and {Fuerst}, S.~V. and {Branduardi-Raymont}, G. and {Ashton}, C.~E.},
        title = "{The nature and origin of Seyfert warm absorbers}",
      journal = {\aap},
         year = 2005,
        month = feb,
       volume = {431},
        pages = {111-125},
          doi = {10.1051/0004-6361:20041775},
archivePrefix = {arXiv},
       eprint = {astro-ph/0411297},
 primaryClass = {astro-ph},
       adsurl = {https://ui.adsabs.harvard.edu/abs/2005A&A...431..111B}
}

@ARTICLE{Elvis2000,
       author = {{Elvis}, Martin},
        title = "{A Structure for Quasars}",
      journal = {\apj},
         year = 2000,
        month = dec,
       volume = {545},
       number = {1},
        pages = {63-76},
          doi = {10.1086/317778},
archivePrefix = {arXiv},
       eprint = {astro-ph/0008064},
 primaryClass = {astro-ph},
       adsurl = {https://ui.adsabs.harvard.edu/abs/2000ApJ...545...63E}
}

@ARTICLE{Sok2026,
       author = {{Sok}, Visal and {Nelson}, Erica J. and {Begelman}, Mitchell C. and {Dexter}, Jason and {D'Eugenio}, Francesco and {Greene}, Jenny E. and {Leja}, Joel and {Whitaker}, Katherine E. and {Bunker}, Andrew J. and {P{\'e}rez-Gonz{\'a}lez}, Pablo G. and {Rinaldi}, Pierluigi and {Torralba}, Alberto and {{\"U}bler}, Hannah},
        title = "{Constraints on the Gas Geometry Surrounding Little Red Dots through Narrow-Line Diagnostics}",
      journal = {arXiv e-prints},
         year = 2026,
        month = jun,
          eid = {arXiv:2606.23778},
        pages = {arXiv:2606.23778},
          doi = {10.48550/arXiv.2606.23778},
archivePrefix = {arXiv},
       eprint = {2606.23778},
 primaryClass = {astro-ph.GA},
       adsurl = {https://ui.adsabs.harvard.edu/abs/2026arXiv260623778S}
}




\appendix

\section{H$\alpha$ absorption fits}

Here we show the \Has profile fits to sources not displayed in \autoref{fig:sample_example}.

\begin{figure*}
    \centering
    \includegraphics[width=0.32\linewidth]{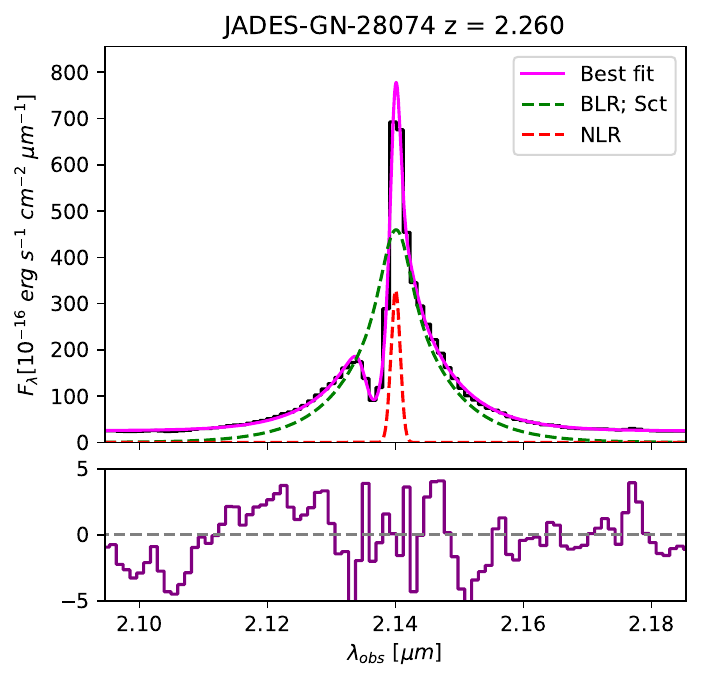}
    \includegraphics[width=0.32\linewidth]{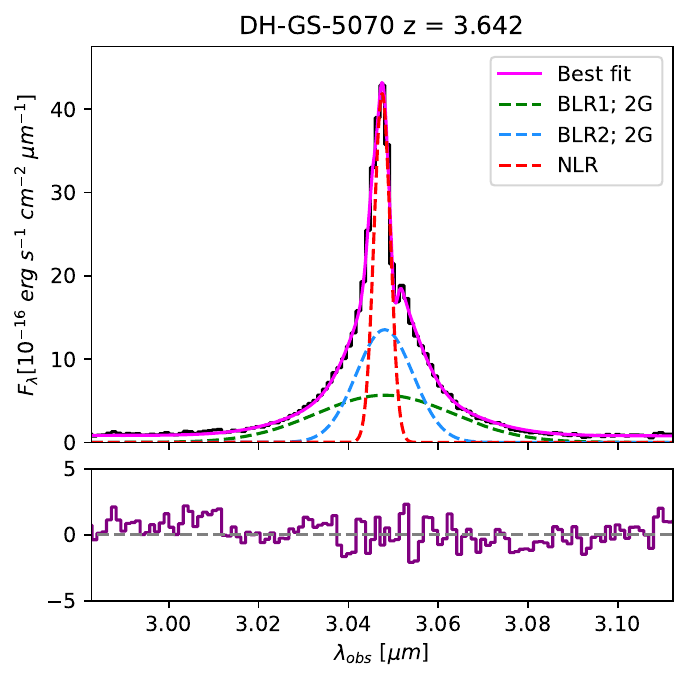}
    \includegraphics[width=0.32\linewidth]{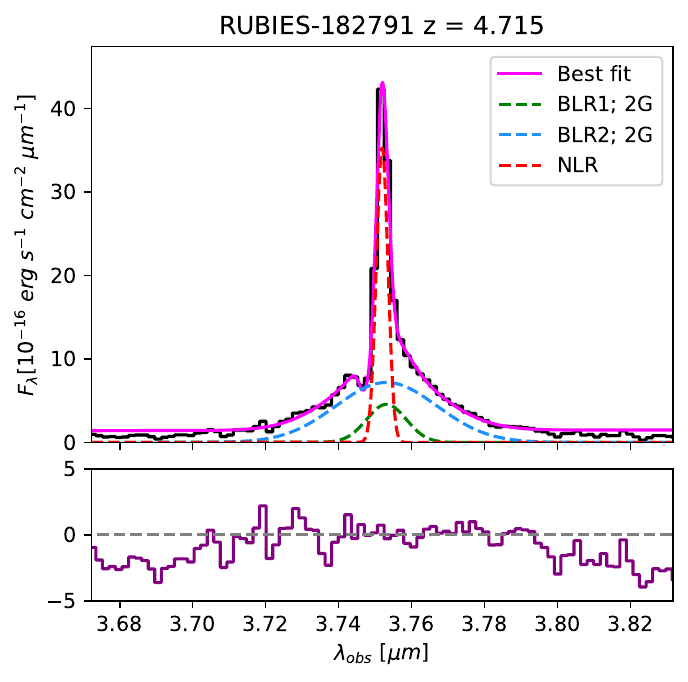}
    \includegraphics[width=0.32\linewidth]{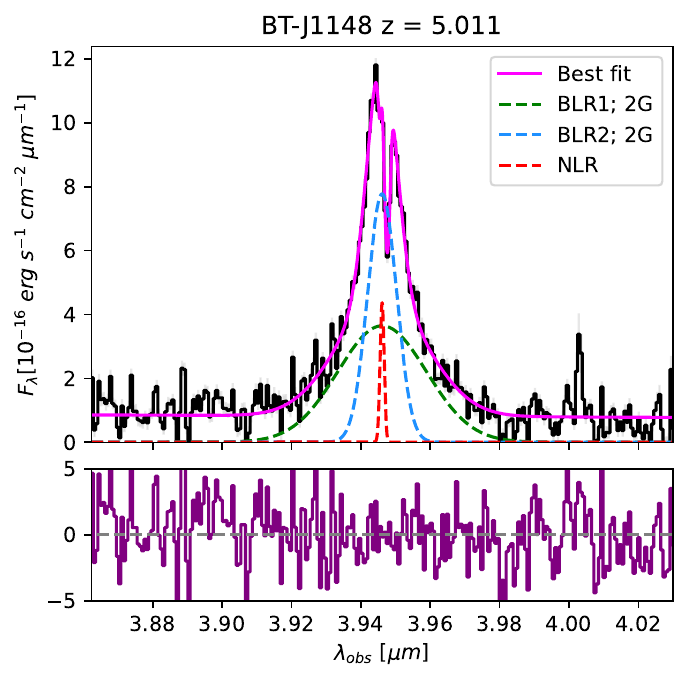}
    \includegraphics[width=0.32\linewidth]{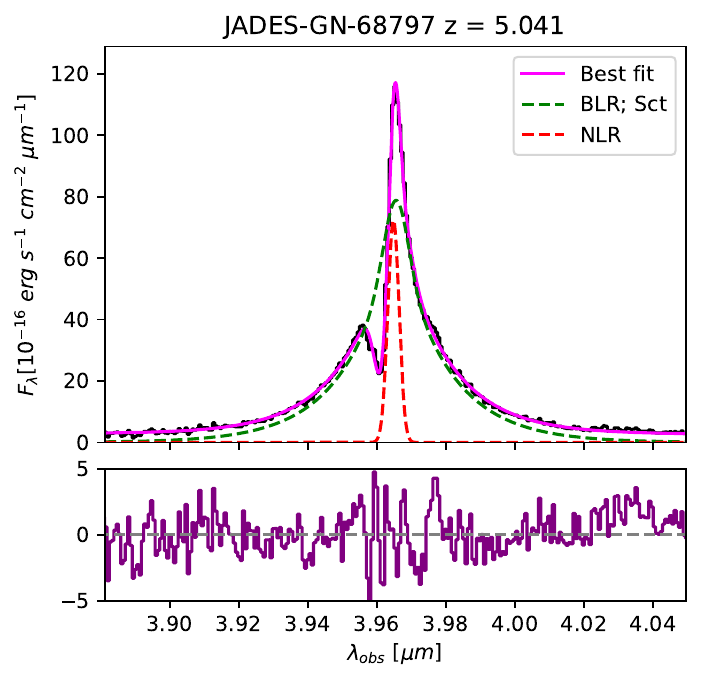}
    \includegraphics[width=0.32\linewidth]{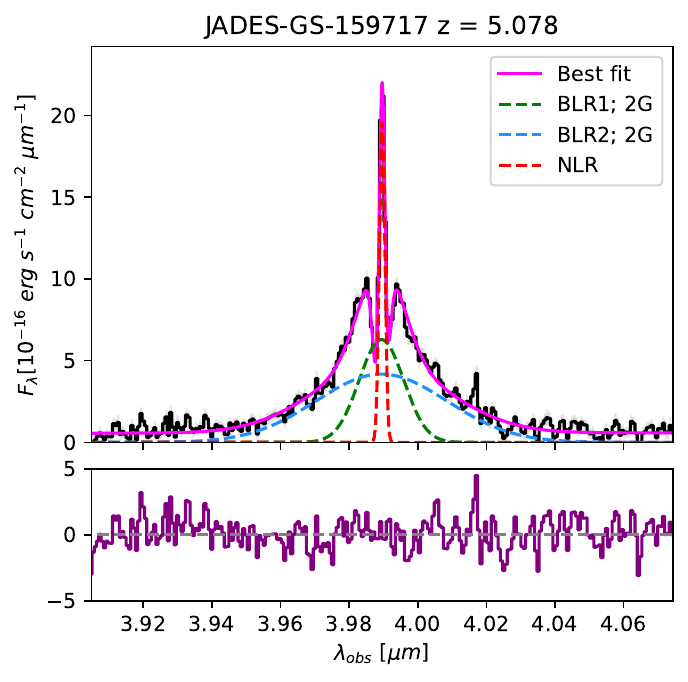}
    \includegraphics[width=0.32\linewidth]{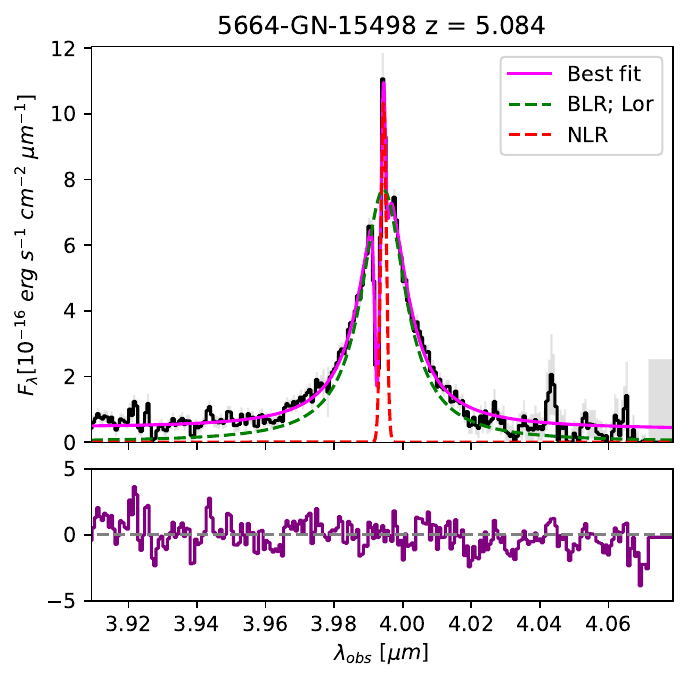}
    \includegraphics[width=0.32\linewidth]{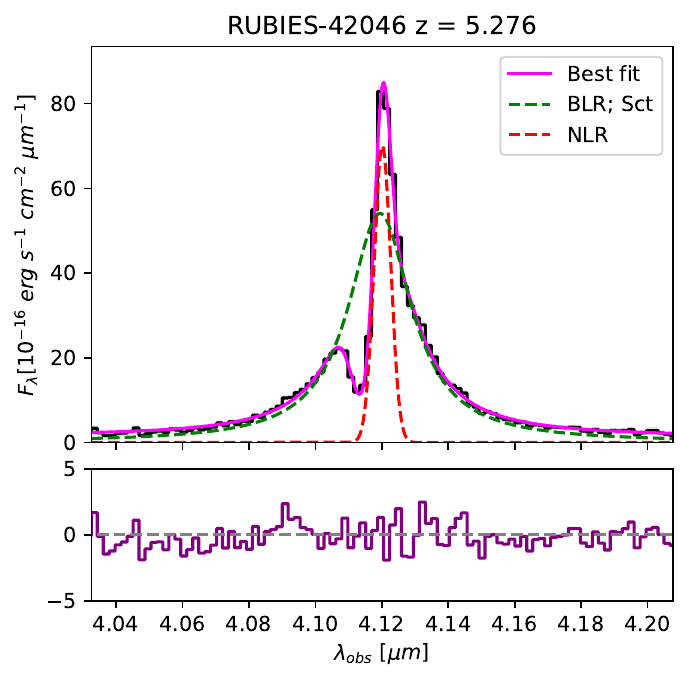}
    \includegraphics[width=0.32\linewidth]{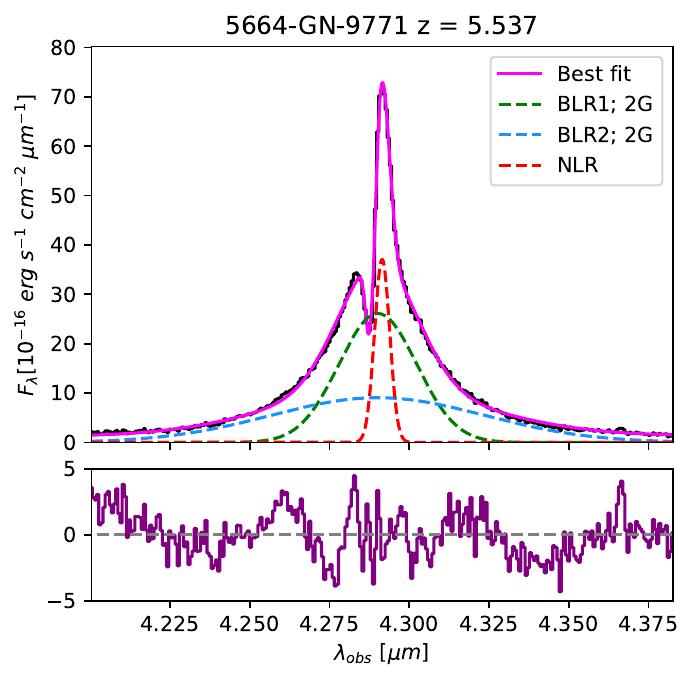}
    \includegraphics[width=0.32\linewidth]{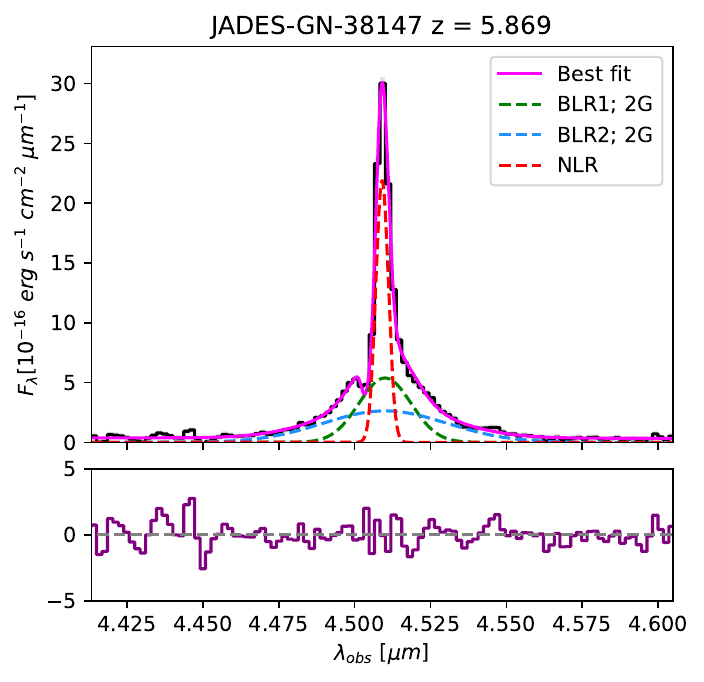}
    \includegraphics[width=0.32\linewidth]{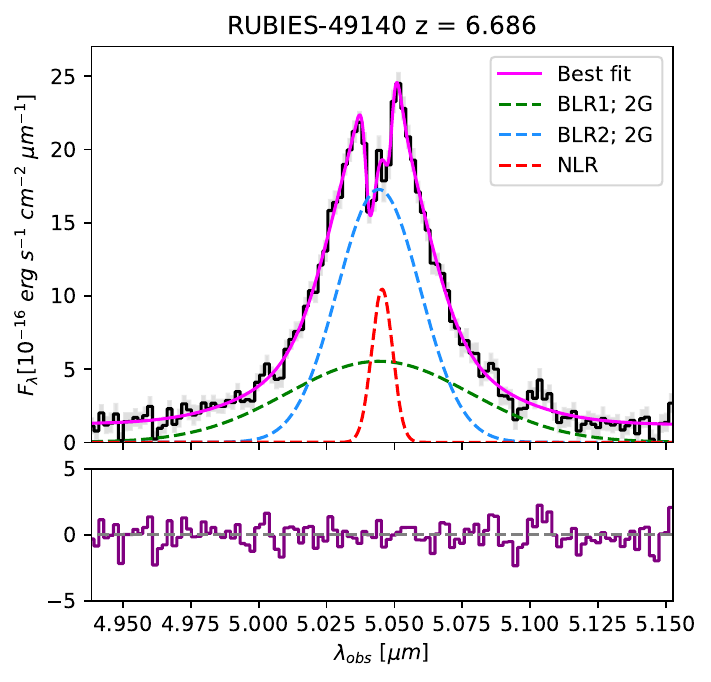}
    \includegraphics[width=0.32\linewidth]{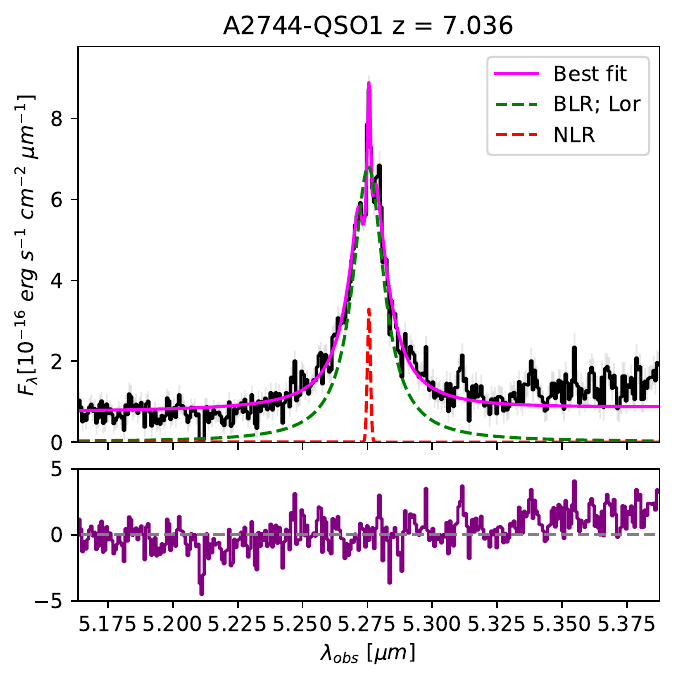}
    \caption{Showcase of the fits to the sample sources not shown in \autoref{fig:sample_example}. The notation is the same as in the latter figure.}
    \label{fig:sample_fits_Ha}
\end{figure*}

\begin{figure}
    \centering
    \includegraphics[width=\linewidth]{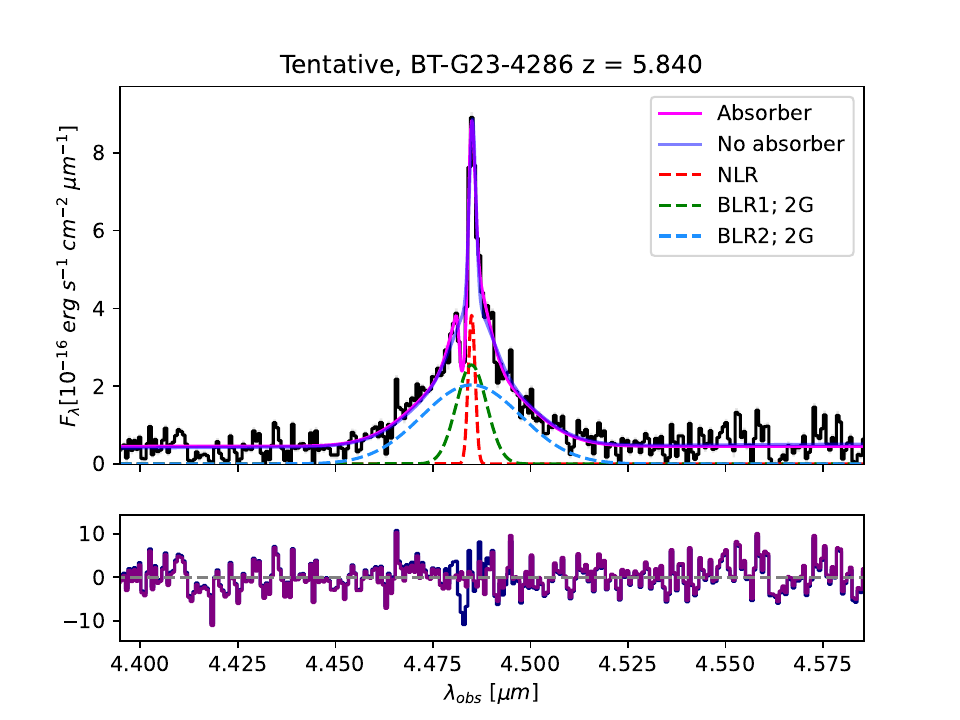}
    \caption{Broad \Has profile of BT-G23-4286, the tentative absorber found in our sample. The lines have the same meaning as in \autoref{fig:sample_fits_Ha} except the translucent blue line shows the fit without an absorbtion feature with the corresponding residuals shown in dark blue in the panel below. While the no absorber fit residual appears significant, the overall scale of the residuals is a factor of 2 larger than in \autoref{fig:sample_fits_Ha}, due to underestimation of errors by the IFU pipeline.}
    \label{fig:tentative_absorber}
\end{figure}

\section{H$\beta$ absorption fits}
Here we show the combined \Hbs and \OIIIs doublet fits for the sample sources with \Hbs absorption.

\begin{figure*}
    \centering
    \includegraphics[width=0.32\linewidth]{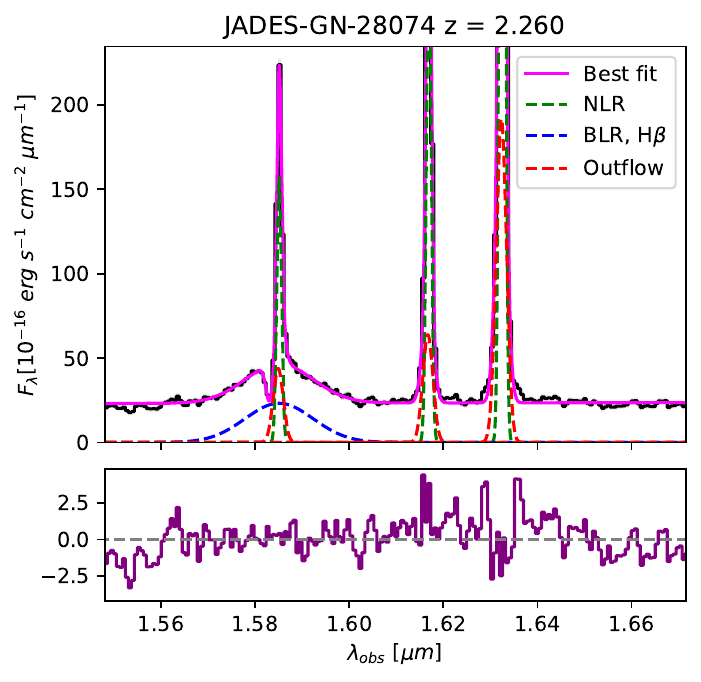}
    \includegraphics[width=0.32\linewidth]{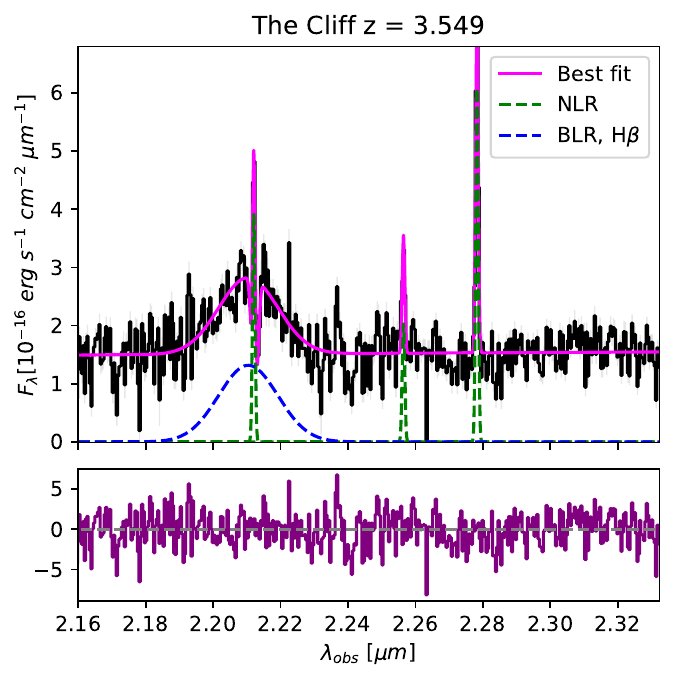}
    \includegraphics[width=0.32\linewidth]{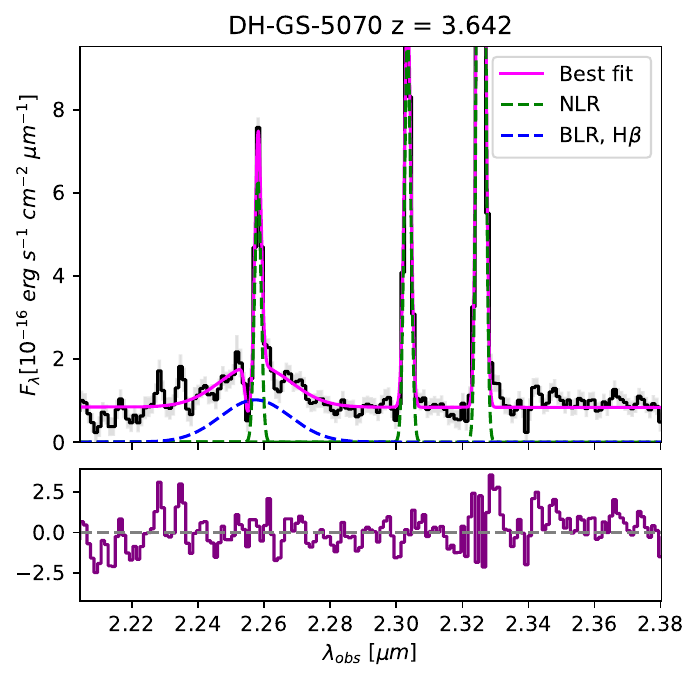}
    \includegraphics[width=0.32\linewidth]{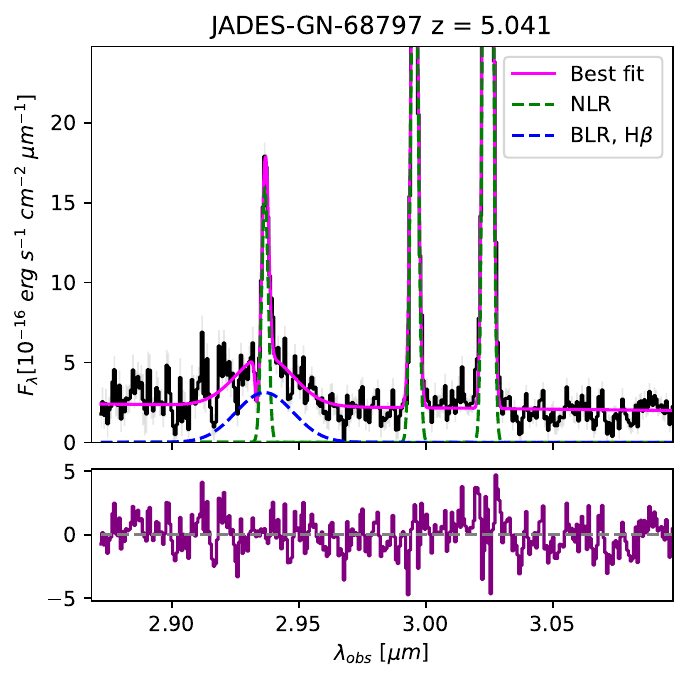}
    \includegraphics[width=0.32\linewidth]{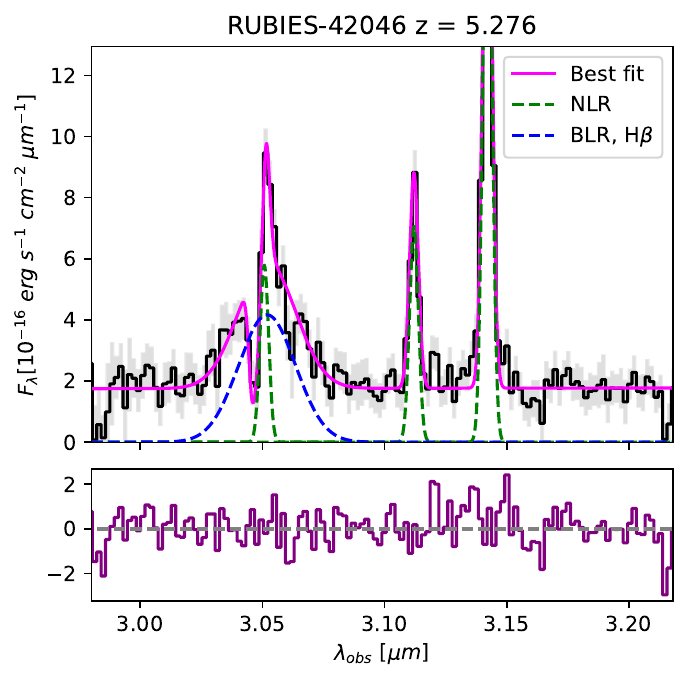}
    \includegraphics[width=0.32\linewidth]{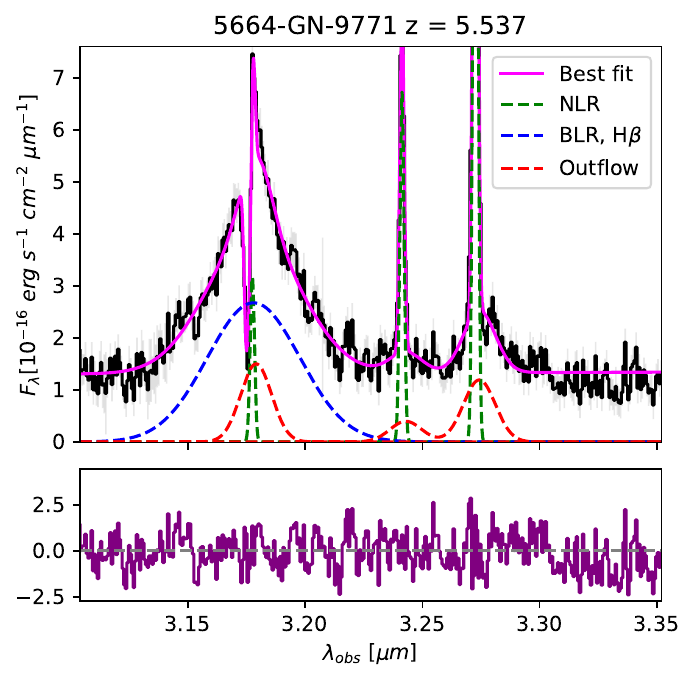}
    \includegraphics[width=0.32\linewidth]{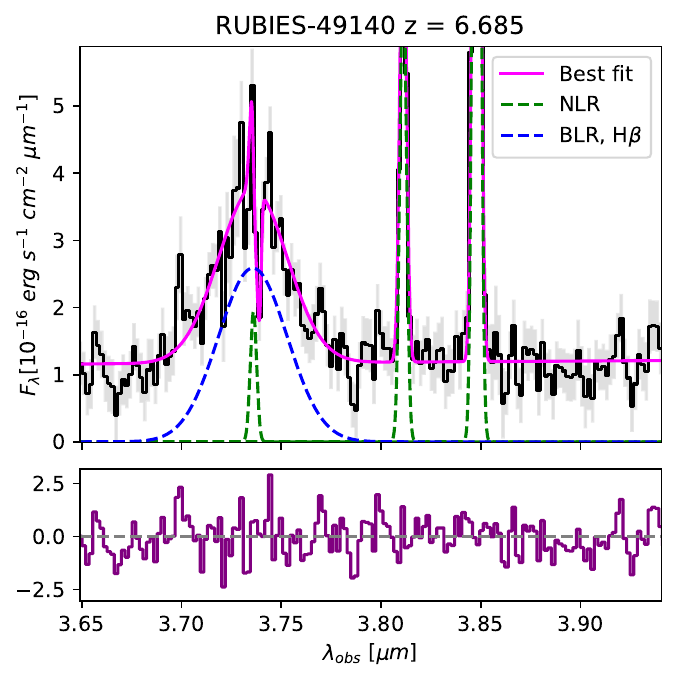}
    \includegraphics[width=0.32\linewidth]{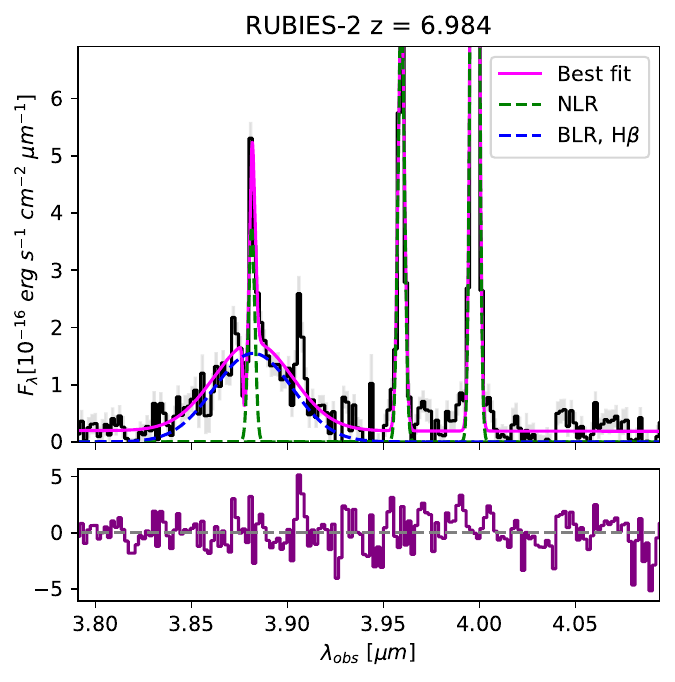}
    \includegraphics[width=0.32\linewidth]{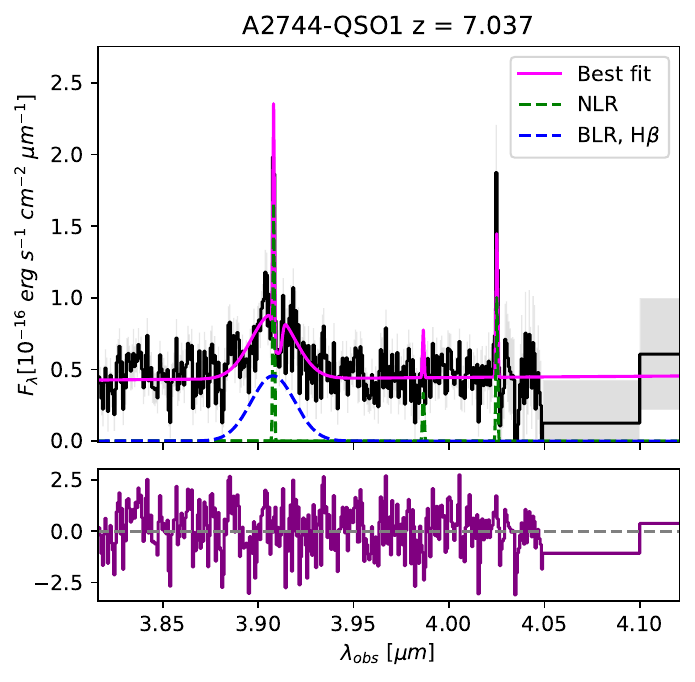}
    \caption{Showcase of the fits to the broad \Hbs and \OIII\ doublet of the sample sources with \Hbs absorption. The combined fit is shown in magenta, the broad \Hbs -- in blue, the narrow lines are plotted in green, while outflows, if present, are plotted in red.}
    \label{fig:sample_fits_Hb}
\end{figure*}


\bsp	
\label{lastpage}
\end{document}